\documentclass[12pt,a4paper]{article}
\usepackage[T1]{fontenc}

\usepackage[english]{babel}

\usepackage[%showframe,
a4paper]{geometry}
\usepackage[backend=biber, style=numeric-comp,sorting=none,maxbibnames=999, maxcitenames=2]{biblatex}

\usepackage[bf,sf]{titlesec} 
\titleformat{\subsubsection}[runin]{\bfseries\sffamily\normalsize}{\thesubsubsection}{1em}{}[]			% in-line subsubsection

\usepackage{amssymb,amsthm,amsfonts,amstext,textcomp,mathrsfs,thmtools,nccmath,mathtools,easybmat,booktabs,tabu,graphicx,enumitem,physics,faktor,xspace,suffix,twoopt,authblk}

\usepackage[dvipsnames]{xcolor}

\definecolor{cblue}{HTML}{2A52BE}
\definecolor{cred}{HTML}{B31B1B}
\definecolor{orangia}{HTML}{FF6600}
\definecolor{marun}{HTML}{800000}
\definecolor{bluu}{HTML}{006699}
\definecolor{blu}{HTML}{0099FF}
\definecolor{rosso}{HTML}{FF0033}
\definecolor{verte}{HTML}{006633}
\definecolor{vertino}{HTML}{148526}
\definecolor{marunin}{HTML}{796E6F}
\definecolor{pesca}{HTML}{f6af87}
\definecolor{battleship}{HTML}{808F87}
\definecolor{equa}{HTML}{D32F2F}
\definecolor{equa}{HTML}{D32F2F}
\definecolor{azu}{RGB}{0, 140, 215}

\usepackage[unicode=true,bookmarks=true,bookmarksopen=false,breaklinks=false,pdfborder={0 0 0},colorlinks=true]{hyperref}

\hypersetup{linkcolor=cred,citecolor=azu,urlcolor=black}

\usepackage[figurename=Fig.,tablename=Tab.]{caption}

\DeclareCaptionLabelFormat{bla}{#1~{#2}}
\usepackage[font={footnotesize}]{subcaption}

\newcommand{\secref}[1]{{\bfseries\sffamily\ref{#1}}}

\newcommand{\figref}[1]{{\sffamily Fig.~\ref{#1}}}
\newcommandtwoopt\figrefs[2][][]{{\sffamily Figs.~\ref{#1}--\ref{#2}}}

\newcommand{\tabref}[1]{{\sffamily Tab.~\ref{#1}}}

\usepackage[scaled=1.15,mathscr]{urwchancal}
\DeclareMathAlphabet{\rsfs}{U}{rsfs}{m}{n}

\newcommand{\e}{\mathrm{e}}
\newcommand{\iu}{\mathrm{i}\mkern1mu}
\newcommand{\I}{\vb i\mkern1mu}
\newcommand{\J}{\vb j\mkern1mu}
\newcommand{\K}{\vb k\mkern1mu}
\newcommand{\id}{\mathbb I}  %identity matrix
\DeclareMathOperator{\diag}{\mathrm{diag}}

\newcommand{\SU}[1][2]{{\rm SU}(#1)}
\newcommand{\U}[1][d]{{\rm U}(#1)}
\newcommand{\PU}[1][2]{{\rm PU}(#1)}
\newcommand{\SO}[1][3]{{\rm SO}(#1)}

\newcommandtwoopt{\Ut}[2][U][t]{#1^{\otimes #2}}
\newcommandtwoopt{\Udt}[2][U][t]{{#1^\dag }^{\otimes #2}}
\newcommandtwoopt{\Udtt}[2][U][t]{%\cramped
	{#1^{\dag \otimes #2}}}

\newcommandtwoopt\cdtt[2][t][d]{\cramped{\qty(\mathbb{C}^{#2}){}^{\otimes #1}}}
\newcommand\cd[1][d]{\mathbb{C}^{#1}}

\newcommandtwoopt{\dU}[2][U][\mu_d]{\dd{#2(#1)}}

\newcommand{\Choi}{{\sf C}}

\newcommandtwoopt{\M}[2][d][t]{\rsfs{M}_{#1}^{#2}}

\DeclareMathOperator{\SW}{{\sf F}}

\newcommand{\PR}[1]{
	\Pi_{#1}
}
\newcommand{\PS}{\PR{\rm s}}
\newcommand{\PT}{\PR{\rm t}}

\newcommand{\f}[1]{f_{#1}}
\newcommand{\fs}{\f{\rm s}}
\newcommand{\ft}{\f{\rm t}}

\newcommand{\B}{\mathcal{P}}
\WithSuffix\newcommand\B*{\B'}

\newcommand{\DD}{\mathcal{D}}
\WithSuffix\newcommand\DD*{\DD'}

\newcommand{\N}{\mathscr{N}}
\newcommand{\Ss}{\mathcal{S}}
\WithSuffix\newcommand\Ss*{\Ss'}

\newcommand{\Bb}{\mathcal{\bar{P}}}
\newcommand{\Db}{\mathcal{\bar{D}}}

\newcommand{\Clif}{\mathsf{\hat{C}}}

\newcommand\ketp[1]{\left|\mkern1mu #1 \mkern1mu\right)}
\newcommand\brap[1]{\left( \mkern1mu #1\mkern1mu \right|}

\newcommand\inner[2]{
	\left(\mkern1mu #1 
	\mkern1mu ,\mkern1mu 
	#2 \mkern1mu \right)
}
\WithSuffix\newcommand\inner*[2]{
	\left(\mkern1mu #1 
	\mkern1mu \middle\vert \mkern1mu 
	#2 \mkern1mu \right)
}

\newcommand\D{di\-men\-sion\-al\xspace}
\newcommand\dep{de\-pend\-ent\xspace}
\newcommand\q{qu\-bit\xspace}
\newcommand\des{de\-sign\xspace}
\newcommand\dess{de\-signs\xspace}
\newcommand\dia{di\-ag\-on\-al\xspace}
\newcommand\rep{rep\-re\-sen\-ta\-tion\xspace}

\usepackage{tikz}
\usetikzlibrary{shapes.geometric,decorations.markings} 

\newcommand{\rossolinea}{%
	\raisebox{2pt}{\tikz{\draw[ultra thick,rosso](0,0) -- (3mm,0);}}%
}

\newcommand{\vertepalla}{%
	\raisebox{0pt}{\tikz{\node[scale=0.5,circle,thick,draw=verte, fill=verte](){};}}%
}

\newcommand{\aranpalla}{%
	\raisebox{0pt}{\tikz{\node[scale=0.4,regular polygon, regular polygon sides=3, semithick,draw=orangia, fill=orangia,fill opacity=0.2](){};}}%
}

\newcommand{\aranlinea}{%
	\raisebox{2pt}{\tikz{\draw[very thick,orangia](0,0) -- (3mm,0);}}%
}

\newcommand{\maruntri}{%
	\raisebox{0pt}{\tikz{\node[scale=0.4,regular polygon, semithick,regular polygon sides=3, draw=marun, fill=marun,fill opacity=0.2](){};}}%
}

\newcommand{\maruninpalla}{%
	\raisebox{0pt}{\tikz{\node[scale=0.5,circle,thick,draw=marunin, fill=marunin](){};}}%
}

\newcommand{\marunpalla}{%
	\raisebox{0pt}{\tikz{\node[scale=0.5,circle,thick,draw=marun, fill=marun](){};}}%
}

\newcommand{\bluupalla}{%
	\raisebox{0pt}{\tikz{\node[scale=0.5,circle,thick,draw=bluu, fill=bluu](){};}}%
}

\newcommand{\singlo}{%
	\raisebox{0pt}{\tikz{\node[scale=0.5,circle,thick,draw=vertino, fill=vertino](){};}}%
}

\newcommand{\neropalla}{%
	\raisebox{0pt}{\tikz{\node[scale=0.5,circle,thick,draw=black, fill=black](){};}}%
}

\newcommand{\azurpalla}{%
	\raisebox{0pt}{\tikz{\node[scale=0.5,circle,thick,draw=blu, fill=blu](){};}}%
}

\newcommand{\oranst}{%
	\raisebox{0pt}{\tikz{\node[scale=0.5, circle, thick, draw=orangia, fill=orangia](){};}}%
}

\newcommand{\sepia}{%
\raisebox{2pt}{\tikz{\draw[very thick, Sepia](0,0) -- (2.5mm,0);}}%
}

\newcommand{\pesca}{%
\raisebox{2pt}{\tikz{\draw[very thick, pesca](0,0) -- (2.5mm,0);}}%
}

\newcommand{\battle}{%
\raisebox{2pt}{\tikz{\draw[very thick, battleship](0,0) -- (2.5mm,0);}}%
}

\newcommand{\bricco}{%
\raisebox{0pt}{\tikz{\node[scale=0.5, circle, thin, draw=black, fill=BrickRed](){};}}%
}
	
\newcommand{\oranbru}{%
\raisebox{0pt}{\tikz{\node[scale=0.5, circle, thin, draw=black, fill=BurntOrange](){};}}%
}
	
\newcommand{\forest}{%
\raisebox{0pt}{\tikz{\node[scale=0.5, circle, thin, draw=black, fill=ForestGreen](){};}}%
}
	
\newcommand{\midnight}{%
\raisebox{0pt}{\tikz{\node[scale=0.5, circle, thin, draw=black, fill=MidnightBlue](){};}}%
}

\newcommand{\equa}{%
	\raisebox{0pt}{\tikz{\node[scale=0.5, circle, thick,draw=equa, fill=equa, fill opacity=0.1](){};}}%
}

\newcommandtwoopt{\tetra}[2][1.9ex][-.75pt]{%
	\raisebox{#2}{\includegraphics[height=#1]{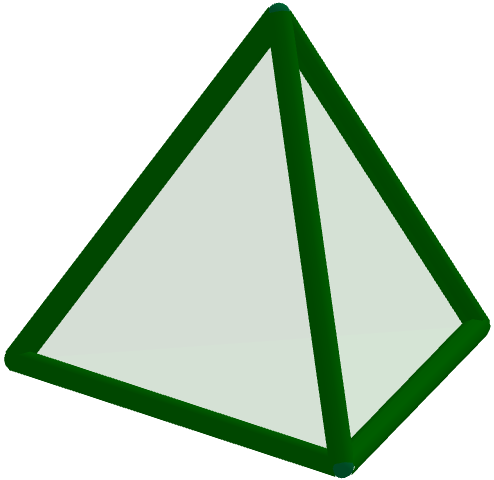}}%
}

\usepackage{tocloft}
\cftsetindents{sec}{0em}{1.8em}
\cftsetindents{subsec}{1.8em}{2.2em}

\DeclareCiteCommand{\citazione}
{\usebibmacro{prenote}}
{\printnames[labelname][1-1]{author}% Imposta il formato "Bennett et al"
	\setunit{\addspace}% Spazio prima della parentesi
	\printtext[parens]{% Apre la parentesi tonda
		\printdate% Anno (1996)
		\setunit{\addspace}%
		\printfield{journaltitle}% Rivista (Phys. Rev. A)
		\setunit{\addspace}%
		\mkbibbold{\printfield{volume}}% Volume in GRASSETTO (54)
		\setunit{\addspace}%
		\printfield{pages}% Pagine (3824–51)
}}
{\multicitedelim}
{\usebibmacro{postnote}}

\begin{document}

\renewcommand\Affilfont{\small}

\author[1,2,*]{Rocco Maggi}
\author[3,1,2]{Cosmo Lupo}
\author[1,2]{Saverio Pascazio}

\affil[1]{Dipartimento Interateneo di Fisica, Universit\`a di Bari, 70126, Bari, Italy}
\affil[2]{INFN, Sezione di Bari, 70126 Bari, Italy}
\affil[3]{Dipartimento  Interateneo di Fisica, Politecnico di Bari, 70126, Bari, Italy}
\affil[*]{Email: \texttt{rocco.maggi@uniba.it}}

\date{\today}

\title{\bfseries\sffamily 
Explicit construction of $2$-\dess of $\U[2]$\\ from the theory of angular momentum}

\maketitle

\begin{abstract}
The main aim of this work is to present an explicit construction of $2$-\dess of $\U[2]$, relying only on a tool that belongs to every physicist's toolbox: the theory of angular momentum.
Unitary designs are a rich and fundamental mathematical topic, with numerous fruitful applications in quantum information science and technology.
In this work we take a peek under the hood. We begin with a minimal set of definitions and properties. Then we derive all $1$-\dess of $\U[2]$ of minimum size. Finally, we set out, step by step, a completion procedure extending such $1$-\dess to $2$-\dess.
In particular, starting from the Pauli basis --- the prototypical unitary $1$-\des\ --- one ``naturally'' obtains the $2$-\des originally employed in~\citazione{Bennett}.
The present work also serves as a gentle and largely self-contained introduction to the subject.
\end{abstract}

\tableofcontents

\section{Introduction}\label{sec:intro}
In this work we take a gentle peek under the hood of unitary designs.

The term ``unitary'' is straightforward. It refers to the unitary group $\U$, a rich mathematical %(algebraic, topological, and differentiable) 
structure --- a compact and connected Lie group, and one of Weyl's classical groups. Owing to the interplay between its algebraic and topological properties, $\U$ is endowed with a unique Borel probability measure, the normalized Haar measure $\mu_d$, which is invariant under left and right multiplication and inversion. That is, if $\Ss\subset \U$ is a measurable set, and $V\in\U$, then the sets $V\Ss=\{VU \,\vert\, U\in \Ss\}$, $\Ss V=\{UV \,\vert\, U\in \Ss\}$ (the left and right translates of $\Ss$ by $V$), and $\Ss^{-1}=\{U^{-1} \,\vert\, U\in \Ss\}$ (the set of inverses of the elements of $\Ss$) are measurable, and have the same measure as $\Ss$, namely, $\mu_d(\Ss) = \mu_d(V\Ss)
= \mu_d(\Ss V) = \mu_d\qty(\Ss^{-1})$. %The measure $\mu_d$ cannot distinguish between $E$, and its translates, or its symmetric, precisely like the Lebesgue measure on the line cannot --- the obvious difference being that the composition law on $\U$ is not commutative for $d>1$, hence the need to distinguish between left and right multiplication. 
These invariance properties are inherited by integration: if $f$ is an integrable function on $\U$, then the functions $U\mapsto f(VU)$, $U\mapsto f(UV)$, and $U\mapsto f\qty(U^{-1})= f\qty(U^\dag)$ are all integrable, and have the same integral as $f$. Continuous functions on $\U$ are measurable (since $\mu_d$ is a Borel measure), hence integrable (because $\U$ is compact). In other words, continuous functions can be averaged over the unitary group.

The term ``design'' is more subtle. For instance, even in the over one thousand pages of~\citetitle*{Handbook}~\cite{Handbook}, no attempt is made to ``define'' such notion --- a choice that is hardly unintentional. Following such example, we content ourselves with recalling the closest classical relative of a unitary design: a spherical $t$-\des in $d$ dimensions is a finite subset of the unit sphere in $\mathbb{R}^{d}$ such that the integral average over $\mathbb{S}^{d-1}$ of any polynomial $p(x_1,\ldots,x_d)$ of degree at most $t$ equals its average over the subset~\cite{Bannai2009}.

The concept of unitary design can be conveniently introduced by an example. Let us consider a one-\q  state $\rho$, singled out by its polarization vector $\vb*r=\tr (\vb*X \rho)$, which lies in the 3-\D real unit ball (the Bloch ``sphere''). Given a unitary matrix $U\in \U[2]$, the transformation $\rho\mapsto U\rho U^\dag$ has the sole effect of rotating $\vb*r$. Our intuition tells us that, for any possible input polarization, the probability distribution of the direction of the output polarization should be uniform. %on $U$.
Because there is no privileged output direction, the average of the states $U\rho U^\dag$ over all unitaries should be completely depolarized. Indeed,
\begin{equation}
\int_{\U[2]} U\rho U^\dag \dU[U][\mu_2]=\frac{\id_2}{2}\,,
\end{equation}
since, by left invariance of the Haar integral, the left-hand side is invariant under conjugation by every unitary, hence a scalar. On the other hand, one has the well-known identity
\begin{equation}\label{eq:WellKnown}
\frac{1}{4}
\qty(\rho+X\rho X+Y\rho \,Y+ Z\rho \,Z)
=
\frac{\id_2}{2}
\,,
\end{equation}
arising, for instance, when discussing the Kraus representations of the completely depolarizing channel on a single qubit. Consequently, the average of the function $U\mapsto U\rho U^\dag$ over the entire unitary group (with respect to the Haar measure) coincides with its discrete average over the Pauli basis $\{\id_2, X,Y,Z\}\subset \U[2]$. 
In general, a unitary $t$-\des of $\U$ is required to play the same role played by the Pauli basis for a single ($t=1$) qubit ($d=2$): a finite set $\Ss\subset\U$ is a $t$-\des of $\U$ if and only if
\begin{equation}\label{eq:DesignIntroTwirl}
\frac{1}{\abs{\Ss}} \sum_{U\in \Ss} \Ut\rho \,\Udt
=\int_{\U} \Ut\rho\Udt \dU\,,
\end{equation}
for all states $\rho$ of $\cdtt$. Both sides of the above equation are called \emph{twirls} of $\rho$. 
Twirling a state over $\U$ yields a ``more symmetric'' version of the input --- in particular, the output is completely mixed if and only if $t=1$. Therefore, a unitary $t$-\des of $\U$ is any finite subset of $\U$ that twirls an arbitrary input state of $\cdtt$ in the same way as the entire unitary group. 

Finally, by considering the defining identity~\eqref{eq:DesignIntroTwirl} in terms of matrix elements in the standard basis of $\cdtt$, it is easy to see that it is equivalent to
\begin{equation}\label{eq:DesignIntroPoly}
\frac{1}{\abs{\Ss}} \sum_{U\in \Ss} p(U)
=\int_{\U} p(U) \dU,
\end{equation}
for all polynomial functions $U\mapsto p(U)$ on $\U$ that are homogeneous of degree $t$ in both the entries of $U$ and their complex conjugates. It is also easy to observe that a unitary $(t+1)$-\des is also a $t$-\des. But then a unitary design is a finite set of unitaries that exactly replicates the averages of the Haar measure on homogeneous polynomials up to a given degree, and it may therefore be regarded as an analogue of a spherical design.

Let us provide some historical context by reviewing  the main developments in the theory of unitary designs. 

As far as the early stages are concerned, we rely on the careful reconstruction presented in~\cite{gross}. For quite a long time, the problem was essentially considered in the formulation of Eq.~\eqref{eq:DesignIntroTwirl}, and for the case $t=2$. Twirling operations on bipartite states became relevant to quantum information theory in the 1980s --- especially after a celebrated paper by Werner which led to the notion of Werner state~\cite{Werner}. A set with property~\eqref{eq:DesignIntroTwirl} for $d=2$ was employed by Bennett, DiVincenzo, Smolin, and Wootters~\cite{Bennett}. Subsequently, DiVincenzo, Leung, and Terhal showed that the $n$-\q  Clifford group (defined as the normalizer of the $n$-\q  Pauli group, and already well-known in quantum information theory, due to its prominent role in quantum error correction~\cite{Gottesman1998}) yields a set with property~\eqref{eq:DesignIntroTwirl} for $d=2^n$~\cite{DiVincenzo}. This result was later generalized by Chau to $d=p^n$, for any prime integer $p$~\cite{Chau}.

Since the first half of the 2000s, several authors had begun adapting the concept of spherical design to quantum mechanics, in the guise of complex spherical designs (obtained by replacing the unit sphere of $\mathbb{R}^d$ with that of $\cd$, whose ``natural'' measure is invariant under $\U$) and complex projective designs (obtained by also identifying points of the unit sphere of $\cd$ that differ by a phase factor, yielding the $(d-1)$-\D complex projective space). This line of work led to the study of mutually unbiased bases and symmetric informationally complete POVMs (see~\cite{gross} for references).

Dankert went a step further when, in his master's thesis, he replaced the complex sphere or the complex projective space with the unitary group itself~\cite{Dankert}. While studying complex projective 2-\dess, he was led to consider the problem of finding a finite subset $\Ss\subset \U$ satisfying
\begin{equation}\label{eq:DesignIntroChan}
\frac{1}{\abs{\Ss}}\sum_{U\in \Ss}
\hat U^\dag\rsfs{E}\hat U
=\int_{\U} \hat U^\dag\rsfs{E}\hat U\dU,
\end{equation}
for all superoperators $\rsfs{E}$ on $\cd$, where $\hat U$ is the unitary channel associated with $U\in \U[d]$, namely, $\rho\mapsto U\rho U^\dag$. The above equation involves twirls of superoperators, rather than of states, and its right-hand side had already been studied in~\cite{Emerson_2005}. Dankert observed that his problem is equivalent to those set by Eqs.~\eqref{eq:DesignIntroTwirl} and~\eqref{eq:DesignIntroPoly} for $t=2$, and interpreted these equivalent problems as defining a new kind of design, which he called \emph{unitary design}. These considerations were later generalized in~\cite{Dankert2009}.

Once the connection with spherical designs was unearthed, attempts were made to borrow some of their standard mathematical techniques. In particular, Gross, Audenaert, and Eisert successfully adapted the concept of frame potential to unitary $2$-\des, a feature soon after generalized to arbitrary $t$ by Scott~\cite{gross,Scott}. Intuitively, any two unitaries repel each other, and the repulsive energy, at order $t$, is proportional to the $t$-th power of the squared modulus of their Hilbert--Schmidt overlap. The average repulsion of a finite set of unitaries, its \emph{frame potential} of order $t$, is bounded from below by the frame potential of the entire unitary group, defined as the corresponding Haar average over $\U$ --- an integer depending only on $t$ and $d$. A finite set of unitaries is a unitary $t$-\des if and only if its frame potential of order $t$ saturates this inequality, and, more generally, the difference between the two potentials is a measure of how far that set is from being a unitary $t$-\des. In this way, unitary designs can be conceived as finite nets of evenly distributed unitaries, just as spherical designs can be regarded as finite nets of evenly distributed points on the unit sphere. Simple considerations from group representation theory then suggest looking for unitary designs of $\U$ among the images of finite groups under $d$-\D unitary representations. The introduction of the frame potential was not the only merit of~\cite{gross}: although specifically devoted to 2-\dess, this work represented a comprehensive attempt to systematize the study of unitary designs, drawing particular attention to fundamental aspects such as existence, explicit construction, and determination of the minimum size.

Scott immediately pointed out (on the grounds of an old result on spherical designs~\cite{seymour}) that the problem of existence had already been settled: for all possible values of $d$ and $t$ unitary designs do exist, and of almost every size~\cite{gross2}. Then Roy and Scott used $\U$ representation theory to derive a lower bound on the size of a unitary $t$-\des of $\U$. In  particular, the Clifford bound --- conjectured in~\cite{gross}, stating that a unitary 2-\des of $\U$ must contain at least $d^4-d^2$ elements --- followed as a special case~\cite{RoyScott}.

Although the case $t>2$ began attracting attention in the second half of the 2000s, it took another decade before Zhu showed that the Clifford group yields 3-\dess for $d=2^n$~\cite{Zhu}. However, he also presented some negative results: in contrast to the case of 2-\dess, the generalized Clifford group does not yield a 3-\des, unless the qudit dimension $d$ is a power of 2. Moreover, Zhu, Kueng, Grassl, and Gross showed that the Clifford group does not yield 4-\dess, although it ``fails gracefully'' to do so~\cite{ZhuGracefully}. 

The problem of explicit construction was addressed, at least in principle, even more recently, when Bannai, Nakata, Okuda, and Zhao presented an iterative procedure for constructing unitary designs for all possible values of $t$ and $d$~\cite{Bannai2022}.

Even if the present work has a rather different focus, a few words are in order about the relevance of unitary designs to applications. A quick glance at the references of any paper on the topic suffices to realize that unitary designs are ``a ubiquitous tool of quantum information science''~\cite{Zhu}, with applications across such diverse contexts as  randomized benchmarking, quantum process tomography, quantum state tomography, decoupling, quantum cryptography, data hiding, tensor networks (see, e.g.,~\cite{Haferkamp2022,Bannai2022,ZhuGracefully,Zhu} for references), and, more recently, entanglement detection~\cite{stair}. 
But why is it so? Because of the symmetries of the Haar measure, many algorithms and protocols rely on the ability to generate Haar-random unitaries. However this process is inefficient. Quantum circuits are constructed from universal gate sets, which are finite. Any transformation realized by a circuit is an element of the group generated by the elementary gates, and such group is dense in $\U$. The number of elementary gates required to implement most unitaries grows exponentially with the number of qubits. 
The intuition behind unitary $t$-\des is that they are finite subsets of $\U$ that replicate the behavior of the Haar measure, provided that the task at hand does not exceed a certain threshold of complexity, quantified by the parameter $t$. In other words, as long as this task is simple enough, in order to get sensible results we do not need to sample from $\epsilon$-nets of $\U$ (i.e., finite sets of unitaries whose $\epsilon$-neigh\-bor\-hoods cover the whole group), but only from much sparser nets, whose unitaries only need to be ``evenly distributed'' (cf., e.g.,~\cite{Dankert2009,Webb,Haferkamp2022}).

This work adopts a bottom-up, constructive approach. Its structure is as follows. We start by considering $1$-\dess of $\U[2]$, and obtain the general form of those of minimum size, whose best-known instance is the Pauli basis. Then we lay down a step-by-step, explicit construction that extends any $1$-\des of minimum size to a $2$-\des of $\U[2]$, also of minimum size. When this procedure is applied to the Pauli basis, it yields precisely the $2$-\des considered in~\cite{Bennett}. The resulting structure --- characterized by remarkable geometric and algebraic properties, which are also briefly illustrated --- is, of course, not new. What is new, to the best of our knowledge, is the derivation itself. By focusing on the qubit case ($d=2$), we are able to develop the entire discussion using a single, yet powerful, tool that is (or should be) in the cultural baggage of every physicist: the theory of angular momentum.

It is often said that all it takes to unveil many of the fundamental traits of quantum mechanics is stepping up from one qubit to two~\cite{Preskill}. The motivations for this work are along the same lines: several general aspects of unitary designs already emerge by working with qubits ($d=2$), and stepping up from $t=1$ to $t=2$. Despite considering just this specific case, the reader should be able to realize first-hand the feasibility of unitary designs, and grasp the reasons behind it.

As far as unitary designs are concerned, nowadays the constructive aspects are largely confined to a few advanced and highly technical mathematical works --- unfortunately, but inevitably so. In much of the literature, the reader is presented with a result to check, rather than a process to follow. Our goal of taking ``a peek under the hood'' allows us to bridge this gap, offering a hands-on construction that complements both the more abstract and the purely result-oriented approaches. In the end, the present work is reminiscent of, and perhaps also close in spirit to, some early papers on the subject, overlapping with~\cite{Bennett} and subsequent articles such as~\cite{Aravind}.

Finally, as to the adjective ``gentle'', this work is conceived to be readable by a broad audience, including those encountering unitary designs for the first time. To this end, the prerequisites have been kept as minimal as possible: essentially, the theory of angular momentum as presented in any good textbook on quantum mechanics (e.g.,~\cite{saku,cohen}), together with some basic notions that are usually covered in the introductory chapters of any good textbook on quantum information theory (e.g.,~\cite{watrous,renes}).

\section{Preliminaries}\label{sec:Pre}

\subsection{Notation and conventions}
Given two finite-\D (complex) Hilbert spaces $\mathscr H_1$ and $\mathscr H_2$, the space of all (linear) operators from $\mathscr H_1$ to $\mathscr H_2$, denoted by ${\rm L}(\mathscr H_1,\mathscr H_2)$, equipped with the Hilbert--Schmidt inner product $\inner{A}{B}= \tr(A^\dag B)$, for $A,B \in {\rm L}(\mathscr H_1,\mathscr H_2)$, is itself a Hilbert space.

For a finite-\D Hilbert space $\mathscr H$, we write ${\rm L}(\mathscr H)$ for ${\rm L}(\mathscr H,\mathscr H)$. This is the associative and unital algebra of all operators on $\mathscr H$. This structure is endowed with a conjugate-lin\-e\-ar and isometric involution, the adjoint operation $A\mapsto A^\dag$, whose interplay with the multiplication and the inner product is expressed by $(LR)^\dag= R^\dag L^\dag$, and $\inner{A}{LR} = \inner{L^\dag A}{R} = \inner{AR^\dag }{L}$, for $A,L,R \in {\rm L}(\mathscr H)$. 
The (convex and compact) set of all states, or density operators, on $\mathscr H$ is denoted by ${\rm D}(\mathscr H)$. Given a state $\rho$, and a family $(\PR{x})$ of orthogonal projections that are pairwise orthogonal and resolve the identity (explicitly, $\PR{x}^\dag \PR{y}  = \delta_{xy} \PR{x} $, and $\sum_x\PR{x} = \id_{\mathscr H}$), we set $\f{x}(\rho) = \inner{\PR{x} }{\rho}=\tr (\PR{x} \rho)$. These numbers form the probability distribution $(\f{x}(\rho))$ associated with the projective measurement with measurement operators $(\PR{x})$, performed on the state $\rho$. Throughout this work, the state is often fixed, and we simply write $\f{x}$. Moreover, we often make use of the following properties of the Hilbert--Schmidt norm: $\norm{U}^2 = \dim \mathscr H$ for a unitary operator $U$; $\norm{\Pi}^2 = \tr\Pi =\dim (\Im \Pi)$ for a projection $\Pi$. For a state $\rho$, $\norm{\rho} \le 1$, and $\rho$ is pure if and only if $\norm{\rho} = 1$.

In this work we will essentially deal with two kinds of finite-\D Hilbert spaces: complex Euclidean spaces of the form $\cdtt$, and the corresponding operator spaces ${\rm L}(\cdtt)$. The standard basis of $\cd$, denoted by $\ket{i}$ with $i=0,\dots,d-1$, induces standard bases $\ket{i_1\dots i_t}$ of $\cdtt$, and $\dyad{i_1\dots i_t}{j_1\dots j_t}$ of ${\rm L}(\cdtt)$, where $i_p,j_q=0,\dots,d-1$, and $p,q=1,\dots,t$. These standard bases are orthonormal, and allow us to use basis-\dep notions such as complex conjugation or transposition.

It will be convenient to use Dirac notation also on ${\rm L}(\cdtt)$: indeed, since an operator space is itself a Hilbert space, it naturally provides bras and kets for its elements. For an operator $A$ on $\cdtt$ we denote its ket and bra by $\ketp{A}$ and $\brap{A}$, respectively. As usual, upon identifying a complex number with its multiplication map, we have $\inner{A}{B} = \inner*{A}{B}$, for $A,B\in{\rm L}(\cdtt)$. Moreover, for a superoperator on $\cdtt$ of the form $\rsfs E = \sum_a \ketp{G_a}\!\brap{F_a}$ with $F_a, G_a  \in {\rm L}(\cdtt)$, that is, $ \rsfs E (\rho) = \sum_a \tr(F_a^\dag \rho) G_a$ --- which is general, since in finite dimensions any linear map can be decomposed into rank-one linear maps --- its Choi operator is $\Choi(\rsfs E)=\sum_a G_a \otimes F_a^\ast$.

\subsubsection{Pauli basis.}
The following operators on the one-\q space $\cd[2]$, 
\begin{align}
X=\dyad{0}{1}+\dyad{1}{0},
&&
Y= -\iu \dyad{0}{1}+\iu\dyad{1}{0},
&&
Z=\dyad{0}{0}-\dyad{1}{1},
\end{align}
are called \emph{Pauli matrices}. The standard basis of $\cd[2]$ is an eigenbasis of $Z$, since $Z\ket*{0}=\ket*{0}$ and $Z\ket*{1}=-\ket*{1}$.
The Pauli matrices satisfy the multiplication rule
\begin{equation}\label{eq:PauliComp}
X_i X_j 
= \delta_{ij} \id_2 
+ \iu  \sum_{k=1}^3 \epsilon_{ijk} X_k\,,
\end{equation}
therefore any two distinct Pauli matrices anticommute.
Setting $\vb*{X} = (X, Y, Z)$ and $X_0 = \id_2$, each \emph{Pauli operator} $X_\mu$ ($\mu=0,1,2,3$) is Hermitian, unitary, and involutive --- in general, any two of these properties imply the third.
The four Pauli operators form an orthogonal basis $\B=\{\id_2,X,Y,Z\}$ of ${\rm L}(\cd[2])$, known as the \emph{Pauli basis}. Specifically, the normalized operators $X_\mu/\sqrt{2}$ form an orthonormal basis, and a one-\q operator $A$ can be expanded as
\begin{equation}\label{eq:PauliDec}
A =
\frac{1}{2}
\sum_{\mu=0}^3 \inner{X_\mu}{A} X_\mu
=\frac{a_0\id_2+ \vb*a \cdot \vb* X}{2}\,,
\end{equation}
where $a_\mu = \inner{X_\mu}{A} = \tr(X_\mu A)$. This is the Bloch decomposition of $A$ into its scalar part (which retains its trace) and its traceless part along the Pauli matrices.
In particular, a one-qubit state has the \emph{Bloch form} $\rho=(\id_2+\vb*r\cdot \vb*X)/2$, hence is completely specified by the expectation values of the Pauli matrices, $r_i = \inner{X_i}{\rho}=\tr(X_i \rho)$. The vector $\vb*{r}$, known as polarization or Bloch vector of $\rho$, lies in the closed unit ball of $\mathbb{R}^3$. One-\q states are in one-to-one correspondence with points of the closed unit ball, with pure states corresponding to its boundary $\mathbb S^2$, the \emph{Bloch sphere}.

\subsubsection{Bell basis.}
The following maximally entangled unit vectors on the two-\q space $\cd[2]\otimes\cd[2]$,
\begin{align}\label{eq:PsiPhi}
\ket{\Phi^\pm}
=\frac{\ket{00}\pm\ket{11}}{\sqrt 2}\,,
&&
\ket{\Psi^\pm}
=\frac{\ket{01}\pm\ket{10}}{\sqrt 2}\,,
\end{align}
are called \emph{Bell vectors}. They are uniform real combination of two standard basis vectors with the same parity. Each vector is a joint eigenstate of the commuting observables $Z\otimes Z$ and $X\otimes X$, with eigenvalues the parity ($\Phi$ or $\Psi$, that is, parallel or antiparallel) and the relative phase ($+$ or $-$), respectively. In other words, the Bell vectors constitute a joint eigenbasis (which is unique, up to overall phases) of the pairwise commuting operators $X_i\otimes X_i$ ($i=1,2,3$),
\begin{align}\label{eq:PauliPhi}
	X \otimes X\ket{\Phi^\pm} &=\pm\ket{\Phi^\pm},&
	Y\otimes Y \ket{\Phi^\pm} &=\mp\ket{\Phi^\pm},&
	Z\otimes Z \ket{\Phi^\pm} &=\ket{\Phi^\pm},
	\\
	\label{eq:PauliPsi}
	X \otimes X\ket{\Psi^\pm}&=\pm\ket{\Psi^\pm},&
	Y\otimes Y  \ket{\Psi^\pm}&=\pm\ket{\Psi^\pm},&
	Z\otimes Z \ket{\Psi^\pm}&=-\ket{\Psi^\pm}.		
\end{align}
This orthonormal basis of $\cd[2]\otimes\cd[2]$ is called the Bell basis.

The projections $\PR{\beta}=\dyad{\beta}$, with $\beta=\Psi^\pm,\Phi^\pm$, are 1-\D, and will be referred to as  \emph{Bell projections} or \emph{Bell states}, depending on the context. Throughout this work, the index $\beta$ is reserved to label either the Bell vectors, or the corresponding states.

\subsection{Generalities on unitary designs}\label{sec:UniDes}
Let $d$ and $t$ be two integers. A qudit system of dimension $d$ is described by $\cd$, and the composite system of $t$ such qudits by the tensor product space $\cdtt$. For each unitary $U\in \U$, $\Ut$ is a unitary operator on $\cdtt$, which induces the unitary channel $\rho \mapsto\Ut \rho\Udtt$ on $\cdtt$, mapping an input state $\rho$ to its conjugate by $\Ut$. Our aim is to express the average of these output states, as $U$ varies over $\U$, as an average over a finite subset of $\U$. Therefore, given $N$ distinct unitaries $U_a$, the set $\Ss=\{U_a\}\subset \U$ is called a \emph{unitary $t$-\des} if and only if the following identity,
\begin{equation}\label{eq:design}
\frac{1}{N}\sum_{a}
\Ut[U_a]\rho \Udt[U_a]
=\int_{\U} \Ut\rho \Udt \dU,
\end{equation}
holds for all states $\rho $ of  $\cdtt$.

\subsubsection{Haar integrals in a nutshell.}
To begin with, we need to gain some familiarity with the right-hand side of the above equation. The object under consideration is first of all an integral with respect to the \emph{Haar probability measure} on $\U$, denoted by $\mu_d$. In general, a locally compact Hausdorff group admits left and right Haar measures, unique up to a scale factor, whose invariance properties are intertwined; in the compact case, left and right invariance coincide. For our purposes, all we need to know about integration of scalar (complex-valued) functions on $\U$ with respect to its Haar probability measure is that: (i) all continuous functions are integrable; (ii)  integration is linear, positive, left invariant, and normalized:
\begin{gather}\label{eq:HaarIntLin}
\int_{\U} \bigl(f(U)+\alpha g(U)\bigr) \dU 
=\int_{\U} f(U) \dU + \alpha \int_{\U} g(U) \dU,
\\
\label{eq:HaarIntPos}
0\le f  \quad \Rightarrow \quad 0\le \int_{\U} f(U) \dU ,
\\
\label{eq:HaarIntLeftInv}
\int_{\U} f(VU) \dU 
=\int_{\U} f(U) \dU,
\\
\label{eq:HaarIntNorm}
\int_{\U} \dU =1\,,
\end{gather}
for all continuous functions $f,g$ on $\U$, all complex numbers $\alpha$, and all unitaries $V$. These properties uniquely characterize the Haar integral as the only functional on the space of continuous functions on $\U$ that satisfies them.

Now, the integrand in Eq.~\eqref{eq:design} is not a scalar function, since its values are linear maps on $\cdtt$. That is, the integral in Eq.~\eqref{eq:design} is of a different nature from those in Eqs.~\eqref{eq:HaarIntLin}--\eqref{eq:HaarIntNorm}. Formally, this is a \emph{Bochner integral}, whose introduction involves tools from functional analysis. 
%\color{red}
However, in our setting the integrand is a continuous function taking values in a finite-\D Hilbert space with a standard basis. Since its continuity is equivalent to the continuity of its components in the standard basis, its integral can be defined componentwise by means of the integration of scalar functions.
In the main case of interest, a continuous map $F$ from $\U$ to ${\rm L}(\cdtt)$, we set
%\color{black}
\begin{equation}\label{eq:IntStand}
\mel{i_1 \dots i_t}
{\qty(\int_{\U} F(U)\dU)}{j_1\dots j_t}
=\int_{\U}
\mel**{i_1 \dots i_t}{F(U)}{j_1\dots j_t}\dU .
\end{equation}
In this way: (i) all continuous maps of $\U$ to ${\rm L}(\cdtt)$ are integrable; (ii) integration inherits linearity, left invariance, and normalization (with $1$ replaced by the identity map); (iii) integration commutes with all operations that act entrywise in the standard basis.
As a consequence of this last observation, familiar matrix manipulations --- such as taking matrix elements, traces, multiplying on either side by a given operator, or taking the Hermitian conjugate, complex conjugate, or transpose --- can all be performed under the integral sign.
Moreover, if $A$ is an operator on $\cdtt[t']$, then $U\mapsto  F(U)\otimes A$  is continuous from $\U$ to ${\rm L}(\cdtt[(t+t')])$, and, working componentwise, we straightforwardly obtain
\begin{equation}\label{eq:adhoc}
\int_{\U} \bigl(F(U)\otimes A\bigr)\dU
=\qty(\int_{\U} F(U)\dU) \otimes A\,.
\end{equation}

\subsubsection{Twirling.}\label{sec:MatEl}
Let $A$ be an operator on $\cdtt$. Each matrix element in the standard basis of the function $U\mapsto\Ut A\Udtt $ is given by

\begin{equation}\label{eq:poly}
\bra{i_1 \dots i_t} \Ut A \Udt \ket{j_1\dots j_t}
=
%\color{red}
 \sum_{i'_r,j'_s}
U_{i_1 i'_1}\dots U_{i_t i'_t}
\bra{i'_1\dots i'_t} A \ket{j'_1\dots j'_t}
U^\ast_{j_1 j'_1}\dots U^\ast_{j_t j'_t}
%\color{black}
\,,
\end{equation}

hence is a \emph{homogeneous polynomial of degree $(t,t)$}, that is, a homogeneous polynomial of degree $t$ in both the entries of $U$ and those of $U^\dag$.
In particular, the function $U\mapsto\Ut A\Udtt $ is continuous, therefore integrable. Its integral,
\begin{equation}\label{eq:BigMom}
\M(A)=\int_{\U} \Ut A \Udt \dU,
\end{equation}
is a well-defined operator on $\cdtt$. Since it represents the average of the function $U\mapsto\Ut A\Udtt$, as $U$ ranges over the entire unitary group, it can be compared with the average over a finite set $\Ss\subset \U$,
\begin{equation}\label{eq:SmallMom}
\M[\Ss](A)=
\frac{1}{N}\sum_{a}\Ut[U_a] A\Udt[U_a]\,.
\end{equation}
The operators $\M(A)$ and $\M[\Ss](A)$ are called \emph{twirls} of $A$: the former is obtained by twirling $A$ over all unitaries, the latter by twirling over the unitaries in $\Ss$.

The properties of the Haar integral determine those of the twirling map $\M$. In particular, linearity~\eqref{eq:HaarIntLin}, positivity~\eqref{eq:HaarIntPos}, and normalization~\eqref{eq:HaarIntNorm} ensure that the Haar average~\eqref{eq:BigMom} behaves  in the same way as the arithmetic mean~\eqref{eq:SmallMom}. As a notable example, observe that $\M[\Ss]$ is a channel, since, by definition, it is a convex combination of unitary channels. The same holds for $\M$. Indeed, $\M$ is a superoperator by linearity of the integral. Trace preservation is straightforward, because we can trace under the integral sign. Finally, by Eq.~\eqref{eq:adhoc},
\begin{equation}\label{eq:CP1}
%\color{red} 
	\qty(\M\otimes \rsfs I_{d^{\otimes t}}) (B)
	=\int_{\U} \qty(\Ut \otimes \id_{d^{\otimes t}}) B \qty(\Ut \otimes \id_{d^{\otimes t}})^\dag \dU,
\end{equation}
for any operator $B$ on $\cdtt[2t]$, therefore complete positivity follows from the positivity of the integral.

The following observation will play a crucial role later on: by left invariance of the integral, $\M(A)$ is invariant under conjugation by $\Ut$, that is,
\begin{equation}\label{eq:commu}
\comm{\M(A)}{\Ut}=0\,,
\end{equation}
for all  operators $A$ on $\cdtt$, and all unitaries $U$. (Indeed, we could go even further: the invariance of the integral under both multiplication and inversion immediatly entails that $\M$ is the projection in ${\rm L}(\cdtt)$ onto the commutant of all operators $\Ut$.)

Finally, expressing both sides of Eq.~\eqref{eq:design} in terms of twirling channels, we see that $\Ss$ is a unitary $t$-\des if and only if, for all states $\rho$ of $\cdtt$,
\begin{equation}\label{eq:SameTwirl}
	\M[\Ss](\rho)=\M(\rho)\,.
\end{equation}
Since states span the operator space, this is equivalent to
\begin{equation}\label{eq:DesMom}
	\M[\Ss]=\M\,.
\end{equation}
As usual with superoperators, working with states or with arbitrary linear operators is, to a large extent, a matter of convenience --- or perhaps even of taste --- and we will often switch between these perspectives.

\subsubsection{Unitary designs as Kraus representations.}
\label{sec:DesVsKraus}
Our definition~\eqref{eq:design} is equivalent to saying that $\Ss=\{U_a\}$ is a $t$-\des if and only if the normalized maps $\smash{\Ut[U_a]/\sqrt{N}}$, acting locally and identically on each subsystem, form a set of Kraus operators for $\M$. In other words, unitary $t$-\dess correspond to a special class of Kraus representations of $\M$, expressing it as a uniform convex combination of local and diagonal unitary channels --- namely, induced by unitary operators of the form $\Ut$, with $U\in \U$.

As a direct and important consequence, since the minimum number of Kraus operators required to represent a channel equals its Choi rank (i.e., the rank of its Choi operator), the rank of $\Choi(\M)$ provides a lower bound on the size of a $t$-\des.

\subsubsection{A $(t+1)$-\des is also a $t$-\des.}\label{pr:incrisi}
The problem of finding a $t$-\des becomes increasingly difficult as $t$ increases. Indeed, it is easy to see that
\begin{align}
\M[d][t+1]\qty(A\otimes \id_d)=
\M\qty(A)\otimes \id_d\,,
&&
\M[\Ss][t+1]\qty(A\otimes \id_d)=
\M[\Ss]\qty(A)\otimes \id_d\,,
\end{align}
for all operators $A$ on $\cdtt$. Hence, if $\Ss$ is a $(t+1)$-\des, then it is also a $t$-\des. In other words, a necessary condition for $\Ss$ to be a $t$-\des is that it be a $t'$-\des for every $t'< t$.

\subsubsection{Unitary designs and homogeneous polynomials.}\label{sec:homopoly}
Let us now elaborate on our definition of a unitary design, trying to understand why it poses an interesting problem. 
The twirl of an operator $A$ over all unitaries amounts to averaging the homogeneous polynomials of degree $(t,t)$ of Eq.~\eqref{eq:poly}. 
Regardless of how complicated the input $A$ may be, its only role is to supply the coefficients of such polynomials: in this sense, the input is merely a spectator in the twirling process.
Because of the invariance properties of the Haar measure, the twirl of $A$ is a more symmetric, ``smoothed'' version of $A$, as expressed by Eq.~\eqref{eq:commu}. 
The output retains only a part of the information encoded in the input
--- namely, the part that is invariant under conjugation by all unitary operators of the form $\Ut$ --- while all other details are ``washed away''  by the averaging. 
This effect will become evident once we compute the twirl over all unitaries of a one-\q  state (in Sec.~\secref{sec:onedes}), and of a two-\q  state (in Sec.~\secref{sec:twirl22}). 
Precisely because the output is such a huge simplification of the input %--- more precisely, the invariance of the Haar integral under multiplication and inversion entails that $\M$ is the projection onto the commutant of all unitary operators of the form $\Ut$ --- 
we can ask whether a finite set of unitaries could produce the same effect.
These considerations also explain the observation above: the larger $t$ is, the more structure of the input survives the twirling, and the more demanding the task a $t$-\des must accomplish. This is why constructing a $t$-\des becomes harder as $t$ increases.

Let us also note that $t$-\dess can be characterized as finite sets of unitaries that reproduce the same averages as the Haar measure for all homogeneous polynomials of degree $(t,t)$.
From the discussion above, if $\Ss$ has this property, then it is a $t$-\des. In fact, the converse is also true. 
Indeed, any homogeneous monomial of degree $(t,t)$ can be obtained by setting $A = \dyad*{i''_1\dots i''_t}{j''_1\dots j''_t}$ in Eq.~\eqref{eq:poly}, for all possible multi-indices with $i''_p,j''_q=1,\dots, d$. Therefore, if $\Ss$ is a $t$-\des, the averages over $\Ss$ and over all unitaries coincide for all such monomials, and, by linearity, for all polynomials.

Finally, this characterization immediately shows that unitary designs are stable under left and right multiplication. 
Namely, if $\Ss=\{U_a\}$ is a $t$-\des, and $V,V'\in \U$, then $V\Ss V'=\{V U_a V'\}$ is also a $t$-\des. 
Indeed, any homogeneous polynomial $p$ of degree $(t,t)$ is of the form $p(U)=\inner{C }{\Ut\otimes \Udtt}$ for some $C\in {\rm L}(\cdtt[2t])$. As a result, $p(V U V')= p'(U)$, where $p'$ is the homogeneous polynomial of degree $(t,t)$ with coefficient matrix $C'=\Udtt[V]\otimes \Ut[V'] \,C\, \Udtt[V']\otimes \Ut[V] $.

\subsubsection{On the projective nature of unitary designs.}\label{sec:proj}
Conjugation by a unitary operator is invariant under arbitrary phase shifts, and a twirling channel over a finite set of unitaries performs an arithmetic mean of such conjugations. However, according to our definition, a twirling channel is not necessarily invariant under arbitrary phase shifts of its elements. The reason is simple: two distinct unitaries can still be proportional to each other, hence phase-shift\-ed to the same unitary. In other words, let us consider $N$ distinct unitaries $U_a$, and phase-shift them as $U'_a=\e^{\iu \phi_a} U_a$. Then $\Ss=\{U_a\}$ consists of $N$ unitaries, whereas $\Ss'=\{U'_a\}$ may contain fewer than $N$. Of course, if the unitaries $U'_a$ are all distinct, then the twirling channels of $\Ss$ and $\Ss'$ coincide (at all orders), 
\begin{equation}\label{eq:Mphaseshift}
	\M[\Ss](A)
	=\frac{1}{N}\sum_{a}\Ut[U_a] A\Udt[U_a]
	=\frac{1}{N}\sum_{a}{U'_a}^{\otimes t} A{{U'_a}^\dag}^{\otimes t}
	=\M[\Ss'](A)\,,
\end{equation}
and, in particular, $\Ss$ is a $t$-\des if and only if $\Ss'$ is. However, in the general case, even if the first two equalities above hold, the third need not, since $\tfrac{1}{N} \sum_{a}{U'_a}^{\otimes t} A {U'_a}^{\dag\otimes t}$ may not be a twirl over $\Ss'$.

A sufficient condition for the invariance of all twirls over a finite set $\Ss \subset \U$ under arbitrary phase shifts is that $\Ss$ contain no two proportional unitaries. 
%In that case, $\Ss$ and its phase-shift\-ed version $\Ss'$ yield the same twirling channels (of any order), so $\Ss$ is a $t$-\des if and only if $\Ss'$ is. 
In other words, this sufficient condition is that the elements of $\Ss$ remain distinct also when regarded up to global phase.
In this way we are led to consider the projective unitary group, which we now recall. Scalar unitaries are of the form $\e^{\iu\phi}\,\id_d$, and constitute a one-parameter subgroup of $\U$ isomorphic to $\U[1]$. By Schur's lemma, this subgroup is precisely the center of $\U$. The resulting quotient structure is the projective unitary group, $\PU[d]=\U/\U[1]$, whose elements are equivalence classes of unitaries modulo multiplication by a phase factor.

Several authors emphasize the projective nature of unitary designs, and some formulate the defining identity~\eqref{eq:design} in $\PU[d]$, rather than in $\U$~\cite{RoyScott}. The resulting notion is narrower than ours, since, in our language, it translates into excluding designs containing proportional unitaries. 
By contrast, other authors define unitary designs as multisets, rather than sets, of unitaries~\cite{Bannai2022}. 
This leads to a broader notion than ours,  since the same unitary may appear in a design with arbitrary multiplicity. In our language, this translates into allowing non-uniform designs (and twirls) with rational weights. Defining twirls on multisets ensures invariance under arbitrary phase shifts, which is often a desirable property.

In this work we adopt the most common definition of design (up to equivalent formulations and generalizations). 
Fortunately, in what follows we will only deal with simple sets containing no proportional unitaries, so phase shifts will not be a significant concern.
The relevant observation on phase shifts for the next two sections is that finding a $t$-\des of $\U$ of size $N$ is equivalent to finding $N$ matrices $V_a\in \SU[d]$ satisfying Eq.~\eqref{eq:design} for all states of $\cdtt$. 
Note that we do not require the matrices $V_a$ to be distinct; if they are not, the set $\{V_a\}\subset \SU[d]$ has fewer than $N$ elements, hence is not necessarily a $t$-\des\ --- Eq.~\eqref{eq:design} is the defining property of a $t$-\des only for a subset of $\U$ of size $N$. 
Let us now show the equivalence. 
Suppose first that $\Ss\subset \U$ is a $t$-\des, consisting of $N$ distinct unitaries $U_a$. 
Then the matrices $V_a = \omega_a^\ast U_a $, where $\omega_a$ is any $d$-th root of $\det U_a$, are $N$ (not necessarily distinct) elements of $\SU[d]$  satisfying Eq.~\eqref{eq:design} --- note, in particular, that the matrices $V_a$ cannot be all distinct whenever $\Ss$ contains $d+1$ proportional unitaries. 
Conversely, suppose $V_a$ are $N$ (not necessarily distinct) special unitary matrices satisfying Eq.~\eqref{eq:design}. 
Then we can phase-shift them so that the unitaries $U_a=\e^{\iu\phi_a}V_a$ are all distinct (e.g., by choosing distinct phase factors), and hence $\Ss=\{U_a\}\subset \U$ is a $t$-\des of size $N$.

\section{1-\dess of \texorpdfstring{$\U[2]$}{U(2)} }\label{sec:onedes}
In this section we consider the simplest nontrivial instance of a unitary design: a 1-\des of $\U[2]$, namely, a set of $N$ distinct matrices $U_a \in \U[2]$ satisfying
\begin{equation}\label{eq:1design}
\frac{1}{N}\sum_{a}U_a\rho U_a^\dag
=\int_{\U[2]} U\rho U^\dag \dU[U][\mu_2],
\end{equation}
for all one-\q states $\rho$. In Sec.~\secref{sec:Pauli1des} we recall that the Pauli basis is a 1-\des of $\U[2]$ of minimum size. Then in Sec.~\secref{sec:SU} we obtain the general form of the 1-\dess of $\U[2]$ of minimum size from the theory of angular momentum.

\subsection{The Pauli basis is a 1-\des of \texorpdfstring{$\U[2]$}{U(2)}}
\label{sec:Pauli1des}
In a sense, the Pauli basis is the oldest and best-known example of a unitary design.

Let us start by computing  the full twirl of a one-\q state $\rho$ (i.e., the right-hand side of Eq.~\eqref{eq:1design}),
\begin{equation}\label{eq:M1rho}
\M[2][1](\rho)=\int_{\U[2]} U\rho  U^\dag \dU[U][\mu_2].
\end{equation}
One way to obtain it is from its invariance property~\eqref{eq:commu}. Since $\M[2][1](\rho)$ commutes with all $U\in \U[2]$, it commutes with the defining representation of $\SU$, generated by the spin-\textonehalf\ angular momentum $\vb*X /2$. This representation is irreducible, hence, by Schur's lemma, $\M[2][1](\rho)$ is a scalar. Since $\M[2][1](\rho)$ is also a state, it must be the completely mixed one,
\begin{equation}\label{eq:M1rhoEv}
\M[2][1](\rho)=\frac{\id_2}{2}\,.
\end{equation}
This is an extreme situation: the twirling operation completely erases the polarization of the input state, and the Bloch sphere collapses to its center. The output has no relation with the input, except for the fact that both are states. 

Then, the well-known formula~\eqref{eq:WellKnown} immediately shows that the Pauli basis is a 1-\des. 
Eq.~\eqref{eq:WellKnown} arises from Pauli anticommutation relations: $X_i X_j X_i $ (i.e., the conjugate of the $j$-th Pauli matrix by the $i$-th Pauli matrix) is $X_j$ for $i=j$, and $-X_j$ for $i\ne j$, that is,
\begin{equation}
X_i X_j X_i =(2\delta_{ij}-1)X_j \,.
\end{equation}
Conjugating a linear combination of Pauli matrices, $\vb*a \cdot \vb*X$, by $X_i$ has a nice geometric interpretation, as the vector $\vb*a$ transforms to its reflection with respect to the $i$-th axis,
\begin{equation}\label{eq:conjPauli}
X_i  \,\vb*a \cdot \vb*X\, X_i 
= 2a_i X_i -\vb*a \cdot\vb*X,
\end{equation}
resulting in
\begin{equation}
\sum_{\mu=0}^3 
X_\mu  \,\vb*a \cdot \vb*X\, X_\mu = 0\,.
\end{equation}
On the other hand, a scalar operator is obviously invariant under conjugation. Therefore, twirling an operator $A=\qty(a_0\id_2 +\vb*a \cdot \vb*X)/2$ over the Pauli basis amounts to extracting its scalar component. In particular, the Pauli twirl of a one-\q state is maximally mixed.

From the standpoint of quantum information theory, Eq.~\eqref{eq:M1rhoEv} is just the observation that $\M[2][1]$ is the one-\q (completely) depolarizing channel. In general, the depolarizing  channel in $d$ dimensions, denoted by $\rsfs{D}_d$, is the replacement map $\rho\mapsto \id_d/d$, that is, the rank one map $\rsfs{D}_d=\ketp{\id_d} \!\brap{\id_d}/d$. Moreover, $\id_d/\sqrt d$ is a unit vector in ${\rm L}(\cd)$, so that $\rsfs{D}_d$ is exactly the projection onto the 1-\D subspace of all scalar operators. Its Choi operator, $\Choi(\rsfs{D}_d)=\id_{d\otimes d}/d$, has full rank. In the case of interest, we have
\begin{equation}
	\M[2][1]
	= \frac{1}{2} \ketp{\id_2} \!\brap{\id_2},
\end{equation}
with Choi rank 4. %--- whereas its rank as a linear map is obviously 1.
Identity~\eqref{eq:WellKnown} is well-known in quantum information theory because it yields a Kraus representation of the completely depolarizing channel on a single qubit. Since the Choi rank of the channel is 4,
such representation is of minimum size.
%the set of its Kraus operators $X_\mu/\sqrt 2$ has minimum size
Therefore, from the observation in Sec.~\secref{sec:DesVsKraus}, the Pauli basis is a 1-\des of minimum size.

\subsection{1-\dess of \texorpdfstring{$\U[2]$}{U(2)} of minimum size}\label{sec:SU}
Several aspects of the above discussion hint at a geometric background, which can be unveiled in the context of $\SU$ representation theory, and the interplay between $\SU$ and $\SO$. Let us start by recalling the fundamental relation
\begin{equation}\label{eq:SO(3)}
U \,\qty(\vb*v\cdot \vb*X)\, U^\dag
= \qty( \mathscr R(U) \vb*v) \cdot \vb*X\,,
\end{equation}
for $U\in\SU$, and $\vb*v\in\mathbb R^3$, where $\mathscr R$ is the representation of $\SU$, acting on $\mathbb{R}^3$, defined by
\begin{equation}\label{eq:SO(3)expl}
\mathscr R_{ij}(U)=\frac{1}{2}\tr(X_i U X_j U^\dag)\,.
\end{equation}
This representation is also a (Lie) group homomorphism of $\SU$ \emph{onto} $\SO$, and will be referred to as the \emph{$\SO$-\rep of $\SU$}. It is two-to-one, as it maps antipodal unitaries to the same 3-\D rotation,
\begin{equation}\label{eq:2to1}
\ker \mathscr R=\qty{\id_2,\,-\id_2}\,.
\end{equation}
In fact, the above equation is just the observation that $\mathscr R$ is a \emph{double covering map} of $\SO$, and Eq.~\eqref{eq:SO(3)} is just an instance of the general relation between representations of the Lie group $\SU$ and the corresponding representations of its Lie algebra ${\rm su}(2)$.

Let $R(\phi \vb*n)$ denote the rotation by $\phi$ about the oriented direction singled out by the unit vector $\vb*n$. This rotation acts on $\vb*x\in \mathbb{R}^3$ according to \emph{Rodrigues formula},
\begin{equation}\label{eq:Rodrigues}
R(\phi\vb*n) \vb*x
=\vb*x\cdot \vb*n\, \vb*n
+ \cos \phi \,\qty(\vb*x -\vb*x\cdot \vb*n\, \vb*n)
+\sin \phi \;\vb*n\times \vb*x\,,
\end{equation}
which defines the \emph{axis--angle parametrization} of $\SO$, $\vb*\theta\mapsto R\qty(\vb*\theta)$. When special unitary matrices are put in exponential form, Eq.~\eqref{eq:SO(3)expl} yields
\begin{equation}\label{eq:RstortovsRdritto}
\mathscr R\qty(\e^{-\iu \vb*\theta\cdot \vb*X/2} )
=R(\vb* \theta)\,,
\end{equation}
and Eq.~\eqref{eq:SO(3)} takes the form
\begin{equation}\label{eq:rotatio}
\e^{-\iu \vb*\theta\cdot \vb*X/2} 
\,\qty(\vb*v\cdot \vb*X) 
\,\e^{\iu \vb*\theta\cdot \vb*X/2}
= R(\vb*\theta)\vb*v \cdot \vb*X \,,
\end{equation}
as can be checked by Euler formula for Pauli matrices. The exponential form makes explicit the two-to-one nature of $\mathscr R$, ultimately relating it to the $2\pi$-pe\-ri\-o\-dic\-i\-ty of the sine and cosine functions in Rodrigues formula~\eqref{eq:Rodrigues},
\begin{equation}
\mathscr R\qty(-\e^{-\iu \phi \vb*n \cdot \vb*X/2})
=\mathscr R\qty(\e^{-\iu (\phi+2\pi) \vb*n \cdot \vb*X/2})
=R\qty((\phi+2\pi) \vb*n)
=R\qty(\phi\vb*n)
%=\mathscr R\qty(\e^{\iu \phi \vb*n \cdot \vb*X/2})
\,.
\end{equation}

Let us now look at Eq.~\eqref{eq:conjPauli}. Since each Pauli matrix is unitary and has determinant $-1$, then the matrices $- \iu X_i=\exp(-\iu \pi X_i/2)$ are special unitary, resulting in
\begin{equation}\label{eq:TransGen}
X_i \,\qty(\vb*v \cdot \vb*X) \,X_i 
= (-\iu X_i) \,\qty(\vb*v \cdot \vb*X) \,(-\iu X_i)^\dag
=\e^{-\iu\pi X_i/2}\,\qty(\vb*v \cdot \vb*X)\,\e^{\iu  \pi X_i/2 }
=\qty(R_i(\pi ) \vb*v)\cdot \vb*X \,,
\end{equation}
where we set $R_i(\phi )=R(\phi\vb e_i)$. Then we just have to observe that an arbitrary rotation by $\pi$ maps a vector to its reflection with respect to the fixed direction,
\begin{equation}\label{eq:RotPi}
R(\pi \vb*n) \vb*x=2\vb*x\cdot \vb*n\, \vb*n -\vb*x\,,
\end{equation}
as can be easily verified by Rodrigues formula.

Now we are ready to show that the problem of finding a 1-\des of $\U[2]$ can be handled in terms of 3-\D rotations. Without loss of generality, we can look for $N$ not necessarily distinct matrices $U_a\in \SU$ ($a=0,\ldots,N-1$) such that, for an arbitrary state with Bloch form $\rho=(\id_2+\vb*r\cdot \vb*X)/2$, we have
\begin{equation}
\frac{\id_2}{2}
= \frac{1}{N} \sum_{a=0}^{N-1} U_a \,\rho U_a^\dag
=\frac{\id_2}{2} + \frac{1}{2N}
\sum_{a=0}^{N-1} \mathscr R(U_a)\vb*r \,.
\end{equation}
The above problem reads
\begin{equation}
\sum_{a=0}^{N-1} \mathscr R(U_a)=0\,.
\end{equation}
Obviously, this problem is relevant only for $N>1$, in which case we set $Q_i=\mathscr R(U_0^{-1} U_i)$ for $i=1,\ldots,N-1$, and then write each of these matrices in axis--angle form, as $Q_i=R(\phi_i\vb*n_i)$. Since the $\SO$-\rep is onto, the problem of finding a unitary 1-\des is reduced to the problem of finding $N-1$ rotation matrices $Q_i$, such that 
\begin{equation}\label{eq:RotCond}
\id_3+\sum_{i=1}^{N-1} Q_i=0\,.
\end{equation}
Applying the above equation to $\vb*n_1$,  we obtain the identity $-2\vb*n_1 = \sum_{i=2}^{N-1} Q_i\vb*n_1$, hence the inequality $2=\norm*{\sum_{i=2}^{N-1} Q_i\vb*n_1}\le N-2$, showing that Eq.~\eqref{eq:RotCond} cannot hold for $N< 4$.

For $N=4$, Eq.~\eqref{eq:RotCond} has obvious solutions, namely, $Q_i=R(\pi\vb*n_i)$, for any orthonormal basis $\{\vb*n_1,\vb*n_2,\vb*n_3\}$ of $\mathbb{R}^3$. Indeed, by Eq.~\eqref{eq:RotPi}, $Q_i \vb*n_j = (2 \delta_{ij}-1)\vb*n_j$, so that Eq.~\eqref{eq:RotCond} is satisfied on all basis vectors. Moreover, these are the only solutions for $N=4$. Indeed, $2=\norm{(Q_2+Q_3)\vb*n_1}$ is equivalent to $(Q_2\vb*n_1)\cdot(Q_3\vb*n_1)=1$, that is, $Q_2\vb*n_1=Q_3\vb*n_1$. As a result, we have $Q_i\vb*n_1=-\vb*n_1$ for $i=2,3$, thus, by Rodrigues formula, it must be $\cos \phi_i=-1$, and 
$\vb*n_i\cdot \vb*n_1=0$. Likewise for the orthogonality of $\vb*n_2$ and $\vb*n_3$.

These considerations entail a classification of the  1-\dess of minimum size: they consist of four unitaries of the form $U_\mu=\e^{\iu \phi_\mu}  V X_\mu  V' $ ($\mu=0,1,2,3$), with arbitrary $V,V'\in \SU$, and $\phi_\mu\in \mathbb{R}$. The Pauli basis is the special case for $V=V'=\id$, and all phase factors equal to 1. In this way we \emph{deduced} --- as opposed to \emph{verified} --- Eq.~\eqref{eq:WellKnown}. 

A 1-\des of minimum size clearly inherits from the Pauli basis the property of being an orthogonal basis of the matrix algebra of order 2. The converse is also true. Indeed, let $U_\mu\in \U[2]$ ($\mu=0,1,2,3$) be an orthogonal basis of ${\rm L}(\cd[2])$. Then the basis vectors can be written $U_\mu=\e^{\iu \theta_\mu} \, V_\mu$, where $\e^{2\iu \theta_\mu}=\det U_\mu$. The matrices $V_\mu\in \SU$ are in turn an orthogonal basis, hence, in particular, $0=\inner{V_0}{V_i}$ ($i=1,2,3$). Since the matrices ${V_0}^\dag V_i \in \SU$ are traceless, they must be of the form $\exp(-\iu \pi \vb*n_i \cdot \vb*X)=-\iu \vb*n_i \cdot  \vb*X$; moreover, we also have $2\delta_{ij}=\inner{V_i}{V_j} = \inner{{V_0}^\dag V_i }{{V_0}^\dag V_i } =2\vb*n_i \cdot  \vb*n_j$. As a result, $\{\vb*n_1,\vb*n_2,\vb*n_3\}$ is an orthonormal basis of $\mathbb{R}^3$. 
After possibly reordering these unit vectors, the ordered basis $(\vb*n_i)_{i=1}^3$ is positive, hence singles out $R\in \SO$ such that $\vb*n_i=R\vb e_i$. Then we have $V_i=  -\iu V_0 \,\qty(\qty(R\vb e_i) \cdot  \vb*X)$, from which the basis vectors $U_\mu$ are of the form $\e^{\iu \phi_\mu} V X_\mu V' $.
We have just shown that the 1-\dess of $\U[2]$ of minimum size are precisely the orthogonal bases of the matrix algebra of order 2 made up of unitary matrices.

\section{Explicit construction of a 
2-\des of \texorpdfstring{$\U[2]$}{U(2)}}
\label{sec:twodes}
In this section we carry out an explicit construction of a 2-\des of $\U[2]$.
That is, we determine $N$ distinct matrices $U_a \in \U[2]$ such that the following identity,
\begin{equation}\label{eq:design22}
	\frac{1}{N} \sum_a
	(U_a\otimes U_a )\rho\qty(U_a^\dag \otimes U_a^\dag)
	=\int_{\U[2]}
	\qty(U\otimes U )\rho\qty(U^\dag \otimes U^\dag)
	\dU[U][\mu_2],
\end{equation}
holds for all two-\q states $\rho$. This construction relies only on considerations from the theory of angular momentum, and on the results established in the previous sections. Let us briefly outline its logic.

We start by fixing a state $\rho$ of two qubits. Then we consider its full twirl (the twirl over all unitaries, that is, the right-hand side of Eq.~\eqref{eq:design22}),
\begin{equation}\label{eq:full}
	\M[2][2](\rho)
	=\int_{\U[2]} (U\otimes U)\rho(U^\dag \otimes U^\dag )\dU[U][\mu_2],
\end{equation}
and its Pauli twirl (the twirl over the Pauli basis, which we know from Sec.~\secref{sec:onedes} to be a 1-\des),
\begin{equation}\label{eq:Pauli}
	\M[{\B}][2](\rho)
	=\frac{1}{4} \sum_{\mu=0}^3 
	(X_\mu\otimes X_\mu)\rho(X_\mu\otimes X_\mu)\,.
\end{equation}
The construction proceeds in three steps.

Denoting the projections onto the singlet and the triplet subspaces by $\PS$ and $\PT$, in Sec.~\secref{sec:twirl22} we show that the full twirl~\eqref{eq:full} takes the form of a convex combination of the singlet state $\PS$ and the completely mixed triplet state $\PT/3$, with coefficients given by the expectation values of $\PS$ and $\PT$,
\begin{align}\label{eq:M22RhoEv}
	\M[2][2](\rho)
	=\fs \,\PS 
	+ \ft\,\frac{\PT}{3}\,,
&&
	\fs=\tr(\PS \rho)\,,
&&
	\ft=\tr(\PT \rho)\,.
\end{align}

Then, in Sec.~\secref{sec:twirl2Pau} we show that the Pauli twirl~\eqref{eq:Pauli} likewise takes the form of a convex combination of the Bell states $\PR{\beta}$ ($\beta = \Psi^\pm,\Phi^\pm$), with coefficients given by the corresponding expectation values,
\begin{align}\label{eq:M2PauRhoEv}
	\M[{\B}][2](\rho)
	=\sum_\beta \f{\beta}\,\PR{\beta} \,,
&&
	\f{\beta}= \tr(\PR{\beta}\rho)\,.
\end{align}

Since $\PR{\Psi^-}$ is the projection onto the singlet subspace, whereas $\PR{\Phi^-}$, $\PR{\Psi^+}$, and $\PR{\Phi^+}$ resolve the projection onto the triplet subspace, we have $\fs = \f{\Psi^-}$, as well as $\ft = \f{\Phi^-} + \f{\Psi^+} + \f{\Phi^+}$. Therefore, the full twirl can be obtained from the Pauli twirl by replacing each of the coefficients $\f{\Phi^-}$, $\f{\Psi^+}$, and $\f{\Phi^+}$ in Eq.~\eqref{eq:M2PauRhoEv} with their arithmetic mean, $\ft/3$. In Sec.~\secref{sec:ave} we show that such averaging %of the triplet components of $\M[{\B}][2](\rho)$ 
can be obtained by an additional twirl,
\begin{equation}\label{eq:MediaTwirl}
\M[2][2](\rho)=
\frac{1}{3} \sum_{k=0,\pm1}
\qty(W^k\otimes W^k) \M[{\B}][2](\rho) 
\qty( W^{k^\dag}\otimes W^{k^\dag})
\,,
\end{equation}
where $W=(\id_2 - \iu(X+Y+Z))/2$.
The above identity and Eq.~\eqref{eq:Pauli} yield
\begin{equation}
\M[2][2](\rho)
=\frac{1}{12} \sum_{k=0,\pm1} 
\sum_{\mu=0}^3
\qty(W^k X_\mu \otimes W^k X_\mu ) 
\rho
\qty( \qty(W^k X_\mu )^\dag\otimes \qty(W^k X_\mu )^\dag)
\,,
\end{equation}
so that the twelve (distinct) unitaries $W^k X_\mu $, with $\mu=0,1,2,3$, and $k=0,1,-1$, form a 2-\des, which is of minimum size as it saturates the Clifford bound. %~\cite{gross,RoyScott}
Moreover, the above procedure also works when the Pauli basis is replaced by any other $1$-\des of minimum size.

The main properties of the resulting structures are finally discussed in Sec.~\secref{sec:rema}.

\subsection{Full twirl of a two-\q  state} \label{sec:twirl22}
Just as in the one-\q  case, the full twirl of the two-\q state $\rho$ can be obtained from the theory of angular momentum.
The starting point is exactly the same: by Eq.~\eqref{eq:commu}, $\M[2][2](\rho)$ commutes with all operators of the form $U\otimes U$, with $U\in \U[2]$, hence, in particular, with the $(\frac{1}{2}\times\frac{1}{2})$-\rep of $\SU$, acting on the two-\q space $\cd[2]\otimes\cd[2]$.

Let us briefly recall how the $(j_1\times j_2)$-\rep of $\SU$, acting on $\cd[2j_1+1] \otimes \cd[2j_2+1]$, is decomposed into irreducible components. This representation is generated by the total angular momentum $\vb*J=\vb*J_1\otimes \id_{2j_2+1} +\id_{2j_1+1}\otimes \vb*J_2$, the sum of a spin-$j_1$ and a spin-$j_2$ angular momentum. Its irreducible components are determined by the spectral properties of $\vb*J^2$, whose eigenvalues are $j(j+1)$, with $j$ taking the $2\min(j_1,j_2)+1$ values ranging in integer steps from $\abs{j_1-j_2}$ to $j_1+j_2$. Indeed, the eigenspace corresponding to a given value of $j$ is $(2j+1)$-\D, and $\vb*J$ acts on it as a spin-$j$ angular momentum. Since the eigenspaces of $\vb*J^2$ are mutually orthogonal and span the entire space, the irreducible components of the representation are precisely the subrepresentations induced on each such eigenspace. In summary, the $(j_1\times j_2)$-\rep is decomposed into $2\min(j_1,j_2)+1$ irreducible components, corresponding to all possible values of the total spin $j = \abs{j_1-j_2},\abs{j_1-j_2}+1, \ldots, j_1+j_2$.

In particular, the $(\frac{1}{2} \times \frac{1}{2})$-\rep of $\SU$, acting on $\cd[2]\otimes\cd[2]$, is generated by the sum of two spin-\textonehalf\ angular momenta,
\begin{equation}\label{eq:totalJ}
\vb*J
=\frac{\vb*X}{2}\otimes \id_2
+\id_2\otimes\frac{\vb*X}{2}\,.
\end{equation}
Its irreducible invariant subspaces are the eigenspaces of
\begin{equation}\label{eq:J2}
\vb*J^2
=\frac{3}{2}\,\id_{2\otimes2}
+\frac{1}{2}\,\qty(X\otimes X+Y\otimes Y+Z\otimes Z)\,,
\end{equation}
and its irreducible components are the subrepresentations on such eigenspaces. The eigenspace with $j=0$ --- the \emph{singlet} subspace --- is 1-\D, and corresponds to the zero eigenvalue. It is the kernel of $\vb*J^2$, as well as the intersection of the kernels of the Cartesian components of $\vb*J$. The eigenspace with $j=1$ --- the \emph{triplet} subspace --- is 3-\D and corresponds to the eigenvalue 2. The irreducible components are the spin-0 subrepresentation, acting on the singlet subspace, and the spin-1 subrepresentation, acting on the triplet subspace. This decomposition is encoded in the properties of the corresponding projections, denoted by $\PS$ and $\PT$, which sum to the identity, multiply to zero, and commute with the entire representation. By this last observation, the expectation values of $\Pi_{\rm s}$ and $\Pi_{\rm t}$ are invariant under the full twirl%
%\color{red}%
: indeed, for $U\in {\rm U}(2)$, $U\otimes U$ acts on the singlet subspace as multiplication by $\det U$%
%\color{black}%
.

Since $\M[2][2](\rho)$ commutes with $\vb*J^2$, it also commutes with $\PS$ and $\PT$, hence is diagonal in the singlet--triplet decomposition, $\M[2][2](\rho) = \PS \M[2][2](\rho) \PS + \PT \M[2][2](\rho) \PT$. Each of the singlet and the triplet components of $\M[2][2](\rho)$ commutes with the corresponding irreducible component of the $(\frac{1}{2} \times \frac{1}{2})$-\rep, hence, by Schur's lemma, is a scalar multiple of the corresponding projection. This yields the spectral decomposition
\begin{equation}\label{eq:M22RhoBis}
\M[2][2](\rho)
=\fs\,\PS+\ft\,\frac{\PT}{3}\,,
\end{equation}
with
\begin{align}\label{eq:fsft}
\fs=\tr(\PS \M[2][2](\rho))=\tr(\PS \rho)
\,,
&&
\ft=\tr(\PT \M[2][2](\rho))=\tr(\PT \rho)
\,.
\end{align}
Summing up, $\M[2][2](\rho)$ is a convex combination of the singlet state $\PS$, and the completely mixed triplet state $\PT /3$, with coefficients given by the expectation values of $\PS$ and $\PT$, respectively. These two nonnegative numbers summing to~1 are the only information about $\rho$ that survives (and can be retrieved after) the full twirl. Observe that $\M[2][2](\rho)$ can be obtained by performing a non selective measurement on $\rho$, with measurement operators $\PS$ and $\PT$ --- e.g., by measuring $\vb*J^2$, and discarding the outcome 
%\color{red}
--- and then completely depolarizing the triplet subspace.
%\color{black}

\subsubsection{Properties of \texorpdfstring{$\M[2][2]$}{the twirling channel over U(2)}.}
\label{par:M22}
 Eqs.~\eqref{eq:M22RhoBis}--\eqref{eq:fsft} can also be written
\begin{equation}\label{eq:M22RhoHS}
\M[2][2](\rho)
=\inner{\PS}{\rho}\PS+\inner{\PT}{\rho}\frac{\PT}{3}\,,
\end{equation}
so that
\begin{equation}\label{eq:BigTwirlCh}
\M[2][2]
=\ketp{\PS}\!\brap{\PS}
+\frac{1}{3}\ketp{\PT}\!\brap{\PT}.
\end{equation}
Since $\PS$ and $\PT/\sqrt{3}$ form an orthonormal set in ${\rm L}\qty(\cd[2]\otimes \cd[2])$, $\M[2][2]$ is the projection onto their span, consisting of all linear combinations of $\PS$ and $\PT$.

\begin{figure}[!tb]
	\centering
%\resizebox{\textwidth}{!}{
%\begin{tikzpicture}
%
%
%
%% Include the image in a node
%\node [above right,inner sep=0] (image) at (0,0) {\includegraphics[width=\textwidth]{1ad}};
%
%%\draw [help lines] (0,0) grid (15,10);
%
%
%\node[circle] at (8.2,10.3){$\PS
%%=\PR{\Psi^-}
%$};
%
%\node[circle] at (8.93,4.7) {$\M[2][2](\rho)$};
%
%\node[circle] at (7.45,3.6){$\id_{2\otimes 2}/4$};
%
%\node[circle] at (12.95,2.8){$\PR{\Phi^-}$};
%\node[circle] at (1.75,1.9){$\PR{\Psi^+}$};
%
%\node[circle] at (8.2,1.25){$\PT/3$};
%
%\node[circle] at (10,-0.3){$\PR{\Phi^+}$};
%
%
%
%
%%\node[circle] at (7,6.5){{\small $ x_2$}};
%
%
%\end{tikzpicture}
%}
\includegraphics[width=\textwidth]{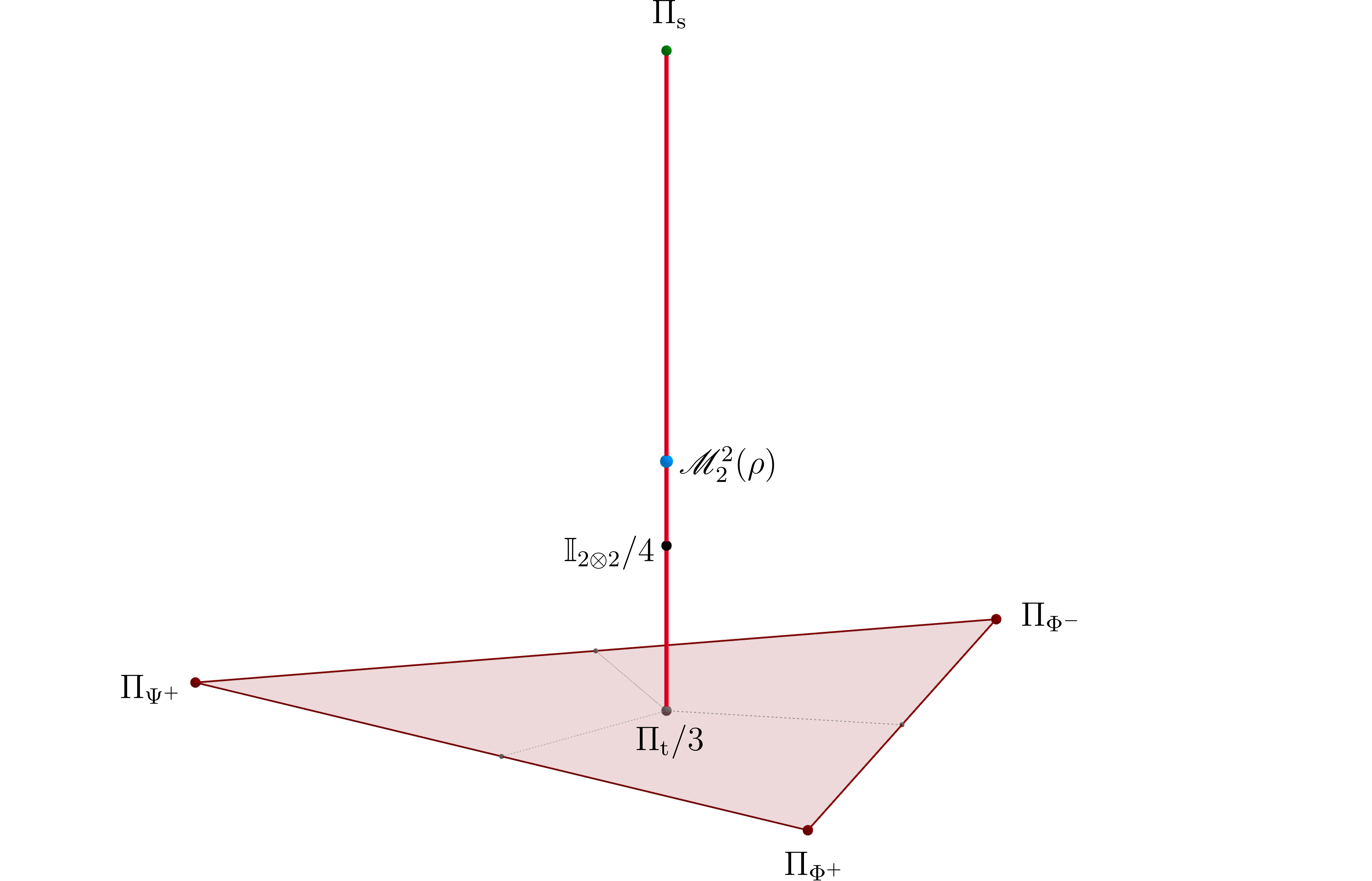}
\vspace{1pt}
\caption[LoF entry]{--- The full twirl of a two-\q state (\azurpalla) --- explicitly, $\M[2][2](\rho)=\fs \,\PS +\ft\,\PT/3$, where $f_a=\tr(\Pi_a \rho)$, with $a={\rm s}, {\rm t}$ --- lies on the segment (\rossolinea) connecting the singlet state $\PS$ (\singlo) and the completely mixed triplet state $\PT/3$ (\maruninpalla), which sits at the center of the equilateral triangle (\maruntri) with vertices the triplet Bell states (\marunpalla). The segment is divided by $\M[2][2](\rho)$ in the ratio $\ft \!:\! \fs$.  Since $\PS$ and $\PT$ resolve the identity, the segment also passes through the completely mixed state of two qubits $\id_{2\otimes2}/4$ (\neropalla), dividing it in the ratio $3  \!:\! 1$.} \label{fig:segmento}
\end{figure}

The set of all two-\q states is mapped onto the convex hull of the singlet and the completely mixed triplet states, that is, onto the segment with extreme points $\PS$ and $\PT/3 $ (see~\figref{fig:segmento}). The coefficients $\fs$ and $\ft=1-\fs$ of Eq.~\eqref{eq:fsft} are proportional to the distance of $\M[2][2](\rho)$ from $\PT/3$ and $\PS$, respectively. The proportionality factor is the length of the segment, $2/\sqrt{3}$, which follows from $\norm {\PS-\PT/3}^2 = \norm{\PS}^2 + \norm{\PT}^2/9 = 4/3$. More precisely, $\M[2][2](\rho)$ is the point on the segment connecting $\PS$ and $\PT/3$ whose distances from the endpoints are $2\ft/\sqrt{3}$, and $2\fs/\sqrt{3}$, respectively. In particular, all states with $\fs=1/4$ and $\ft=3/4$ are mapped to the completely mixed state $\id_{2\otimes2}/4$.

The amount of information that is averaged out by the full twirl is huge: $\M[2][2]$ maps the 16-\D algebra ${\rm L}\qty(\cd[2]\otimes \cd[2])$ of all two-\q operators onto a 2-\D subspace, and the real $15$-\D submanifold ${\rm D}\qty(\cd[2]\otimes \cd[2])$ of all two-\q states onto a 1-\D simplex.

\subsubsection{Triplet and singlet subspaces as symmetric and antisymmetric subspaces.}\label{sec:TripSingSymAsym}
Let us now reinterpret the singlet--triplet decomposition independently of the angular-momentum formalism.
In all generality, the symmetric and antisymmetric subspaces of $\cd \otimes \cd$ can be defined from the swap operator on two $d$-\D qudits. The swap is singled out by its action on separable vectors, $\psi\otimes \phi\mapsto \phi\otimes \psi$, and decomposes on the standard basis as $\SW_{d\otimes d}
= \sum_{i,j=0}^{d-1}\dyad{ij}{ji}$. Therefore, it is unitary, Hermitian and involutive, hence its spectrum is $\{+1, -1\}$. The symmetric and antisymmetric subspaces can be defined as its eigenspaces relative to the eigenvalues $+1$ and $-1$, respectively. Taking the trace of the projections $(\id_{d\otimes d}\pm\SW_{d\otimes d})/2$, one immediately obtains that the symmetric subspace has dimension $d(d+1)/2$ and the antisymmetric $d(d-1)/2$.

In the case of interest, $d=2$, the projection on the antisymmetric subspace is
\begin{equation}
\frac{\id_{2\otimes 2}-\SW_{2\otimes 2}}{2}
=\frac{\dyad{01}{01}+\dyad{10}{10}
-\dyad{01}{10}-\dyad{10}{01}}{2}
=\dyad{\Psi^-}.
\end{equation}
By Eq.~\eqref{eq:PsiPhi}, $\dyad{\Psi^-}$ is the projection onto the singlet subspace. In other words, the singlet and the triplet subspaces are precisely the antisymmetric and symmetric subspaces of the two-\q space, respectively.

Alternatively, we can completely avoid dealing with vectors. The two-\q Pauli operators $X_\mu\otimes X_\nu$ form an orthogonal basis of ${\rm L}(\cd[2]\otimes \cd[2])$, and, since unitary, their Hilbert--Schmidt norms square to $4$. Expanding the swap operator on such basis yields
\begin{equation}\label{eq:swap2qubit}
\SW_{2\otimes 2}
%&=\frac{1}{4}\sum_{\mu,\nu=0}^3
%\inner{X_\mu\otimes X_\nu}{\SW_{2\otimes 2}}
%X_\mu\otimes X_\nu
=\frac{1}{4}\sum_{\mu,\nu=0}^3
\inner{X_\mu\otimes X_\nu}{\SW_{2\otimes 2}}
X_\mu\otimes X_\nu
%=\frac{1}{4}\sum_{\mu,\nu=0}^3
%\inner{X_\mu }{X_\nu}
%X_\mu\otimes X_\nu
%=\frac{1}{4}\sum_{\mu,\nu=0}^4
%\delta_{\mu\nu}X_\mu\otimes X_\nu
=\frac{1}{2}\sum_{\mu=0}^3 X_\mu\otimes X_\mu\,,
\end{equation}
where we used a general property of the swap, $\tr((A\otimes B)\SW_{d\otimes d})=\tr(AB)$, for $A,B\in {\rm L}(\cd)$, specialized for $d=2$, and the orthogonality relation between one-\q Pauli operators, $\inner{X_\mu}{X_\nu} = 2\delta_{\mu\nu}$ --- analogous considerations hold for $n$-\q systems.%, where $d_n = 2^n$.

In general, for the $(j_1\times j_2)$-\rep of $\SU$, acting on $\cd[2j_1+1] \otimes \cd[2j_2+1]$, we have the spectral decomposition $\vb*J^2 = \sum_j j(j+1) \,\Pi_{2j+1}$, where $j$ ranges in integer steps from $\abs{j_1-j_2}$ to $j_1+j_2$, and $\Pi_{2j+1}$ is the projection onto the $(2j+1)$-\D eigenspace $\smash{\ker( \vb*J^2  - j(j+1) )}$. In the case of  interest, $j_1=j_2=1/2$, we have $\vb*J^2=2\PT$, hence Eq.~\eqref{eq:J2} immediately allows us to express $\PT$ and $\PS=\id_{2\otimes 2}-\PT$ in terms of Pauli matrices, as
\begin{gather}\label{eq:PS}
	\PS=\frac{\id_{2\otimes 2} - X\otimes X - Y\otimes Y - Z\otimes Z}{4}
	=\frac{1}{2}\, \qty(\id_{2\otimes 2} 
	-\frac{1}{2} \sum_{\mu=0}^3 X_\mu\otimes X_\mu)
	=
	\frac{\id_{2\otimes 2}-\SW_{2\otimes 2}}{2}\,,
	\\\label{eq:PT}
\PT=\frac{3\id_{2\otimes 2} +X\otimes X+Y\otimes Y+Z\otimes Z}{4}
=\frac{1}{2}\, \qty(\id_{2\otimes 2} 
+\frac{1}{2} \sum_{\mu=0}^3 X_\mu\otimes X_\mu)
=
\frac{\id_{2\otimes 2}+\SW_{2\otimes 2}}{2}\,.
\end{gather}

\subsection{Pauli twirl of a two-\q  state} \label{sec:twirl2Pau}
Building on the previous section, to obtain a 2-\des we need to find $N$ distinct matrices $U_a\in \U[2]$ such that
\begin{equation}\label{eq:task}
\frac{1}{N} \sum_{a}
(U_a\otimes U_a )\rho 
\qty(U_a^\dag \otimes U_a^\dag)
=\tr(\PS\rho)\, \PS
+\tr(\PT\rho)\,\frac{\PT}{3}
\,.
\end{equation}
The obvious question is: where should we look for such unitaries? A good starting point is the observation from Sec.~\secref{pr:incrisi}: every 2-\des must also be a 1-\des. We know from Sec.~\secref{sec:onedes} that the Pauli basis is a 1-\des. Therefore, we can compute the Pauli twirl of $\rho$ and compare it with the full twirl (i.e., the right-hand side of the above equation). Our idea --- and our hope --- is that, even if the two twirls do not coincide, they may still be close enough to suggest how the Pauli basis could be modified into a 2-\des.

In fact, we can already tell that the Pauli basis is too small to be a 2-\des: from the considerations of Sec.~\secref{sec:DesVsKraus}, the size of any 2-\des of $\U[2]$ is bounded from below by the Choi rank of $\M[2][2]$, and, by Eq.~\eqref{eq:BigTwirlCh}, the Choi operator of $\M[2][2]$ is $\Choi(\M[2][2]) = \PS\otimes\PS + \PT\otimes\PT/3$, which has rank 10. Nevertheless, it is worth testing our intuition. Therefore, we now compute the Pauli twirl,
\begin{equation}\label{eq:BarRhoexpl}
	\M[{\B}][2](\rho)
	=\frac{1}{4}\,\qty[\rho
	+(X\otimes X) \rho (X\otimes X)
	+(Y\otimes Y) \rho (Y\otimes Y)
	+(Z\otimes Z) \rho (Z\otimes Z)]\,,
\end{equation}
and then compare it with the full twirl.

Before proceeding, we need the following observation.
Let $A=(a_{\mu\nu})$ be a square matrix, and let $L=\diag \vb*l=(l_\mu \delta_{\mu\nu})$ and $R=\diag \vb*r=(r_\mu \delta_{\mu\nu})$ be two diagonal matrices of the same order as $A$.
Then the product $LAR=(l_\mu a_{\mu\nu} r_\nu)$ can be written as
\begin{equation}\label{eq:HadProd}
LAR =  
%\color{red}
\qty(\vb*l\otimes\vb*r^\top)\odot A
%\color{black}
\,,
\end{equation}
where $\odot$ denotes the \emph{entry-wise} (or \emph{Hadamard}) \emph{product}, which inherits commutativity, associativity and distributivity from ordinary multiplication of real numbers, while $\otimes$ denotes the Kronecker product.

\subsubsection{Computation in the Bell basis.}
A convenient way to compute $\M[\B][2](\rho)$ is to work with matrix representations in the Bell basis~\eqref{eq:PsiPhi} --- see Sec.~\secref{par:BasisInd} for an alternative, more abstract (and perhaps more elegant) approach. This choice is convenient \emph{a priori} for two main reasons. First, since the triplet Bell projections resolve the projection onto the triplet subspace, the full twirl is diagonal in the Bell basis. Second, and more importantly, each operator $X_i\otimes X_i$ is diagonal in this basis, which greatly simplifies the computation of the Pauli twirl of Eq.~\eqref{eq:BarRhoexpl}. Indeed, ordering the basis vectors as
\begin{equation}\label{eq:OrdoBell}
	B_{\rm B}=\qty(\Psi^-,\Phi^-,\Psi^+,\Phi^+)\,,
\end{equation}
the identities~\eqref{eq:PauliPhi}--\eqref{eq:PauliPsi} translate into the matrix representations
\begin{align}
(X\otimes X)_{\rm B}
=\mqty*(\begin{BMAT}(b)[1pt]{cccc}{cccc}
-& &  & \\
& - &  &  \\
&  & + &  \\
&  &  & +
\end{BMAT}),
&&
(Y\otimes Y)_{\rm B}
=\mqty*(\begin{BMAT}(b)[1pt]{cccc}{cccc}
-& &  & \\
& + &  &  \\
&  & + &  \\
&  &  & -
\end{BMAT}),
&&
(Z\otimes Z)_{\rm B}
=\mqty*(\begin{BMAT}(b)[1pt]{cccc}{cccc}
-& &  & \\
& + &  &  \\
&  & - &  \\
&  &  & +
\end{BMAT}).
\end{align}
Equivalently, we have the spectral decompositions
\begin{align}\label{eq:XX}
	X\otimes X &= 
	-\PR{\Psi^-}-\PR{\Phi^-} +\PR{\Psi^+} +\PR{\Phi^+}\,,
	\\\label{eq:YY}
	Y\otimes Y &= 
	-\PR{\Psi^-} +\PR{\Phi^-}  +\PR{\Psi^+}-\PR{\Phi^+}\,,
	\\\label{eq:ZZ}
	Z\otimes Z &= 
	-\PR{\Psi^-} +\PR{\Phi^-}  -\PR{\Psi^+}+\PR{\Phi^+}\,.
\end{align}
Note that the above expressions entail the diagonal nature of $\vb*J^2$, as envisaged by Eq.~\eqref{eq:J2}, namely, $\qty(\vb*J^2)_{\rm B}=\diag (0,2,2,2)$.

In fact, we have more: due to the peculiar way in which the operators $X_i\otimes X_i$ are Bell-\dia, $\M[{\B}][2](\rho)$ turns out to be Bell-\dia as well. Indeed, we can write $(X_\mu\otimes X_\mu)_{\rm B}=\diag \vb*x_\mu$, where
\begin{align}
\vb*x_0=\mqty*(\begin{BMAT}(b)[1pt]{c}{cccc}
+\\+\\+\\+
\end{BMAT}),
&&
\vb*x_1=\mqty*(
\begin{BMAT}(b)[1pt]{c}{cccc}
-\\-\\+\\+
\end{BMAT}),
&&
\vb*x_2=\mqty*(\begin{BMAT}(b)[1pt]{c}{cccc}
+\\-\\-\\+
\end{BMAT}),
&&
\vb*x_3=\mqty*(\begin{BMAT}(b)[1pt]{c}{cccc}
-\\+\\-\\+
\end{BMAT}),
\end{align}
so that 
\begin{align}
	\vb*x_0\otimes \vb*x_0^\top
	&=\mqty*(\begin{BMAT}(b)[1pt]{cccc}{cccc}
		+&+ &  +&+ \\
		+&+ &  +&+ \\
		+&+ &  +&+ \\
		+&+ &  +&+ 
	\end{BMAT}),
	\\
	\vb*x_1\otimes \vb*x_1^\top
	&=\mqty*(\begin{BMAT}(b)[1pt]{cccc}{cccc}
		+&+ &  -&- \\
		+& + &  -&-  \\
		-&- &  +& + \\
		-& - &  +&+ 
	\end{BMAT}),
	\\
	\vb*x_2\otimes \vb*x_2^\top
	&=\mqty*(\begin{BMAT}(b)[1pt]{cccc}{cccc}
		+&- &  -&+ \\
		-& + &  +&-  \\
		-& + &  +&- \\
		+&- &  -&+ 
	\end{BMAT}),
	\\
	\vb*x_3\otimes \vb*x_3^\top
	&=\mqty*(\begin{BMAT}(b)[1pt]{cccc}{cccc}
		+&- &  +&- \\
		-& + &  -&+  \\
		+&- &  +&- \\
		-& + &  -&+ 
	\end{BMAT}),
\end{align}
resulting in
\begin{equation}
\frac{1}{4}\sum_{\mu=0}^3  
\vb*x_\mu\otimes \vb*x_\mu^\top
=\mqty*(\begin{BMAT}(b)[1pt]{cccc}{cccc}
+&\phantom{-} &  & \\
& + &  &  \\
& &  +& \\
&  &  &+ 
\end{BMAT})
=\id_4\,.
\end{equation}
By Eq.~\eqref{eq:HadProd},  averaging the conjugates of an arbitrary matrix $A=(a_{\mu\nu})_{\mu,\nu=0}^3$ by the (special orthogonal) matrices $\Xi_\mu=\diag \vb*x_\mu$ yields
\begin{equation}
\frac{1}{4}\sum_{\mu=0}^3  
\Xi_\mu\, A\, \Xi_\mu
=\qty(\frac{1}{4}\sum_{\mu=0}^3  
\vb*x_\mu\otimes \vb*x_\mu^\top)\odot A
=\id_4\odot A
=\mqty*(\begin{BMAT}(b){cccc}{cccc}
a_{00}& &  & \\
& a_{11}&  &  \\
& &  a_{22}& \\
&  &  & a_{33}
\end{BMAT}).
\end{equation}
In other words, $A \mapsto \tfrac{1}{4} \sum_{\mu=0}^3 \Xi_\mu \,A\, \Xi_\mu$ is a pinch map, reducing each matrix to its main diagonal.

Let us now specialize this result to the case of interest. (From now on, matrix representations in the Bell basis will make the singlet-triplet decomposition explicit.) The input state is represented by
\begin{equation}
\rho_{\rm B} =\mqty*(
\begin{BMAT}(b){c1ccc}{c1ccc}
\f{\Psi^-}
& \bra{\Psi^-}{\rho}\ket{\Phi^-} 
& \bra{\Psi^-}{\rho}\ket{\Psi^+} &\bra{\Psi^-}{\rho}\ket{\Phi^+}  \\
\bra{\Phi^-}{\rho}\ket{\Psi^-} 
& \f{\Phi^-}
& \bra{\Phi^-}{\rho}\ket{\Psi^+} 
& \bra{\Phi^-}{\rho}\ket{\Phi^+}\\
\bra{\Psi^+}{\rho}\ket{\Psi^-}  
& \bra{\Psi^+}{\rho}\ket{\Phi^-} 
& \f{\Psi^+}
& \bra{\Psi^+}{\rho}\ket{\Phi^+}\\
\bra{\Phi^+}{\rho}\ket{\Psi^-}  
& \bra{\Phi^+}{\rho}\ket{\Phi^-} 
& \bra{\Phi^+}{\rho}\ket{\Psi^+} 
& \f{\Phi^+}
\end{BMAT}),
\end{equation}
and its Pauli twirl by
\begin{equation}\label{eq:M2PauRhoMat}
\qty(\M[{\B}][2](\rho))_{\rm B}
=\mqty*(\begin{BMAT}{c1ccc}{c1ccc}
\f{\Psi^-}& & & \\
& \f{\Phi^-}& & \\
& & \f{\Psi^+}& \\
\phantom{m}& &  & \f{\Phi^+}
\end{BMAT}).
\end{equation}
Equivalently, we can expand $\rho$ on the Bell basis,
\begin{equation}\label{eq:RhoBell}
\rho
=\f{\Psi^-}\,\PR{\Psi^-} 
+\f{\Phi^-}\,\PR{\Phi^-} 
+\f{\Psi^+}\,\PR{\Psi^+} 
+\f{\Phi^+}\,\PR{\Phi^+} 
+ \sum_{\beta\ne\beta'} 
\mel{\beta}{\rho}{\beta'} \dyad{\beta}{\beta'},
\end{equation}
and the Pauli twirl extracts its diagonal part,
\begin{equation}\label{eq:M2PauRhoBis}
\M[{\B}][2](\rho)
=\f{\Psi^-}\,\PR{\Psi^-} 
+\f{\Phi^-}\,\PR{\Phi^-} 
+\f{\Psi^+}\,\PR{\Psi^+} 
+\f{\Phi^+}\,\PR{\Phi^+} 
\,.
\end{equation}
Summing up, twirling $\rho$ over the Pauli basis has the sole effect of averaging out all off-diagonal entries in the Bell basis, while leaving the diagonal ones untouched (that is,  preserving the expectation values of the Bell projections).
As a result, the Pauli twirl is a convex combination of the Bell states, with coefficients given by their respective expectation values in the input state $\rho$. These four nonnegative numbers, summing to~1, are the only information about $\rho$ that survives (and can be retrieved after) the twirl.

From the standpoint of quantum information theory, this is just the observation that $\M[\B][2]$ is the (completely) dephasing channel on two qubits with respect to the Bell basis, namely, $\rho\mapsto \sum_\beta \mel**{\beta}{\rho}{\beta} \dyad{\beta}{\beta}$. Observe also that $\M[\B][2](\rho)$ results from a non selective measurement on $\rho$ with measurement operators given by the Bell projections.

\subsubsection{Comparison with the full twirl.}\label{sec:comparatio}
Let us now compare the full twirl $\M[2][2](\rho)$ with the Pauli twirl $\M[{\B}][2](\rho)$. Since $\Psi^-$ spans the singlet subspace, while the other three Bell vectors form an orthonormal basis of the triplet subspace, we have
\begin{align}\label{eq:STvsBell}
\PS=\PR{\Psi^-}\,,
&&
\PT=\PR{\Psi^+}+\PR{\Phi^-}+\PR{\Phi^+}\,,
\end{align}
which entails
\begin{align}\label{eq:STvsBellComp}
\fs=\f{\Psi^-}\,,
&&
\ft=\f{\Psi^+}+\f{\Phi^-}+\f{\Phi^+}
=1-\f{\Psi^-}\,.
\end{align} 
Then the full twirl has the matrix representation
\begin{equation}\label{eq:M22RhoBellMat}
\qty(
\M[2][2](\rho))_{\rm B}
=\mqty*(
\begin{BMAT}{c1ccc}{c1ccc}
\fs&& & \\
& \ft/3& & \\
& &\ft/3& \\
& &  &\ft/3
\end{BMAT}
)
=\mqty*(
\begin{BMAT}{c1ccc}{c1ccc}
\f{\Psi^-}& & & \\
&\qty(1-\f{\Psi^-})/3&&\\
&&\qty(1-\f{\Psi^-})/3&\\
&&&\qty(1-\f{\Psi^-})/3
\end{BMAT}
),
\end{equation}
or, equivalently, the spectral decomposition
\begin{equation}\label{eq:M22RhoBell}
\M[2][2](\rho)= 
\f{\Psi^-}\,\PR{\Psi^-}
+\frac{1-\f{\Psi^-}}{3}
\,\qty( \PR{\Psi^+}+\PR{\Phi^-}+\PR{\Phi^+})\,.
\end{equation}
These expressions can be immediately compared with those for the Pauli twirl, given in Eqs.~\eqref{eq:M2PauRhoMat} and~\eqref{eq:M2PauRhoBis}.
Both states are Bell-\dia; however, the restriction of the full twirl to the triplet subspace is both diagonal and uniform --- more precisely, it is proportional to the identity on that subspace, with the proportionality factor given by the trace of the input state on the triplet. In other words, while the Pauli twirl removes all off-\dia components and leaves the diagonal ones unchanged, the full twirl not only suppresses the off-\dia terms but also completely mixes the diagonal components within the triplet subspace.

Thus, in contrast with the one-\q case, twirling a two-\q state over the Pauli operators is not enough to erease all the details averaged out by the full twirl. To obtain the latter from the former, the diagonal triplet components of the Pauli twirl must be averaged: each of the coefficients $\f{\Phi^-}$, $\f{\Psi^+}$, and $\f{\Phi^+}$ should be replaced by their arithmetic mean, $(\f{\Phi^-} + \f{\Psi^+} + \f{\Phi^+}) / 3 = (1-\f{\Psi^-})/3=\ft/3$. This will be the starting point of Sec.~\secref{sec:ave}.

\subsubsection{Properties of \texorpdfstring{$\M[{\B}][2]$}{the twirling channel over the Pauli basis}.
}\label{par:cannot}
Eq.~\eqref{eq:M2PauRhoBis} can also be written
\begin{equation}\label{eq:M2PauRhoHS}
	\M[{\B}][2](\rho)
%	=\sum_\beta 
%	\f{\beta}\PR{\beta}
	=\sum_\beta 
	\inner{\PR{\beta}}{\rho}\PR{\beta}
	%=\inner{\PR{\Psi^-}}{\rho}\PR{\Psi^-}\\
	%+\inner{\PR{\Phi^-}}{\rho}\PR{\Phi^-}
	%+\inner{\PR{\Psi^+}}{\rho}\PR{\Psi^+}
	%+\inner{\PR{\Phi^+}}{\rho}\PR{\Phi^+}
	\,,
\end{equation}
so that
\begin{equation}\label{eq:M2PauHS}
	\M[{\B}][2]
	=\sum_\beta \ketp{\PR{\beta}}\!\brap{\PR{\beta}}
	%=\ketp{\PR{\Psi^-}}\!\brap{\PR{\Psi^-}}
	%+\ketp{\PR{\Phi^-}}\!\brap{\PR{\Phi^-}}
	%+\ketp{\PR{\Psi^+}}\!\brap{\PR{\Psi^+}}
	%+\ketp{\PR{\Phi^+}}\!\brap{\PR{\Phi^+}}
	.
\end{equation}
The Bell projections form an orthonormal set in ${\rm L}\qty(\cd[2]\otimes \cd[2])$, thus $\M[{\B}][2]$ is the projection onto their span, consisting of all Bell-\dia operators.

\begin{figure}[!tb]
\centering
%\resizebox{\textwidth}{!}{
%\begin{tikzpicture}
%
%% Include the image in a node
%\node [above right,inner sep=0] (image) at (0,0) {\includegraphics[width=\textwidth]{2ad}};
%
%\node[circle] at (8.2,10.2){$\PR{\Psi^-}$};
%
%\node[circle] at (7.3,4.45){$\M[{\B}][2](\rho)$};
%
%\node[circle] at (8.66,3.25){$\id_{2\otimes 2}/4$};
%
%\node[circle] at (12.95,2.8){$\PR{\Phi^-}$};
%\node[circle] at (1.75,1.9){$\PR{\Psi^+}$};
%\node[circle] at (10,-0.3){$\PR{\Phi^+}$};
%
%
%%\node[circle] at (7,6.5){{\small $ x_2$}};
%
%
%\end{tikzpicture}
%}
\includegraphics[width=\textwidth]{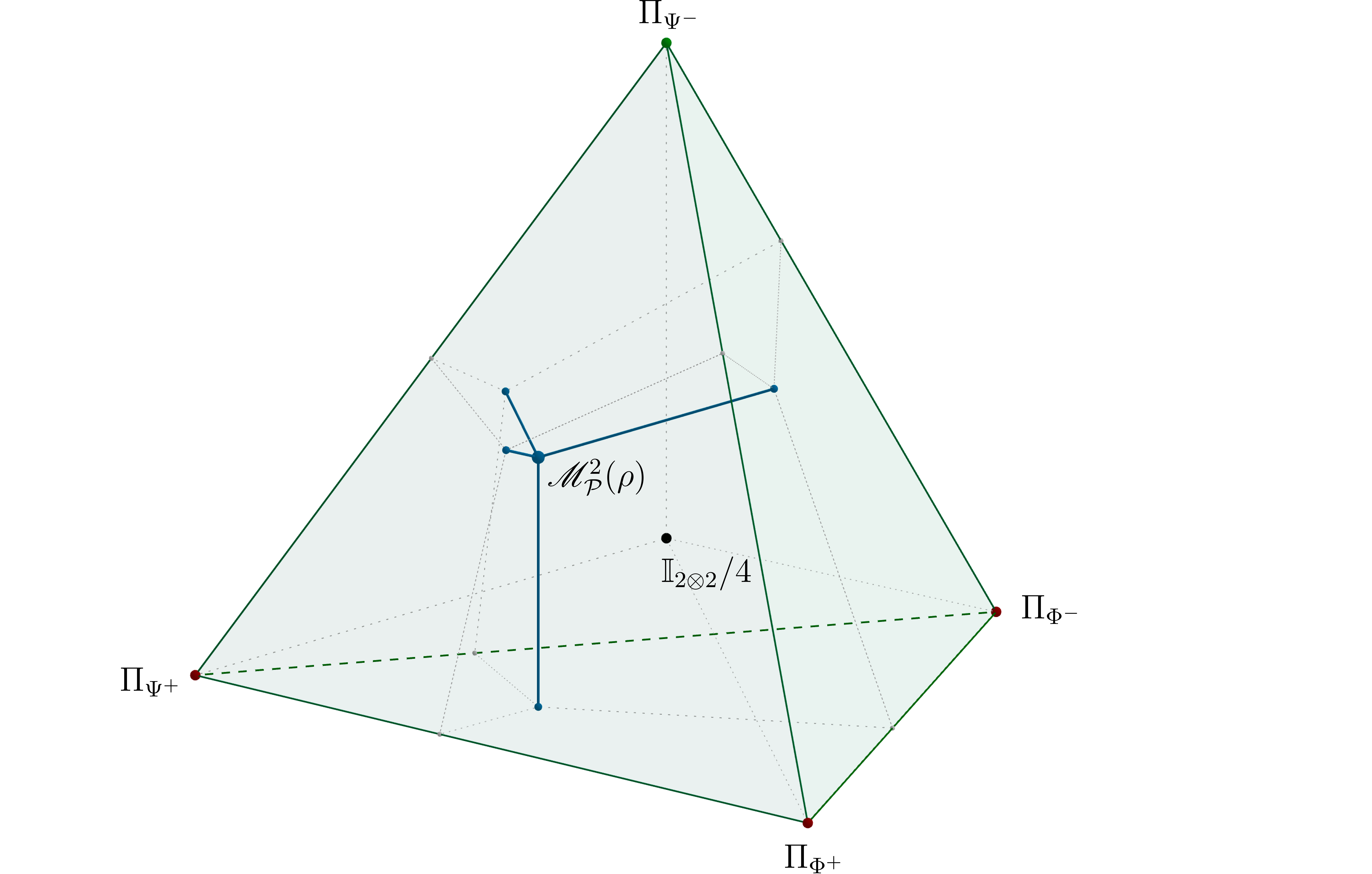}
\vspace{1pt}
\caption[LoF entry]{--- The Pauli twirl of a two-\q state (\bluupalla) --- explicitly,  $\M[{\B}][2](\rho)=\sum_\beta \f{\beta}\,\PR{\beta}$, where $\f{\beta}=\tr (\PR{\beta}\rho)$, with $\beta = \cramped{\Psi^\pm}, \cramped{\Phi^\pm}$ --- lies in the regular tetrahedron (\tetra) with vertices the Bell states (\singlo, \marunpalla). Its distances from the faces opposite to any two Bell states, $\PR{\beta_1}$ and $\PR{\beta_2}$, are in the ratio $\f{\beta_1} \!:\! \f{\beta_2}$. The completely mixed state (\neropalla) sits at the center of the tetrahedron, dividing each altitude in the ratio $3\!:\!1$. As a sidenote, observe that all surface points of the tetrahedron are boundary states, but only its vertices (i.e., the Bell states) are pure.
}\label{fig:tetra}
\end{figure}

In particular, the set of all two-\q  states is mapped onto the convex hull of the Bell states, that is, the regular tetrahedron with vertices the Bell states (see~\figref{fig:tetra}). Since the Bell states are orthogonal unit vectors in ${\rm L}\qty(\cd[2]\otimes \cd[2])$, the length of the edges is equal to $\sqrt 2$, hence the height of the tetrahedron is $2/\sqrt{3}$, consistently with what we found in Sec.~\secref{par:M22}. Indeed, the altitude through each Bell state is the segment connecting it to the center of the opposite face, which is exactly the completely mixed stated of the other three Bell states. In particular, the completely mixed triplet state, $\PT/3=\sum_{\beta\ne {\Psi^-}}\PR{\beta}/3$, sits in the center of the face opposite to the singlet state, hence the segment connecting $\PS$ and $\PT/3$ is the altitude through $\PR{\Psi^-}=\PS$. Each coefficient $\f{\beta}$ %($\beta=\Psi^\pm,\Phi^\pm$) 
is proportional to the distance of $\M[{\B}][2](\rho)$ from the face opposite to $\PR{\beta}$, the proportionality factor being the height of the tetrahedron. In other words, $\M[{\B}][2](\rho)$ is the point in the tetrahedron whose distance from the face opposite to each Bell state $\PR{\beta}$ is equal to $ 2\f{\beta}/\sqrt{3}$.
In particular, all states with $\f{\beta}=1/4$ are mapped to the completely mixed one.

\begin{figure}[!tb]
\centering
%	\resizebox{\textwidth}{!}{
%		\begin{tikzpicture}
%			
%			% Include the image in a node
%			\node [above right,inner sep=0] (image) at (0,0) {\includegraphics[width=\textwidth]{3ad}};
%			
%			%\draw [help lines] (0,0) grid (15,10);
%			
%			
%			\node[circle] at (8.32,10.2){$\PS=\PR{\Psi^-}$};
%			
%			\node[rectangle,%fill=white,opacity=.5,scale=1,inner sep=0.6,text opacity=1
%			] at (8.93,4.95) {$\M[2][2](\rho)$};
%			
%			%\node[circle] at (8.983,4.8){$\M[2][2](\rho)$};
%			
%			
%			\node[circle] at (5.9,4.3){
%				$\M[{\B}][2](\rho)$};
%			
%			
%
%			
%			%\node[circle] at (13.55,5.4){$\f{\Psi^-}\PR{\Psi^-} + (1-\f{\Psi^-})\PR{\Phi^-}$};
%			%
%			%\node[circle] at (3,5.1){$\f{\Psi^-}\PR{\Psi^-} + (1-\f{\Psi^-})\PR{\Psi^+}$};
%			
%			
%			\node[circle] at (7.5,3.4){$\id_{2\otimes 2}/4$};
%			
%			
%\node[circle] at (12.95,2.8){$\PR{\Phi^-}$};
%\node[circle] at (1.75,1.9){$\PR{\Psi^+}$};
%
%\node[circle] at (8.2,1.25){$\PT/3$};
%
%\node[circle] at (10,-0.3){$\PR{\Phi^+}$};
%			
%			
%			
%			
%			
%			%\node[circle] at (7,6.5){{\small $ x_2$}};
%			
%			
%		\end{tikzpicture}
%}
\includegraphics[width=\textwidth]{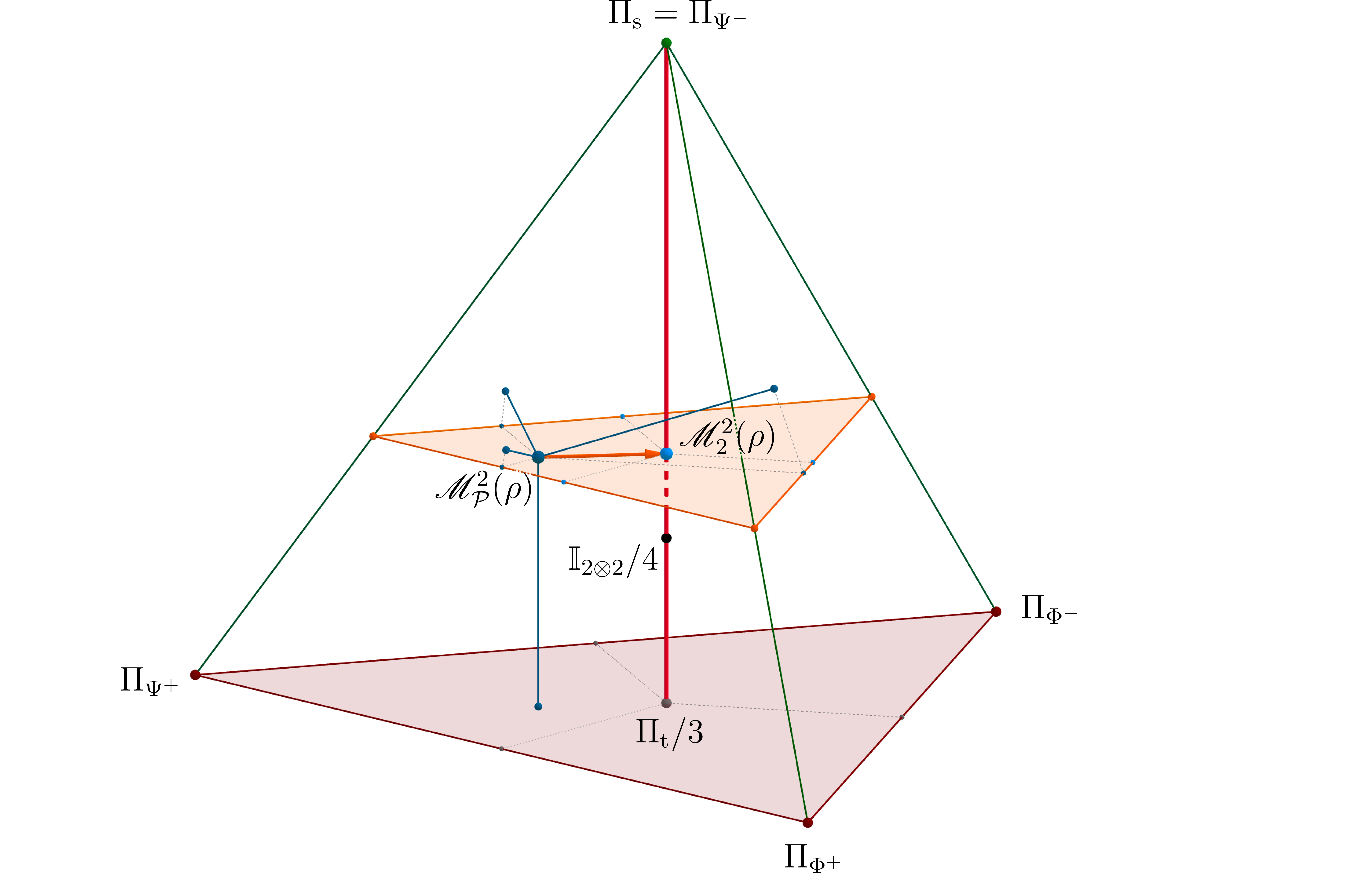}
\vspace{1pt}
\caption[LoF entry]{--- The full twirl $\M[2][2](\rho)$ (\azurpalla), and the Pauli twirl $\M[\B][2](\rho)$ (\bluupalla) are at the same distance from the states in the triplet subspace (\maruntri). Such distance is proportional to $\fs=\tr(\PS\rho)$ by the length of the segment (\rossolinea) connecting $\PS$ and $\PT/3$. While the Pauli twirl can be any point of the equilateral triangle (\aranpalla) with vertices the states $\fs\PR{\Psi^-} + \ft\PR{\beta}$ ($\beta\ne \Psi^-$) (\oranst), the full twirl sits exactly at its center.
}
\label{fig:tetraComp}
\end{figure}

The loss of information occurring in a Pauli twirl is still huge, but less pronounced than for the full twirl (cf.~\figref{fig:tetraComp}): now the 16-\D algebra ${\rm L}\qty(\cd[2]\otimes \cd[2])$ of all two-\q operators is mapped onto a 4-\D (as opposed to 2-\D) subspace, and the real $15$-\D submanifold ${\rm D}\qty(\cd[2]\otimes \cd[2])$ of all two-\q states is mapped onto a 3-\D (as opposed to 1-\D) simplex.

\subsubsection{Bell states and \texorpdfstring{$X_i\otimes X_i$}{X_i\otimes X_i} operators.}\label{par:BasisInd} 
Let us delve deeper into the relationship between Bell states and the operators $X_i\otimes X_i$. This will allow us to recover all the previous results through a basis-in\-de\-pend\-ent approach
%\color{red}
(along the lines of the stabilizer formalism)%
%\color{black}
.

First of all, each operator $X_i\otimes X_i$ is Hermitian, involutive, and unitary, as well as traceless. As a consequence, its eigenvalues are $\pm1$; both are regular, and have multiplicity 2. In particular, the determinant of $X_i\otimes X_i$ is 1. 

Next, let $\mathscr{X}_+$ and $\mathscr{X}_-$ denote the eigenspaces of $X\otimes X$ corresponding to the eigenvalues $1$ and $-1$, respectively, that is,  $\mathscr{X}_\pm=\ker(X\otimes X\mp 1)$; $\mathscr{Y}_\pm$ and $\mathscr{Z}_\pm$ are introduced likewise. These subspaces are 2-\D, and mutually orthogonal. Thus we can write, e.g., $\mathscr{X}_+=(\mathscr{X}_+\cap \mathscr  Z_+)\oplus(\mathscr{X}_+\cap \mathscr  Z_-) $, as $\mathscr  Z_+\oplus \mathscr  Z_-$ is the entire space. If the subspace $\mathscr{X}_+\cap \mathscr  Z_+$ were trivial --- likewise for $\mathscr{X}_+\cap \mathscr  Z_-$ --- then we would have $\mathscr{X}_+ = \mathscr  Z_-$, hence also $\mathscr{X}_-= \mathscr  Z_+$. In this way, $X\otimes X$ and $- Z\otimes Z$ would have the same eigenspaces, hence the same spectral decomposition, and therefore coincide. Consequently, $Y\otimes Y$ would be a scalar, which is a contradiction. Therefore the intersections between eigenspaces of different observables, namely, $\mathscr{X}_{\xi}\cap \mathscr  Y_{\eta}$, $\mathscr{X}_{\xi}\cap \mathscr  Z_{\zeta}$, and $\mathscr{Y}_{\eta}\cap \mathscr  Z_{\zeta}$, for $\xi,\eta,\zeta\in\{1,-1\}$, are all 1-\D. 

Now the multiplication rule~\eqref{eq:PauliComp} --- a more fundamental identity than the anticommutation relations --- entails that, if  $X\otimes X\ket{\Theta}=\xi \ket{\Theta}$, and $Z\otimes Z\ket{\Theta}=\zeta\ket{\Theta}$, then $Y\otimes Y\ket{\Theta}=-\xi \zeta \ket{\Theta}$. This observation shows that
\begin{align}
\mathscr{Y}_+
=\qty(\mathscr{X}_+\cap \mathscr  Z_-) \oplus \qty(\mathscr{X}_-\cap \mathscr  Z_+)\,,
&&
\mathscr{Y}_-
=\qty(\mathscr{X}_+\cap \mathscr  Z_+) \oplus \qty(\mathscr{X}_-\cap \mathscr  Z_-)\,,
\end{align}
and, more generally,
\begin{equation}
\mathscr X_{\xi}\cap \mathscr Y_{\eta}\cap\mathscr  Z_{\zeta}=
\begin{cases}
\qty{\vb*0}\,,
&\text{if~}\,\xi\eta\zeta=1\,,
\\[5pt]
\mathscr X_{\xi}\cap \mathscr Y_{\eta}
=\mathscr X_{\xi}\cap \mathscr Z_{\zeta}
=\mathscr Y_{\eta}\cap \mathscr Z_{\zeta}\,,
\quad&\text{if~}\, \xi\eta\zeta=-1\,.
\end{cases}
\end{equation}
As a result, the entire two-\q space decomposes into four 1-\D and mutually orthogonal subspaces,
\begin{equation}
\cd[2]\otimes \cd[2]
=\qty( \mathscr{X}_-\cap\mathscr{Y}_-\cap\mathscr{Z}_-)
\oplus\qty( \mathscr{X}_-\cap\mathscr{Y}_+\cap\mathscr{Z}_+)
\oplus\qty( \mathscr{X}_+\cap\mathscr{Y}_+\cap\mathscr{Z}_-)
\oplus\qty( \mathscr{X}_+\cap\mathscr{Y}_-\cap\mathscr{Z}_+)
\,.
\end{equation}
These subspaces can be labelled by $\Psi^-,\Phi^-,\Psi^+,\Phi^+$, and define four pairwise orthogonal 1-\D projections, yielding the spectral decompositions~\eqref{eq:XX}--\eqref{eq:ZZ}, which, together with the resolution of the identity, yield the inverse relations
\begin{gather}
\PR{\Psi^-}
=\frac{1}{4} \,\qty(\id_{2\otimes 2} -X\otimes X -Y\otimes Y -Z\otimes Z )\,,\\
\PR{\Phi^-}
=\frac{1}{4} \,\qty(\id_{2\otimes 2} -X\otimes X+Y\otimes Y +Z\otimes Z)\,,\\
\PR{\Psi^+}
=\frac{1}{4} \,\qty(\id_{2\otimes 2}+X\otimes X+Y\otimes Y -Z\otimes Z) \,,\\
\PR{\Phi^+}
=\frac{1}{4} \,\qty(\id_{2\otimes 2}+X\otimes X-Y\otimes Y +Z\otimes Z)\,.
\end{gather}

We can now bring angular momentum into the picture. By Eqs.~\eqref{eq:PS}--\eqref{eq:PT} we immediately obtain $\PS=\PR{\Psi^-}$, as well as $\PT=\PR{\Phi^-}+\PR{\Psi^+}+\PR{\Phi^+}$, hence the usual ensuing formulae hold. Observe, in particular, that $\mathscr{X}_- \cap\mathscr{Y}_- \cap\mathscr{Z}_-$ is the kernel of $\vb*J^2$, hence each component $J_i$ vanishes on this subspace. Moreover, by Eq.~\eqref{eq:totalJ}, we immediately get
\begin{align}
J_x \PR{\Phi^-}=0\,,
&&
J_y \PR{\Phi^+}=0\,,
&&
J_z \PR{\Psi^+}=0\,,
\end{align}
that is, $J_x$ also vanishes on $\mathscr{Y}_+\cap\mathscr{Z}_+$ (the direction of $\Phi^-$), $J_y$ on $\mathscr{X}_+\cap\mathscr{Z}_+$ (the direction of $\Phi^+$), and $J_z$ on $\mathscr{X}_+\cap\mathscr{Y}_+$ (the direction of $\Psi^+$).

We can now characterize Bell-di\-ag\-o\-nal\-i\-ty: a two-\q operator is Bell-\dia if and only if it commutes with all operators $X_i\otimes X_i$, if and only if it commutes with any two of them. As a consequence, each projection $\PR{\beta}$ is invariant under a Pauli twirl, hence $\ev**{\rho}{\beta}=\ev**{\M[{\B}][2](\rho)}{\beta}$.

Finally, for $A\in {\rm L}(\cd[2]\otimes \cd[2])$ we have
\begin{gather}
\comm{(X\otimes X) A (X\otimes X)}{X\otimes X}
=\comm{X\otimes X}{A}\,,\\
\comm{(Y\otimes Y) A (Y\otimes Y)}{X\otimes X}
=(Y\otimes Y) A (Z\otimes Z)
-(Z\otimes Z) A (Y\otimes Y) \,,\\
\comm{(Z\otimes Z) A (Z\otimes Z)}{X\otimes X}
=(Z\otimes Z) A (Y\otimes Y)
-(Y\otimes Y) A (Z\otimes Z)\,.
\end{gather}
from which $\M[{\B}][2](A)$ commutes with $X\otimes X$, and, likewise, with $Z\otimes Z$, hence is Bell-\dia.

\subsection{Extending the Pauli basis to a 2-\des}\label{sec:ave}
In the previous section we saw that averaging the components of the Pauli twirl along the triplet Bell states yields the full twirl. Now the interesting question is: can this transformation be implemented via a twirl? In other words, we can look for $N$ distinct matrices $U_k\in \U[2]$ such that, for any 2-\q operator $A$ of the form
\begin{equation}\label{eq:diag}
A
=a_0\,\PR{\Psi^-}
+a_1\,\PR{\Phi^-}
+a_2\,\PR{\Psi^+}
+a_3\,\PR{\Phi^+}\,,
\end{equation}
we have
\begin{equation}\label{eq:TripletAver}
\frac{1}{N}\sum_k
\qty(U_k \otimes U_k) A 
\qty(U_k^\dag \otimes U_k^\dag )
%=a_{\Psi^-}\,\PR{\Psi^-}
=a_0\,\PR{\Psi^-}
+\bar a\,\qty( \PR{\Psi^+}+\PR{\Phi^-}+\PR{\Phi^+})
%=a_{\Psi^-}\,\PS+ \bar a\, \PT
%=a_0\,\PS+ \bar a\, \PT
\,,
\end{equation}
where $\bar a =(a_1+a_2+a_3)/3$.
Note that we are not concerned with what happens to operators that are not Bell-\dia. Our only request is that, when twirling a Bell-\dia operator, the singlet component is preserved, and the triplet component is replaced by the scalar with the same trace. If we succeed, then the $4N$ unitaries $U_k X_\mu$ would yield a 2-\des.

In the following, it is still convenient to work in the Bell basis, where the problem translates into finding $N$ matrices $U_k\in \U[2]$, such that, if a 2-\q operator $A$ has the Bell-\rep
\begin{equation}\label{eq:diagBell}
A_{\rm B}=\mqty*(\begin{BMAT}(b){c1ccc}{c1ccc}
a_0& & &\\
& a_1& & \\
& & a_2& \\
& &  & a_3
\end{BMAT}),
\end{equation}
then
\begin{equation}\label{eq:Ave3Bell}
\frac{1}{N}\sum_k
\qty(U_k\otimes U_k)_{\rm B}\, A_{\rm B}\, \qty(U_k^\dag \otimes U_k^\dag )_{\rm B}
=\mqty*(
\begin{BMAT}(b){c1ccc}{c1ccc}
a_0& & & \\
& \bar a & & \\
& & \bar a & \\
& &  &\bar a 
\end{BMAT}).
\end{equation}

We can immediately check whether a solution is at hand. For instance, we can consider the right (or forward) and left (or backward) shift operators,
\begin{align}\label{eq:shift3}
\Sigma_{\rm r}
= \mqty*(\begin{BMAT}(b){ccc}{ccc}
\mathmakebox[\widthof{$m$}]{0}
& 0&1\\
1& 0 &0\\
0&1&0
\end{BMAT})
=\sum_{i=1}^3 \vb e_{i\oplus 1}\otimes 
%\color{red}
\vb e_i^\top 
%\color{black}
\,,
&&
\Sigma_{\rm l}
= \mqty*(\begin{BMAT}(b){ccc}{ccc}
\mathmakebox[\widthof{$m$}]{0}
& 1&0\\
0& 0 &1\\
1&0&0
\end{BMAT})
%=\sum_{i=1}^3 \vb e_{i\ominus 1}\otimes \color{red}
\vb e_i^\top 
%\color{black}
\,,
\end{align}
where $\oplus$ and $\ominus$ denote addition and subtraction modulo 3, respectively. Each operator performs a cyclic permutation of the standard basis vectors and the inverse permutation of the standard components: $\Sigma_{\rm r}\vb e_i=\vb e_{i\oplus 1}$, and $\Sigma_{\rm r} (x_i) = \qty(x_{i\ominus 1})$; and $\Sigma_{\rm l}\vb e_i=\vb e_{i\ominus 1}$, and $\Sigma_{\rm l}(x_i = \qty(x_{i\oplus 1})$.
Moreover, each is the transpose, the inverse, and the square of the other, hence is a rotation of algebraic order~3 (i.e., by an angle $2\pi/3$, cf.\ Sec.~\secref{sec:W}).
Given a diagonal matrix
\begin{equation}
M=\mqty*(\begin{BMAT}{ccc}{ccc}
m_1&& \\
& m_2& \\
& &  m_3
\end{BMAT})
=\sum_{i=1}^3 m_i \,\vb e_i\otimes \vb e_i^\top 
\,,
\end{equation}
conjugating $M$ by $\Sigma_{\rm l}$ --- likewise for $\Sigma_{\rm r}$ --- amounts to left-shift\-ing its diagonal entries,
\begin{equation}
\Sigma_{\rm l} M \Sigma_{\rm l}^\dag
= \sum_{\mathclap{i,j,k=1}}^3 \,\delta_{ij} m_j \delta_{jk} \,
\vb e_{i\ominus 1} \otimes \vb e_{k\ominus 1}^\top
=\sum_{i=1}^3 m_{i\oplus 1}\, \vb e_i \otimes 
\vb e_i^\top 
=\mqty*(\begin{BMAT}{ccc}{ccc}
m_2&& \\
& m_3& \\
& &  m_1
\end{BMAT}).
\end{equation}
The same result can be achived, more generally, by any unitary of the form
\begin{equation}\label{eq:projshift}
Q=\mqty*(\begin{BMAT}(b){ccc}{ccc}
0& \e^{\iu\varphi_2} & 0\\
0& 0 &\e^{\iu\varphi_3}\\
\e^{\iu\varphi_1}&0&0
\end{BMAT})
=\sum_{i=1}^3 \e^{\iu\varphi_i}\,\vb e_{i\ominus 1}\otimes \vb e_i^\top 
\,,
\end{equation}
since its action on the basis vectors differs from that of $\Sigma_{\rm l}$ only by a phase shift, which is immaterial to conjugation. Its cube is still a scalar, namely, $Q^3 = \exp(\iu\qty(\varphi_1 +\varphi_2 +\varphi_3))\,\id_3 =\det(Q)\, \id_3$, so that $Q$ is of algebraic order 3 if and only if it is special unitary. Averaging the conjugates of $M$ by the unitaries $\id_3$, $Q$, and $Q^\dag=Q^{-1}$ yields
\begin{equation}
\frac{1}{3}\qty(M + Q M Q^\dag+ Q^\dag M Q)
%\sum_{k=0,\pm1}
%Q^k M {Q^k}^\dag
=\sum_{i=1}^3
\frac{m_i + m_{i\oplus 1} + m_{i\ominus 1}}{3}\, 
\vb e_i \otimes \vb e_i^\top 
=\bar m\,\id_3
=\mqty*(\begin{BMAT}(b){ccc}{ccc}
\bar m&&\\
& \bar m& \\
& &  \bar m
\end{BMAT})
\,,
\end{equation}
where $\bar m=(m_1+m_2+m_3)/3$.
So the idea is to examine the general form of the Bell-\rep of $U\otimes U$, and check whether, among these matrices, we can find one in the block-\dia form $1\oplus Q$, with $Q$ of the form~\eqref{eq:projshift}.

As usual, there is no loss of generality in assuming $U\in \SU$.
Then the theory of angular momentum provides a standard way to decompose the $(\frac{1}{2}\times\frac{1}{2})$-\rep into spin-0 and spin-1 components. Let $\ket{(j,m)}$ denote the angular momentum basis vectors (this notation should prevent confusion between the standard and the angular momentum bases),
\begin{align}\label{eq:amBasisI}
\ket{(0,0)}
&=\frac{\ket{01}-\ket{10}}{\sqrt 2}=\ket{\Psi^-},\\
\ket{(1,1)}
&=\ket{00}=\frac{\ket{\Phi^+}+\ket{\Phi^-}}{\sqrt 2}\,,\\
\ket{(1,0)}
&=\frac{\ket{01}+\ket{10}}{\sqrt 2}=\ket{\Psi^+},
\\\label{eq:amBasisF}
\ket{(1,-1)}
&=\ket{11}=\frac{\ket{\Phi^+}-\ket{\Phi^-}}{\sqrt 2}\,.
\end{align}
When dealing with matrix representations, the angular momentum basis will be ordered as above. The singlet--triplet decomposition of the $(\frac{1}{2}\times\frac{1}{2})$-\rep is block-\dia in this basis, and the matrix elements are provided by the Wigner $D$-matrices,
\begin{equation}\label{eq:UUam}
\qty(U\otimes U)_{\rm am}
=
\mqty*(\begin{BMAT}{c1ccc}{c1ccc}
1& &  &\\
\phantom{m}& & &\vphantom{n} \\
&&D^{(1)}(U)  & \\
& & & \vphantom{n}
\end{BMAT}),
\end{equation}
where $\mel**{(1,m)}{D^{(1)}(U)}{(1,m')}=
D^{(1)}_{m m'}(U)$, with $m,m'=1,0,-1$.

When $U$ is parametrized by Euler angles,
\begin{equation}
U(\alpha\beta\gamma)
=\e^{-\iu\alpha Z/2}\,
\e^{-\iu\beta Y/2}\, 
\e^{-\iu\gamma Z/2}\,,
\end{equation}
we write $D^{(1)} (\alpha\beta\gamma)$ for $D^{(1)} (U(\alpha\beta\gamma))$, which takes the form
\begin{equation}\label{eq:spin1}
D^{(1)}(\alpha\beta\gamma)
%=\mqty*(D^{(1)}_{mm'} (U(\alpha\beta\gamma)))
=\mqty*(
\begin{BMAT}[2.5pt]{ccc}{ccc}
\e^{-\iu (\alpha +\gamma )} {\rm c}_{\beta/2}^2
&\;\,- \e^{-\iu\alpha} {\rm s}_\beta/\sqrt{2}\;\,
& \e^{-\iu(\alpha -\gamma )} {\rm s}_{\beta/2}^2
\\
\e^{-\iu\gamma} {\rm s}_\beta /\sqrt{2}
& {\rm c}_\beta
& - \e^{\iu\gamma} {\rm s}_\beta /\sqrt{2}
\\
\e^{\iu (\alpha -\gamma )} {\rm s}_{\beta/2}^2
& \e^{\iu\alpha} {\rm s}_\beta /\sqrt{2}
& \e^{\iu (\alpha +\gamma )} {\rm c}_{\beta/2}^2
\end{BMAT}),
\end{equation}
where ${\rm c}_x=\cos x$ and ${\rm s}_x=\sin x$. 
The Bell-\rep of $U\otimes U$ is obtained by a change of basis,
\begin{equation}
\qty(U(\alpha\beta\gamma)\otimes U(\alpha\beta\gamma))_{\rm B}
= \braket{B_{\rm B}}{B_{\rm am}} \,
\qty(U(\alpha\beta\gamma)\otimes U(\alpha\beta\gamma))_{\rm am}\,
\braket{B_{\rm am}}{B_{\rm B}}.
\end{equation}
By Eqs.~\eqref{eq:amBasisI}--\eqref{eq:amBasisF}, the transition matrix from the Bell to the angular momentum basis is
\begin{equation}\label{eq:transAMBell}
\braket{B_{\rm am}}{B_{\rm B}}
%=\mqty*(
%\braket{(j,m)}{\beta}
%)
=\mqty*(
1&0 &0  &0\\
0& 1/\sqrt 2&  0& 1/\sqrt 2 \\
0&0&\mathmakebox[\widthof{n}]{1} & 0\\
0& -1/\sqrt 2& 0& 1/\sqrt 2
),
\end{equation}
from which a straightforward computation yields
\begin{equation}\label{eq:UUBell}
\qty(U(\alpha\beta\gamma)\otimes U(\alpha\beta\gamma))_{\rm B}
=\mqty*(
\begin{BMAT}[2.5pt]{c1ccc}{c1ccc}
\mathmakebox[\widthof{$-1$}]{1}&0&0& 0\\
0
& {\rm c}_ \alpha  \,{\rm c}_ \beta  \,{\rm c}_ \gamma -{\rm s}_ \alpha  \,{\rm s}_\gamma  
& \;-{\rm c}_ \alpha  \,{\rm s}_ \beta \;
&-\iu ({\rm c}_ \alpha  \,{\rm c}_ \beta  \,{\rm s}_ \gamma +{\rm s}_ \alpha  \,{\rm c}_ \gamma  )
\\
0
&{\rm s}_ \beta  \,{\rm c}_ \gamma  
& {\rm c}_ \beta 
&-\iu{\rm s}_ \beta  \,{\rm s}_ \gamma  
\\
0
&  -\iu ({\rm s}_ \alpha  \,{\rm c}_ \beta  \,{\rm c}_ \gamma +{\rm c}_ \alpha  \,{\rm s}_ \gamma)  
& \iu {\rm s}_ \alpha  \,{\rm s}_ \beta
&- {\rm s}_ \alpha  \,{\rm c}_ \beta  \,{\rm s}_ \gamma  +{\rm c}_\alpha  \,{\rm c}_\gamma
\end{BMAT}).
\end{equation}

We now look for Euler angles such that the triplet block of the above matrix takes the form~\eqref{eq:projshift}. Within the standard range $(\alpha,\beta,\gamma) \in [0,2\pi\mathclose{[}\times [0,\pi]\times [0,4\pi\mathclose{[}$, the solutions to $\sin \beta \,\cos\gamma=0$, $\cos \beta=0$ and $\sin\alpha\,\sin\beta=0$ are 
%\color{red}
$\alpha =0$, $\beta=\pi/2$, and $\gamma=\pi/2$, up to multiples of $\pi$.
%\color{black}
Let us choose
\begin{equation}\label{eq:Wdef}
W= U\qty(0,\frac{\pi}{2},\frac{\pi}{2})
=\e^{-\iu\pi Y/4}\,\e^{-\iu\pi Z/4}\,,
\end{equation}
which gives
\begin{equation}\label{eq:WWBell}
\qty(W\otimes W)_{\rm B}=
\mqty*(\begin{BMAT}(b){c1ccc}{c1ccc}
1&0&0& 0\\
0& 0& -1 &0\\
0 &0&0 &-\iu\\
0& -\iu& 0 &0
\end{BMAT}).
\end{equation}
Then Eq.~\eqref{eq:Ave3Bell} is satisfied by $U_k=W^k$, with $k=0,1,-1$,
\begin{equation}
\M[2][2](\rho)
=\frac{1}{3}\qty[ \M[{\B}][2](\rho)+
\qty(W \otimes W) \M[{\B}][2](\rho)
\qty(W^\dag \otimes W^\dag)
+\qty(W^\dag \otimes W^\dag)\M[{\B}][2](\rho)
\qty(W \otimes W) 
]\,.
%=\frac{1}{3} \sum_{k=0,\pm1}
%\qty(W^k \otimes W^k) \M[{\B}][2](\rho)
%\qty(W^{k\dag} \otimes W^{k\dag})\,,
\end{equation}
Summing up, we have just shown that
\begin{equation}\label{eq:task3}
\M[2][2](\rho)
=\frac{1}{12}
\sum_{k=0,\pm1}
\sum_{\mu=0}^3
\qty (W^k X_\mu \otimes W^k X_\mu) 
\,\rho\qty((W^k X_\mu)^\dag\otimes (W^k X_\mu)^\dag)
=\M[\DD][2](\rho)
\,,
\end{equation}
where the set
\begin{equation}\label{eq:D}
\DD =
\qty{\id_2,\,  X,\, Y,\, Z, \,
W, \, WX,\, WY, \, WZ,\,
W^\dag,  \,W^\dag X,  \,W^\dag Y,\,  W^\dag Z }\,,
\end{equation}
consisting of the twelve distinct unitaries $W^k X_\mu$ with $k=0,1,-1$ and $\mu=0,1,2,3$, extends the Pauli basis to a 2-\des. In other words, the 1-\des $\B$ has been extended to the 2-\des $\DD$ by adding its left translates by $W$ and $W^\dag$,
\begin{equation}\label{eq:DCompl}
\DD
=\B \cup W\B \cup W^\dag\B
=\bigcup_{k=0,\pm1} W^k \B
\,.
\end{equation}
This design has minimum size, insofar as it saturates the Clifford bound, stating that  a unitary 2-\des of $\U$ has no less than $d^4-d^2$ elements~\cite{gross, RoyScott}.

\subsubsection{Extending any 1-\des of minimum size to a 2-\des of minimum size.}\label{sec:phase}
The above procedure, encapsulated in Eqs.~\eqref{eq:Wdef} and~\eqref{eq:DCompl}, can be used to extend  any 1-\des of minimum size to a 2-\des, also of minimum size. From Sec.~\secref{sec:SU} we know that a 1-\des of minimum size is of the form $\B'=\{\e^{\iu \phi_\mu}  V X_\mu  V'\}$, with $V,V'\in \SU$, and $\phi_\mu\in \mathbb{R}$. 
Moreover, from the observations in Sec.~\secref{sec:homopoly}, we know that the set $\DD'$, consisting of the twelve matrices $\e^{\iu \phi_\mu} V X_\mu  V'$, $\e^{\iu \phi_\mu} V W  X_\mu  V'$, and $\e^{\iu \phi_\mu} V W^\dag X_\mu  V'$ ($\mu=0,1,2,3$), is a 2-\des. Therefore, by setting 
\begin{equation}
\tilde W=VWV^\dag\,,
\end{equation}
we obtain
\begin{equation}\label{eq:D'}
\DD'
=\B'\cup \tilde W \B'\cup {\tilde W}^\dag\B'
=\bigcup_{k=0,\pm1} \tilde W^k \B'
\,.
\end{equation}
These equations generalize our construction to an arbitrary 1-\des of minimum size.

\subsubsection{Interpretation of $W$ and $W^\dag$.}\label{sec:W}
By Eq.~\eqref{eq:Wdef}, $W$ has the following Euler and exponential forms:
\begin{equation}\label{eq:W}
W
=\frac{\id_2-\iu Y}{\sqrt 2}\,
\frac{\id_2-\iu Z}{\sqrt 2}
=\frac{1}{2}\qty[\id_2 - \iu(X+Y+Z)]
=\frac{1}{2} \,\id_2 -
\iu \frac{\sqrt 3}{2} \,\vb*e\cdot \vb*X
=\exp (-\iu \frac{\pi}{3}\, \vb*e\cdot \vb*X)\,,
\end{equation}
where we set
\begin{equation}\label{eq:E}
	\vb*e=\frac{1}{\sqrt 3} (1,1,1)\,.
\end{equation}

From an algebraic standpoint, $W$ generates the cyclic group $\left\langle W \right\rangle$ of order 6, made up of three antipodal pairs: $\id_2$ and $W^3=-\id_2$; $W$ and $W^4=-W$; $W^2=-W^\dag$ and $W^5=W^\dag=W^{-1}$. 
\begin{figure}[!h]
\centering
\begin{tikzpicture}
	[decoration =
	{markings
		,mark=at position 0.55 with {\arrow{stealth'}}
	}
	]
	\newdimen\R
	\R=1.5cm
	%	\node {$\times W$
		%	};
	%	\draw[->] (0:{0.7*\R}) arc (0:330:{0.7*\R});
	\foreach \x/\l in
	{ 60/,
		120/,
		180/,
		240/,
		300/,
		360/
	}
	\draw[thick,Sepia
	%BrickRed!80!black
	%,postaction={decorate}
	] ({\x-60}:\R) -- node[auto,swap]{\l} (\x:\R);

	\foreach \x/\l/\p in
	{ 0/{$\mathmakebox[\widthof{$W^3=-\id_2$}][l]{\id_2}%=(-\id_2)^2
			$}/right,
		60/{$W%={W^\dag}^5
			$}/above right,
		120/{$W^2=-W^\dag%={W^{4\dag}^4
				$
			}/above left,
			180/{$W^3=-\id_2$}/left,
			240/{$W^4=-W%={W^{2\dag}
					$
				}/below left,
				300/{$W^5=W^\dag$}/below right
			}
			\node[inner sep=1.8pt,circle,draw,fill=BrickRed,label={\p:\l}] at (\x:\R) {};

			\foreach \x/\l in
			{ 120/,
				240/,
				360/
			}
			\draw[thick,pesca,%postaction={decorate},
			inner sep=1.8pt,circle,draw] ({\x-120}:\R) -- node[auto,swap]{\l} (\x:\R);
			
			\foreach \x in
			{ 0,
				120,
				240
			}
			\node[inner sep=1.8pt,circle,draw,fill=BurntOrange] at (\x:\R) {};
			
			\foreach \x/\l in
			{
				180/
			}
			\draw[thick,battleship%,postaction={decorate}
			] ({\x-180}:\R) -- node[auto,swap]{\l} (\x:\R);
			
			\foreach \x in
			{ 180
			}
			\node[inner sep=1.8pt,circle,draw,fill=ForestGreen] at (\x:\R) {};
			
			\foreach \x in
			{ 0
			}
			\node[inner sep=1.8pt,circle,draw,fill=MidnightBlue] at (\x:\R) {};
			
		\end{tikzpicture}
%\includegraphics[width=0.45\textwidth]{DiagFull.png}		%png file of the diagram
%\vspace{5pt}		%add some vertical space when using the png file
\caption[LoF entry]{--- Cyclic groups generated by $W$ or $W^5$ (\sepia), $W^2$ or $W^4$ (\pesca), and $W^3$ (\battle). The order of each element is $1$ (\midnight), $2$ (\forest), $3$ (\oranbru), and $6$ (\bricco). Hermitian conjugate pairs are along the same vertical. This diagram can also be interpreted as lying in the real subspace of the $2\times 2$ matrix algebra generated by $W$ and $W^\dag$ (the null vector is at the midpoint of each antipodal pair). Since the regular hexagon is inscribed in the 3-sphere, its sides have unit length. Remarkably, this is also the edge length of the inscribed hypercube in 4 dimensions.
}
\label{fig:sixroot}
\end{figure}
This group contains one element of order 2 ($-\id_2$), two of order 3 ($-W$ and $-W^\dag$), and two of order 6 ($W$ and $W^\dag$). The fact that $-W$ is of order 3 is also evident from its exponential form, $-W=\exp (-\iu 4\pi \,\vb*e \cdot \vb*X/3)$, and likewise for $-W^\dag$. 
Note that each antipodal pair has one element with twice the order of the other (cf.~\figref{fig:sixroot}). Observe also that $W$ has been defined in Eq.~\eqref{eq:Wdef} as $U(0,\pi/2,\pi/2)$; however, even if we retain the choices $\alpha = 0$ and $\beta = \pi/2$, choosing $\gamma = 5\pi/2$ instead of $\gamma=\pi/2$ gives  $U(0,\pi/2,5\pi/2) = -W$, a matrix of order $3$ rather than $6$.

The $\SO$-\rep (cf.\ Sec.~\secref{sec:SU}) of $W$ can be computed via Eq.~\eqref{eq:RstortovsRdritto}, starting either from its definition~\eqref{eq:Wdef}, which yields the right shift operator, 
\begin{equation}\label{eq:Wrot}
\mathscr R\qty(W)
=R_y\qty(\frac{\pi}{2})\,R_z\qty(\frac{\pi}{2})
=\mqty*(
\begin{BMAT}(b){ccc}{ccc}
0& 0& 1 \\
0&1& 0\\
-1& 0 & 0
\end{BMAT})
\mqty*(
\begin{BMAT}(b){ccc}{ccc}
0& -1& 0 \\
1&0& 0\\
0& 0 & 1
\end{BMAT})
=\mqty*(
\begin{BMAT}(b){ccc}{ccc}
\mathmakebox[\widthof{$m$}]{0}\vphantom{-1}& 0& 1 \\
1&0& 0\\
0& 1 & 0
\end{BMAT})
=\Sigma_{\rm r}\,,
\end{equation}
or from its exponential form~\eqref{eq:W}, which yields the rotation by $2\pi/3$ about $\vb*e$. Recalling Eq.~\eqref{eq:2to1}, the following identities hold:
\begin{align}\label{eq:RW}
\mathscr R\qty(\pm W)
= R\qty(\frac{2\pi}{3}\vb*e)
=\Sigma_{\rm r}
\,,
&&
\mathscr R\qty(\pm W^\dag)
= R\qty(-\frac{2\pi}{3}\vb*e)
=\Sigma_{\rm l}\,.
\end{align}
In other words, $\mathscr R\qty(\pm W)$ and $\mathscr R\qty(\pm W^\dag)$ are the two rotations by $2\pi/3$ (order 3) about the axis $x=y=z$, one for each orientation. Each such rotation performs a cyclic permutation of the standard basis vectors (i.e., of the coordinate axes), and the inverse permutation of the standard components.

\subsubsection{Yet another property of the Bell states.}
In the end, our strategy worked. However, this success raises an obvious yet interesting question: why? One way to answer this question is by showing a (quite curious) transformation property of the Bell states.

A convenient starting point is the triplet block of matrix~\eqref{eq:UUBell}. Remarkably, its coefficients are either purely real or purely imaginary. More precisely, the purely imaginary entries are those involving $\Phi^+$ either as a bra or as a ket, whereas the expectation value $\ev**{U \otimes U}{\Phi^+}$, involving $\Phi^+$ on both sides, is real. In other words, multiplying $\Phi^+$ by an imaginary phase yields a real matrix. Even more remarkably, the triplet block~\eqref{eq:UUBell} closely resembles the expression of a rotation matrix in terms of Euler angles,
\begin{equation}\label{eq:RotEuler}
R_z(\alpha)\, R_y(\beta)\, R_z(\gamma)
=\mqty*(\begin{BMAT}[2.5pt]{ccc}{ccc}
{\rm c}_\alpha   \,{\rm c}_ \beta\,{\rm c}_\gamma - {\rm s}_ \alpha \,{\rm s}_ \gamma  
& \,\,- {\rm c}_\alpha \,{\rm c}_ \beta\,{\rm s}_\gamma -{\rm s}_ \alpha \,{\rm c}_ \gamma  \,\,
&  {\rm c}_ \alpha  \,{\rm s}_ \beta  
\\
{\rm s}_\alpha   \,{\rm c}_ \beta\,{\rm c}_\gamma +{\rm c}_ \alpha \,{\rm s}_ \gamma 
& -{\rm s}_ \alpha  \,{\rm c}_ \beta  \,{\rm s}_ \gamma +{\rm c}_ \alpha  \,{\rm c}_\gamma  
& {\rm s}_ \alpha  \,{\rm s}_ \beta \,\,
\\
-{\rm s}_ \beta  \,{\rm c}_ \gamma  
& {\rm s}_ \beta  \,{\rm s}_ \gamma  
& {\rm c}_ \beta 
\end{BMAT}).
\end{equation}

By Eq.~\eqref{eq:RstortovsRdritto}, the above equation yields $\mathscr R(\alpha\beta\gamma) = \mathscr R (U(\alpha\beta\gamma))$. In this way, we are led to examine the relation between $\mathscr R$, the $\SO$-\rep of $\SU$, and $D^{(1)}$, the spin-1 Wigner $D$-matrix. These two representations of $\SU$ --- both two-to-one, since they map $-\id_2$ to the identity matrix of order 3 --- are related by a unitary transformation~\cite{biede},
\begin{equation}\label{eq:DvsR}
D^{(1)}(U) =P^\dag\, \mathscr R(U) \,P\,,
\end{equation}
where
\begin{align}\label{eq:cartesioVSsphere}
P=\mqty*(
\begin{matrix}%[2.5pt]{ccc}{ccc}
-1/\sqrt 2& 0&  1/\sqrt 2 
\\
-\iu/\sqrt 2&  0& -\iu/\sqrt 2 
\\
0&\,\,1\,\,&0
\end{matrix}),
&&
P^\dag=\mqty*(
\begin{matrix}%[2.5pt]{ccc}{ccc}
-1/\sqrt 2&  \iu/\sqrt 2 &0 \\
0& 0&  1\\
1/\sqrt 2&\iu/\sqrt 2 &0
\end{matrix}).
\end{align}
The above relations can be checked by parametrizing $U$ in terms of Euler angles, and using expressions~\eqref{eq:spin1} of $D^{(1)}(\alpha\beta\gamma)$, and~\eqref{eq:RotEuler} of $\mathscr R(\alpha\beta\gamma)$. These formulae are well-known in the theory of angular momentum. Indeed, the \emph{spherical basis} $(\vb*\xi_m)_{m=1,0,-1}$ is expressed in terms of the Cartesian basis $(\vb e_i)_{i=1,2,3}$ as
\begin{align}\label{eq:spherical}
\vb*\xi_1=\frac{-\vb e_1-\iu\vb e_2}{\sqrt 2}\,,
&&
\vb*\xi_0=\vb e_3\,,
&&
\vb*\xi_{-1}=\frac{\vb e_1-\iu\vb e_2}{\sqrt 2}\,.
\end{align}
These definitions, and in particular the sign convention, are modeled on the spherical harmonics with $\ell=1$. Then $P$ is the transition matrix from the spherical to the Cartesian basis, and Eqs.~\eqref{eq:DvsR}--\eqref{eq:cartesioVSsphere}  state that, when the Cartesian basis transforms under $\mathscr R(U)$,
\begin{equation}
\vb e'_i=\sum_{i=1}^3 
\mathscr R_{ji}(U) \,\vb e_j\,,
\end{equation}
the spherical basis transforms under $D^{(1)}(U)$ (i.e., as the spherical harmonics with $\ell=1$),
\begin{equation}
\vb*{\xi}'_m
=\,\sum_{\mathclap{m'=0,\pm1}} \,
D^{(1)}_{m' m}(U) \,\vb*{\xi}_{m'}\,.
\end{equation}

As a result, Eq.~\eqref{eq:DvsR} entails
\begin{equation}\label{eq:BlocRD}
\mqty*(\begin{BMAT}{c1ccc}{c1ccc}
1& &  &\\
\phantom{m}& & &\vphantom{n} \\
\vphantom{D^{(1)}(U)} &&\mathscr R(U)& \\
& & & \vphantom{n}
\end{BMAT})=
\mqty*(\begin{BMAT}{c1ccc}{c1ccc}
1& &  &\\
\vphantom{n}\phantom{m}& & & \\
\vphantom{D^{(1)}(U)}&&\mathmakebox[\widthof{$R(U)$}]{P}& \\
\vphantom{n}& & & 
\end{BMAT})
\mqty*(
\begin{BMAT}{c1ccc}{c1ccc}
1& &  &\\
\phantom{m}& & &\vphantom{n} \\
&&D^{(1)}(U)  & \\
& & & \vphantom{n}
\end{BMAT})
\mqty*(\begin{BMAT}{c1ccc}{c1ccc}
1& &  &\\
\vphantom{n}\phantom{m}& & & \\
\vphantom{D^{(1)}(U)}&&\mathmakebox[\widthof{$R(U)$}]{P^\dag}& \\
\vphantom{n}& & & 
\end{BMAT}),
\end{equation}
with
\begin{equation}\label{eq:transitio}
\mqty*(\begin{BMAT}{c1ccc}{c1ccc}
1& &  &\\
\vphantom{n}\phantom{m}& & &\\
\vphantom{D^{(1)}(U)}&&\mathmakebox[\widthof{$R(U)$}]{P^\dag}& \\
\vphantom{n}& & & 
\end{BMAT})
=
\mqty*(\begin{matrix}
1&0&0&0\\
0&-1/\sqrt 2&  \iu/\sqrt 2&0 \\
0&0& 0 &  1\\
0&1/\sqrt 2&\iu/\sqrt 2 &0
\end{matrix}).
\end{equation}
Comparing the above expression with the transition matrix $\braket{B_{\rm am}}{B_{\rm B}}$ in Eq.~\eqref{eq:transAMBell}, we can adapt the Bell basis~\eqref{eq:OrdoBell} into
\begin{equation}\label{eq:ModiBell}
B_{\rm \tilde    B}
=\qty(\tilde   \Psi^-,\tilde   \Phi^-,\tilde   \Phi^+,\tilde   \Psi^+)
=\qty(\Psi^-,-\Phi^-,\iu\Phi^+,\Psi^+)\,,
\end{equation}
so that the transition matrix $ \braket{B_{\rm am}}{B_{\rm \tilde    B}}$ --- from the adapted Bell basis to the angular momentum basis --- is exactly the one in Eq.~\eqref{eq:transitio}. But then, by Eq.~\eqref{eq:BlocRD}, the matrix representation of $U\otimes U$ in the adapted Bell basis has precisely the form
\begin{equation}\label{eq:UUBellmodi}
\qty(U\otimes U)_{\rm \tilde    B}
=\mqty*(\begin{BMAT}{c1ccc}{c1ccc}
1& &  &\\
\phantom{m}& & &\vphantom{n} \\
&&\mathscr R(U)  & \\
& & & \vphantom{n}
\end{BMAT}).
\end{equation}
To sum up, it is well known that, under a transformation of the form $U\otimes U$, the triplet part of the angular momentum basis transforms according to $D^{(1)}(U)$, as expressed in Eq.~\eqref{eq:UUam}. In contrast to this ``spherical behavior,'' the
the triplet part of the adapted Bell basis exhibits a ``Cartesian behavior,''  transforming according to $\mathscr R(U)$, as expressed in Eq.~\eqref{eq:UUBellmodi}.

Now, the ordered bases $B_{\rm B}$ and $B_{\rm \tilde  B}$ are related by a reshuffling and phase-shift\-ing of the triplet basis vectors.
%In other words, they not only share the same notion of diagonality, but also assign the same matrix representation to a diagonal operator.
Therefore, all our previous considerations in the Bell basis regarding the problem formulated in Eqs.~\eqref{eq:diag}--\eqref{eq:TripletAver} apply equally well in the adapted Bell basis.
However, the matrix representation of $U\otimes U$ in the adapted basis is now given by Eq.~\eqref{eq:UUBellmodi}. 
This means that we have no need to parametrize $U$ in terms of Euler angles, and play with them so as the triplet block mimics a shift operator:
since the $\SO$-\rep is onto, and the shift operators~\eqref{eq:shift3} are in $\SO$, we can choose, right from the start, $U\in \SU$ such that $\mathscr R(U)$ is a shift operator.
By Eq.~\eqref{eq:RW}, $\pm W$ are the solutions of $\mathscr R(U)=\Sigma_{\rm r}$, and $\pm W^\dag$ of $\mathscr R(U)=\Sigma_{\rm l}$.

\subsection{Discussion of the results}\label{sec:rema}
In this section, we shift our focus from the 2-\des constructed in the previous sections to its ``special unitary versions,'' obtained by individually phase-shift\-ing each of its elements to some special unitary matrix. This choice not only enables a direct comparison with the 2-\des of~\cite{Bennett}, but also reveals the geometric and algebraic structures underlying our results.

Let us now go through the main tools we shall rely on in the following pages. 
The axis--angle parametrization of $\SO$, $\vb*\theta \mapsto R\qty(\vb*\theta)$, introduced in Sec.~\secref{sec:SU}, establishes a continuous one-to-one correspondence between $\SO$ and the closed ball of radius $\pi$ with antipodal identifications on its boundary. 
Consequently, each $U\in \SU$ can be visualized as a point in 3-\D space via the axis--angle vector of $\mathscr R (U)$, its image under the $\SO$-\rep.
This picture, however, has inherent limitations. First, distances in $\SO$ cannot be handled in terms of distances between axis--angle vectors: for instance, when $\phi$ approaches $\pi$ from below, $R(\pm \phi\vb*n)$ tend to the same rotation (the reflection with respect to the direction of $\vb*n$), whereas the distance between their axis--angle vectors tends to $2\pi$ (the diameter of the ball). Furthermore, the $\SO$-representation is two-to-one, so that antipodal matrices in $\SU$ have the same $\SO$-representation, although their Hilbert--Schmidt distance is $2\sqrt 2$ --- the diameter of $\U[2]$, i.e., the maximum distance between any two unitaries.

These issues can be overcome by using the \emph{Cayley--Klein parametrization} of $\SU$, which can be conveniently expressed as
\begin{equation}\label{eq:CK}
\left.\begin{aligned}
	\mathbb C^2\ni(a,b)\\
	\abs{a}^2+\abs{b}^2=1
\end{aligned}\,\right\}
\mapsto U(a,b)
=
\mqty*(a&-b^\ast\\
b&a^\ast)\in\SU
\,.
\end{equation}
Switching to real parameters,
\begin{equation}\label{eq:isoS3}
\mathbb S^3\ni(s,x,y,z)
\mapsto  U (s-\iu z,y -\iu x)
=s\id_2 -\iu\qty( x X+ y  Y+ z Z)\in\SU\,,
\end{equation}
gives a continuous one-to-one correspondence between $\SU$ and the 3-sphere. Moreover, 4-\D vectors can be identified with quaternions,
\begin{equation}\label{eq:Quater}
\mathbb R^4\ni(s,x,y,z)
\cong
s+ x\I + y\J +z\K \in\mathbb H\,,
\end{equation}
where $\I$, $\J$, $\K$ are the \emph{imaginary units}, and $s$ and $x\I + y\J +z\K$ are the \emph{scalar} (or \emph{real}) and the \emph{vector} (or \emph{imaginary}) \emph{part} (or \emph{component}) of the quaternion, respectively. In this way a group isomorphism is established between $\SU$ and the unit quaternions, a multiplicative subgroup of the quaternion algebra $\mathbb{H}$. Remarkably, this is also a Lie group isomorphism and an isometry --- more precisely, the Hilbert–Schmidt distance between matrices corresponds to the Euclidean distance between their representative points in $\mathbb{R}^4$, scaled by a factor $\sqrt 2$, the norm of any $2\times 2$ unitary. Moreover, quaternion conjugation, namely, $ \mathbb{H}\ni q=s+ x\I + y\J +z\K\mapsto \bar q = s- x\I - y\J -z\K\in  \mathbb{H}$, coincides with inversion in the multiplicative group of unit quaternions, that is, $q^{-1}=\bar q$ for $\abs{q}=1$. As a result, Hermitian conjugation in $\SU$ corresponds exactly to quaternion conjugation of the associated unit quaternion.
With this correspondence in mind, we may now freely switch between the viewpoints of a special unitary matrix, a point on $\mathbb{S}^3$, and a unit quaternion.

The conventions adopted above lead to $\I\cong -\iu X$, $\J\cong -\iu Y$, $\K\cong -\iu Z$, entailing
\begin{equation}\label{eq:EulQuat}
\e^{-\iu \phi \vb*n\cdot \vb*X/2}
\cong \cos\frac{\phi}{2}
+\sin\frac{\phi}{2}\,
\qty(n_x \I+n_y \J+n_z \K)
\,.
\end{equation}
In other words, by setting
\begin{equation}
I_i=-\iu X_i=\e^{-\iu \pi X_i/2}\,,
\end{equation}
along with $(I_i)_{i=1}^3=(I,J,K)$, we have $\I\cong I$, $\J\cong J$, $\K\cong K$. These matrices satisfy the composition rule $I_i I_j=-\delta_{ij}\id_2 +  \sum_{k=1}^3 \epsilon_{ijk}I_k$, equivalently expressed by
\begin{align}\label{eq:IiComp}
I_i^2=-\id_2\,,
&&
I_i I_{i\oplus1}=I_{i\ominus 1}\,,
&&
I_i I_{i\ominus 1}=-I_{i\oplus 1}\,.
\end{align}
They are anti-in\-vo\-lu\-tive and anti-Her\-mi\-tian. Any two of them anticommute, and generate a realization of the  quaternion algebra. Moreover, Eq.~\eqref{eq:EulQuat} translates into
\begin{equation}\label{eq:EulIi}
\e^{-\iu \phi \vb*n\cdot \vb*X/2}
=\cos\frac{\phi}{2}\,\id_2
+\sin\frac{\phi}{2}\,
\qty(n_x I+n_y J+n_z K)
\,,
\end{equation}
in particular,
\begin{equation}\label{eq:WviaIi}
W
=\frac{1}{2}\qty(\id_2 +I+J +K)
=\frac{1}{2}
\qty(\id_2 +I_{i\ominus 1}+ I_i +I_{i\oplus 1})
\,.
\end{equation}
Conjugation by $W$ and $W^\dag$ on the matrices $I_i$ amounts to a right and a left shift, respectively,
\begin{align}\label{eq:conjWIi}
W I_i W^\dag = I_{i\oplus 1}\,,
&&
W^\dag I_i W = I_{i\ominus 1}\,.
\end{align}
This result can be obtained by direct computation from Eqs.~\eqref{eq:IiComp} and~\eqref{eq:WviaIi}, or by considering conjugation by $W$ and $W^\dag$ on the Pauli matrices from Eqs.~\eqref{eq:rotatio} and~\eqref{eq:Wrot}, or else at the level of  quaternions, by observing that the action by conjugation of $w=\qty(1 + \I + \J + \K)/2 $, the half-integer unit quaternion corresponding to $W$, on the imaginary units is the cyclical permutation $\I\mapsto\J\mapsto \K\mapsto\I$.

\subsubsection{Normalization to $\SU$.}\label{sec:Norma}
Recalling and expanding on our previous considerations in Secs.~\secref{sec:proj} and~\secref{sec:phase}, let us fix some notation and terminology, and collect some simple observations, tailored to the specific cases of interest (a more general perspective will be sketched in Sec.~\secref{sec:AlgeBack}).
Every $U\in \U[2]$ can be phase-shifted to $\SU$ in exactly two ways, as $\pm U'=\pm \omega^\ast U$, with $\omega^2=\det U$; we call these two antipodal matrices the normalizations of $U$ in $\SU$. Observe that two unitaries have the same normalizations in $\SU$ if and only if they are proportional.
For $\Ss \subset \U[2]$, the set of the normalizations of all elements of $\Ss$ will be called the closure of $\Ss$ in $\SU$, and denoted by $\bar \Ss$.
Let us now consider a finite set $\Ss = \{U_a\} \subset \U[2]$ with no proportional elements
--- as in the cases of interest: the Pauli basis $\B$, and its extension $\DD = \B \cup W\B \cup W^\dag \B$ to a 2-\des.
Of course we can write $\bar \Ss= \{\pm U'_a\}$, hence $\bar \Ss$ has twice the size of $\Ss$. Choosing a normalization for each $U_a$ amounts to selecting a sign for each antipodal pair $\pm U'_a$, and yields a subset of $\SU$, of the same size as $\Ss$, which we call a normalization of $\Ss$ in $\SU$. Clearly, two normalizations of $\Ss$ go into each other by individual sign flips. 
If $\Ss*$ is a normalization of $\Ss$, then also $-\Ss*$ is, and  we have $\bar \Ss =\Ss* \cup - \Ss*$, so that, in particular, $\bar \Ss$ is the union of all normalizations of $\Ss$.
For $V\in\SU$, $V\bar \Ss$ is the closure of $V \Ss$, and $\Ss*$ normalizes $\Ss$ if and only if $V\Ss*$ normalizes $V\Ss$. 
Observe that a normalization $\Ss*$ of $\Ss$ produces the same twirls (of any order) as $\Ss$, hence the same is true for $\bar \Ss$. Accordingly, $\Ss$ is a $t$-\des if and only if $\Ss*$ is, if and only if $\bar \Ss$ is.
Since the $\SO$-\rep maps antipodal matrices to the same rotation, the image of any normalization of $\Ss$ coincides with that of $\bar \Ss$. 
Finally, let $(\Ss_\alpha)$  be a partition of $\Ss$. Then
$\bar \Ss = \bigcup_\alpha \bar \Ss_\alpha$. Moreover, a normalization $\Ss*$ of $\Ss$ induces the normalization $\bar \Ss_\alpha \cap \Ss* $ of each $\Ss_\alpha$, and, conversely, normalizing each $\Ss_\alpha$ to $ \alpha$ results in the normalization $\bigcup_\alpha \Ss*_\alpha$ of $\Ss$.

In the cases of interest, the unitary designs $\B$ and $\DD$, we have
\begin{gather}\label{eq:Bnorm}
\Bb
=\qty{\pm\id_2, \,\pm I,\,\pm J,\,\pm K},
\\
\label{eq:Dnorm}
\Db
=\qty{\pm\id_2, \,\pm I,\,\pm J,\,\pm K,
\,\pm W,\,\pm WI,\, \pm WJ, \,\pm WK,\,
\pm W^\dag,\,\pm W^\dag I,\,\pm W^\dag J, \,\pm W^\dag K}.
\end{gather}
By the above observations, $\Bb$ and $\Db$ are a 1- and a 2-\des, respectively, of twice the minimum size. Moreover, both are subgroups of $\SU$, and $\Bb$ is normal in $\Db$, since Eq.~\eqref{eq:conjWIi} entails $W I_i \,\Bb \,(W I_i )^\dag=W\Bb W^\dag=\Bb $, and likewise for $W I_i^\dag$. In particular, $\Bb$, $W\Bb$, and $W^\dag\Bb$ are precisely the cosets of $\Bb$ in $\Db$.
	
A more in depth analysis of these structures is carried out in Sec.~\secref{sec:3S}. Let us now go through each unitary, obtaining its exponential form, and its expansion on the Pauli basis, plugging these results in~\tabref{tab:tabula}.

\begin{table}[!tb]
\fontsize{10pt}{10pt}\selectfont
\tabulinesep=2mm
$
\begin{tabu}to \linewidth{X[$.29c] X[$1.015c] X[$.9c] X[$.66c] X[$.63c] X[$.58c] }
\toprule
&
\text{exp}&
\text{Pauli}& 
\mathbb{H}&
\mathbb{S}^3&
\SO
\\
\midrule
\color{verte}
\pm\id_2& 
%\text{---}
&
\pm\id_2&
\pm1&
(\pm,0,0,0)&
(0,0,0)
\\
\hline
\color{verte}
\pm I& 
\exp(-\iu \qty{\smqty{1\\3}}\frac{\pi}{2} X )& 
\mp\iu X&
\pm \I&
(0,\pm,0,0)&
\pi(1,0,0)
\\
\hline
\color{verte}
\pm J& 
\exp(-\iu \qty{\smqty{1\\3}}\frac{\pi}{2} Y )& 
\mp \iu Y&
\pm \J&
(0,0,\pm,0)&
\pi (0,1,0)
\\
\hline
\color{verte}
\pm K& 
\exp(-\iu \qty{\smqty{1\\3}}\frac{\pi}{2} Z )&
\mp \iu Z& 
\pm \K&
(0,0,0,\pm)&
\pi (0,0,1)
\\
\midrule
\color{marun}
\pm W& 
\exp(-\iu\qty{\smqty{1\\4}} \frac{\pi}{3} \vb*e \cdot \vb*X)&
\pm \tfrac{1}{2}[\id{-}\iu(X{+}Y{+}Z)]&
\pm \tfrac{1}{2}\qty(1{+}\I{+}\J{+}\K)&
\tfrac{1}{2}(\pm,\pm,\pm,\pm)&
\tfrac{2\pi}{3\sqrt 3}(+,+,+)
\\
\hline
\color{marun}
\pm WI &
\exp(-\iu \qty{\smqty{4\\1}} \tfrac{\pi}{3} R_z(\pi)\vb*e\cdot \vb*X)&
\mp \tfrac{1}{2}[\id{+}\iu(X{+}Y{-}Z)]&
\mp \tfrac{1}{2}\qty(1{-}\I{-}\J{+}\K)&
\tfrac{1}{2}(\mp,\pm,\pm,\mp)&
\tfrac{2\pi}{3\sqrt 3}(-,-,+)
\\
\hline
\color{marun}
\pm WJ & 
\exp(-\iu \qty{\smqty{4\\1}} \tfrac{\pi}{3} R_x(\pi)\vb*e\cdot \vb*X)&
\mp \tfrac{1}{2}[\id{-}\iu(X{-}Y{-}Z)]&
\mp \tfrac{1}{2}\qty(1{+}\I{-}\J{-}\K)&
\tfrac{1}{2}(\mp,\mp,\pm,\pm)&
\tfrac{2\pi}{3\sqrt 3}(+,-,-)
\\
\hline
\color{marun}
\pm  WK&
\exp(-\iu \qty{\smqty{4\\1}} \tfrac{\pi}{3} R_y(\pi)\vb*e\cdot \vb*X)&
\mp\tfrac{1}{2}[\id{+}\iu(X{-}Y{+}Z)]&
\mp\tfrac{1}{2}\qty(1{-}\I {+}\J{-}\K)&
\tfrac{1}{2}(\mp,\pm,\mp,\pm)&
\tfrac{2\pi}{3\sqrt 3}(-,+,-)
\\
\midrule
\color{bluu}
\pm W^\dag& 
\exp(\iu \qty{\smqty{1\\4}} \tfrac{\pi}{3} \vb*e\cdot \vb*X)&
\pm \tfrac{1}{2}[\id{+}\iu(X{+}Y{+}Z)]&
\pm \tfrac{1}{2}\qty(1{-}\I{-}\J{-}\K)&
\tfrac{1}{2}(\pm,\mp,\mp,\mp)&
\tfrac{2\pi}{3\sqrt 3}(-,-,-)
\\
\hline
\color{bluu}
\pm W^\dag I& 
\exp(\iu \qty{\smqty{1\\4}} \tfrac{\pi}{3} R_y(\pi)\vb*e\cdot \vb*X)&
\pm \tfrac{1}{2}[\id{-}\iu(X{-}Y{+}Z)]&
\pm\tfrac{1}{2}\qty(1{+}\I{-}\J{+}\K)&
\tfrac{1}{2}(\pm,\pm,\mp,\pm)&
\tfrac{2\pi}{3\sqrt 3}(+,-,+)
\\
\hline
\color{bluu}
\pm W^\dag J& 
\exp(\iu \qty{\smqty{1\\4}} \tfrac{\pi}{3} R_z(\pi)\vb*e\cdot \vb*X)&
\pm \tfrac{1}{2}[\id{-}\iu(X{+}Y{-}Z)]&
\pm\tfrac{1}{2}\qty(1{+}\I{+}\J{-}\K)&
\tfrac{1}{2}(\pm,\pm,\pm,\mp)&
\tfrac{2\pi}{3\sqrt 3}(+,+,-)
\\
\hline
\color{bluu}
\pm W^\dag K&
\exp(\iu \qty{\smqty{1\\4}} \tfrac{\pi}{3} R_x(\pi)\vb*e\cdot \vb*X)&  
\pm \tfrac{1}{2}[\id{+}\iu(X{-}Y{-}Z)]&
\pm \tfrac{1}{2}\qty(1{-}\I{+}\J{+}\K)&
\tfrac{1}{2}(\pm,\mp,\pm,\pm)&
\tfrac{2\pi}{3\sqrt 3}(-,+,+)
\\
\bottomrule
\end{tabu}
$
\caption[LoF entry]{--- Each element of $\Db=\textcolor{verte}{\Bb}\cup\textcolor{marun}{W\Bb}\cup\textcolor{bluu}{W^\dag\Bb}$ is displayed together with its exponential form (where $\qty{\!\smqty{n\\m}\!}$ stands for $n$ in the $+$ case, and $m$ in the $-$ case), its decomposition in the Pauli basis, the associated unit quaternion and point on the 3-sphere, and the axis--angle vector of its $\SO$-\rep (illustrated in~\figref{fig:des}). 
The elements are grouped by cosets: the first four rows correspond to $\Bb$, the next four to $W\Bb$, and the last four to $W^\dag\Bb$.
}
\label{tab:tabula}
\end{table}

First, $\Bb=\{\pm\id_2,\pm I_i\}$ consists of $\id_2$ (the only $2\times 2$ matrix of order 1), $-\id_2$ (an involution, order 2), and the six matrices $\pm I_i$ (all anti-in\-vo\-lu\-ti\-ons, order 4). 

Moving on to $W\Bb=\{\pm W,\pm W I_i\}$ and $W^\dag\Bb=\{\pm W^\dag,\pm W^\dag I_i\}$,  $\pm W$ and $\pm W^\dag$ have been already discussed in Sec.~\secref{sec:W}. As to the other elements, we first note, by Eq.~\eqref{eq:E}, that $R_i(\pi)\vb*e=\qty(-\vb e_{i\ominus 1}+ \vb e_i-\vb e_{i\oplus 1} )/\sqrt 3$, so that, by Eqs.~\eqref{eq:IiComp} and~\eqref{eq:WviaIi}, 
\begin{equation}\label{eq:WIi}
-W I_i 
=\frac{\id_2 +I_{i\ominus 1} - I_i - I_{i\oplus 1}}{2}
=\exp (-\iu \frac{\pi}{3} R_{i\ominus 1}(\pi)\vb*e\cdot \vb*X)
\,.
\end{equation}
Then, by Eq.~\eqref{eq:conjWIi} and the anti-Her\-mi\-ti\-ci\-ty of the matrices $I_i$, we also have $W^\dag I_i =I_{i\oplus 1} W^\dag= (- W I_{i\oplus 1} )^\dag$, resulting in
\begin{equation}\label{eq:WdagIi}
W^\dag I_i 
=\frac{\id_2 -I_{i\ominus 1}  - I_i +I_{i\oplus 1}}{2}
=\exp (\iu \frac{\pi}{3} 
\,R_{i\oplus 1}(\pi)\vb*e\cdot \vb*X)
\,.
\end{equation}
Summing up, the sixteen matrices in $W\Bb \cup W^\dag \Bb$ are of the form $\pm V_{\vb*n}$, where
\begin{align}\label{eq:Vn}
V_{\vb*n}
=\exp(-\iu \frac{\pi}{3} \vb*n \cdot \vb*X)
\,,
&&
\vb*n\in \qty{\pm\vb*e,\,\pm R_i(\pi)\vb*e}\,.
\end{align}
In particular, $W= V_{\vb*e}$ and $W^\dag= V_{-\vb*e}$. Each matrix $V_{\vb*n}$ generates a cyclic group of order 6, with the same group structure as  $\left\langle W\right \rangle$, namely, $V_{\vb*n}^2 = -V_{\vb*n}^\dag$, $V_{\vb*n}^3 = -\id_2$, $V_{\vb*n}^4 = -V_{\vb*n}$, and $V_{\vb*n}^5 = V_{\vb*n}^\dag$ (cf.~\figref{fig:sixroot}).
The eight unit vectors $\pm\vb*e$, $\pm R_i(\pi)\vb*e$, correspond to all possible orientations of the main diagonals of $\mathbb R^3$ (collectively described by $|x|=|y|=|z|$). The overall sign determines the order: 6 for $+$, 3 for $-$. This follows from the group relations, but is also evident from the exponential form (cf.~\tabref{tab:tabula}). The parity of the number of negative components of $\vb*n$ determines the set: $W\Bb$ if this number is even, $ W^\dag\Bb$ if it is odd (cf.~\tabref{tab:tabula}). 

\subsubsection{Some specific normalizations of $\DD$.}\label{sec:SpeciNorma}
Always choosing the $+$ for each antipodal pair in Eqs.~\eqref{eq:Bnorm} and~\eqref{eq:Dnorm} yields the normalizations
\begin{gather}\label{eq:B0}
\B_0=\{\,\id_2,\, I,\,J,\,K\,\},
\\\label{eq:D0}
\DD_0=\qty{\id_2,\, I,\, J,\, K,\, 
\,W,\, WI,\,WJ,\,WK,\,
W^\dag,\, W^\dag I,\, W^\dag J,\, W^\dag K }.
\end{gather}
Observe that, if $\B*$ normalizes $\B$, its extension~\eqref{eq:D'} to a 2-\des in turn normalizes $\DD$, as $\tilde W=W\in\SU$. In other words, Eq.~\eqref{eq:D'} maps normalizations of $\B$ to normalizations of $\DD$,
\begin{equation}\label{eq:N}
\N_0(\B*)
=\B*\cup W\B*\cup W^\dag\B*
=\bigcup_{k=0,\pm1}W^k\B*\,,
\end{equation}
and $\DD_0$ is just a special case of the above class of normalizations, namely, $\DD_0=\N_0(\B_0)$.

Another useful normalization of $\DD$ is 
\begin{equation}\label{eq:D1}
\DD_1=\qty{\id_2,\, I,\, J,\, K,
\,W,\, -WI,\,-WJ,\,-WK,\,
W^\dag,\, W^\dag I,\, W^\dag J,\, W^\dag K },
\end{equation}
differing from $\DD_0$ only by the sign flips $W I_i \mapsto -W I_i $. This choice has a geometric flavor, in the first place, as it amounts to requiring that the first nonzero coordinate of each point on the 3-sphere be positive (cf.~\tabref{tab:tabula}). In other words, the points on the 3-sphere corresponding to the matrices in $\DD_1$ lay all in the same half (e.g., all open hemispheres of the form $w+\epsilon (x+y+z)>0$, with $0<\epsilon%<1/3
$).
However, $\DD_1$ is also characterized by the fact that its induced normalizations of $W\B$ and $W^\dag\B$ 
are related by Hermitian conjugation (i.e., inversion). Indeed, any normalization of the Pauli basis is of the form $\B*=\{\epsilon_0 \id_2, \epsilon_i I_i\}$, with $\epsilon_\mu=\pm 1$. Then, by Eq.~\eqref{eq:conjWIi}, one gets $(W^\dag \B* )^\dag = 
\qty{\epsilon_0 W, -\epsilon_{i\ominus 1} W  I_i }$, so that $(W^\dag \B* )^\dag$ is a normalization of $W\B$. Replacing $W\B*$ with $(W^\dag \B* )^\dag$ in Eq.~\eqref{eq:N} modifies $\N_0$ in such a way that the sets normalizing $W\B$ and $W^\dag\B$ are related by Hermitian conjugation,
\begin{equation}\label{eq:N'}
\N_1\qty(\B*)
=\B*\cup  \qty(W^\dag\B*)^\dag \cup  W^\dag\B*\,,
\end{equation}
and, in particular, $\DD_1=\N_1(\B_0)$. 
Observe that $\DD_1$ can also be obtained as $\B_0 \cup \{V_{\vb*n}\}$, so it has no elements of order 3. On the other hand,  $\B_0\cup -(W^\dag\B_0)^\dag \cup -W^\dag\B_0=\B_0 \cup \{-V_{\vb*n}\}$ is another normalization of $\DD$, normalizing $W\B$ and $W^\dag\B$ to Hermitian conjugate sets, while containing no elements of order 6.

The 2-\des in~\cite{Bennett} is yet another normalization of $\DD$. First, note that their sign convention adopted for the generators of $\SU$ is opposite to ours, hence their ``rotations by $\pi/2$ rad'' about the coordinate axes read $B_i = \exp(\iu\pi X_i/4) = (\id +\iu X_i)/\sqrt{2}$. The 12 elements of their 2-\des are $\id_2$, $B_i^2$, $B_i B_{i\oplus 1}$, $B_y B_x$, $(B_i B_{i\oplus 1})^2$, and $(B_y B_x)^2$ (cf.\ Eq.~(A2) in~\cite{Bennett}). Now, by Eqs.~\eqref{eq:IiComp} and~\eqref{eq:WviaIi}, we have $B_i^2=-I_i$, as well as
\begin{align}
B_i B_{i\oplus 1}
=\frac{\id_2+I_{i\ominus 1}-I_i-I_{i\oplus 1}}
{2}
=-WI_i\,,
&&
B_i B_{i\ominus 1}
=\frac{\id_2-I_{i\ominus 1}-I_i-I_{i\oplus 1}}
{2}=W^\dag\,,
\end{align}
so that $(B_i B_{i\ominus 1})^2 = (W^{\dag})^2 = -W$, and, taking into account Eq.~\eqref{eq:conjWIi},
\begin{equation}
(B_i B_{i\oplus 1})^2=WI_i W I_i
=W^2 I_{i\ominus 1}I_i
=-W^\dag I_{i\oplus 1}\,.
\end{equation}
Collecting these results, the 2-\des in~\cite{Bennett} reads
\begin{equation}\label{eq:DB}
\DD_2=\qty{\id_2,\,- I,\,- J,\,- K,\, 
-W,\,-WI,\,-WJ,\,-WK,\,
W^\dag, \,-W^\dag I,\,-W^\dag J,\,- W^\dag K },
\end{equation}
and is characterized by the fact that $W\B$ and $W^\dag\B$ are normalized to sets related by Hermitian conjugation and a sign flip. Indeed, we are free to change sign to $(W^\dag\B*)^\dag$ in Eq.~\eqref{eq:N'}, and modify the map $\N_1$ in such a way that the resulting normalizations of $W\B$ and $W^\dag\B$ are related by Hermitian conjugation and a sign flip,
\begin{equation}\label{eq:N''}
\N_2\qty(\B*)
=\B*\cup  - \qty(W^\dag\B*)^\dag \cup  W^\dag\B*\,.
\end{equation}
In particular, $\DD_2=\N_2\bigl(\B_0^\dag \bigr)$.

\subsubsection{Interpetation in $\SO$.} \label{sec:3D}
Since the $\SO$-\rep maps opposite matrices to the same rotation, we now discuss the image of each antipodal pair in $\Db$.
Of course, any specific normalization $\DD*$ of $\DD$ yields the same image as $\Db$. By the same token, the normalizations of $\B$, $W\B$, and $W^\dag\B$ induced by $\DD*$ (namely, the intersection of $\DD*$ with the corresponding cosets of $\Bb$) have the same images as $\Bb$, $W\Bb$, and $W^\dag \Bb$, respectively.

\begin{figure}[!tb]
\centering
\includegraphics[width=\textwidth]{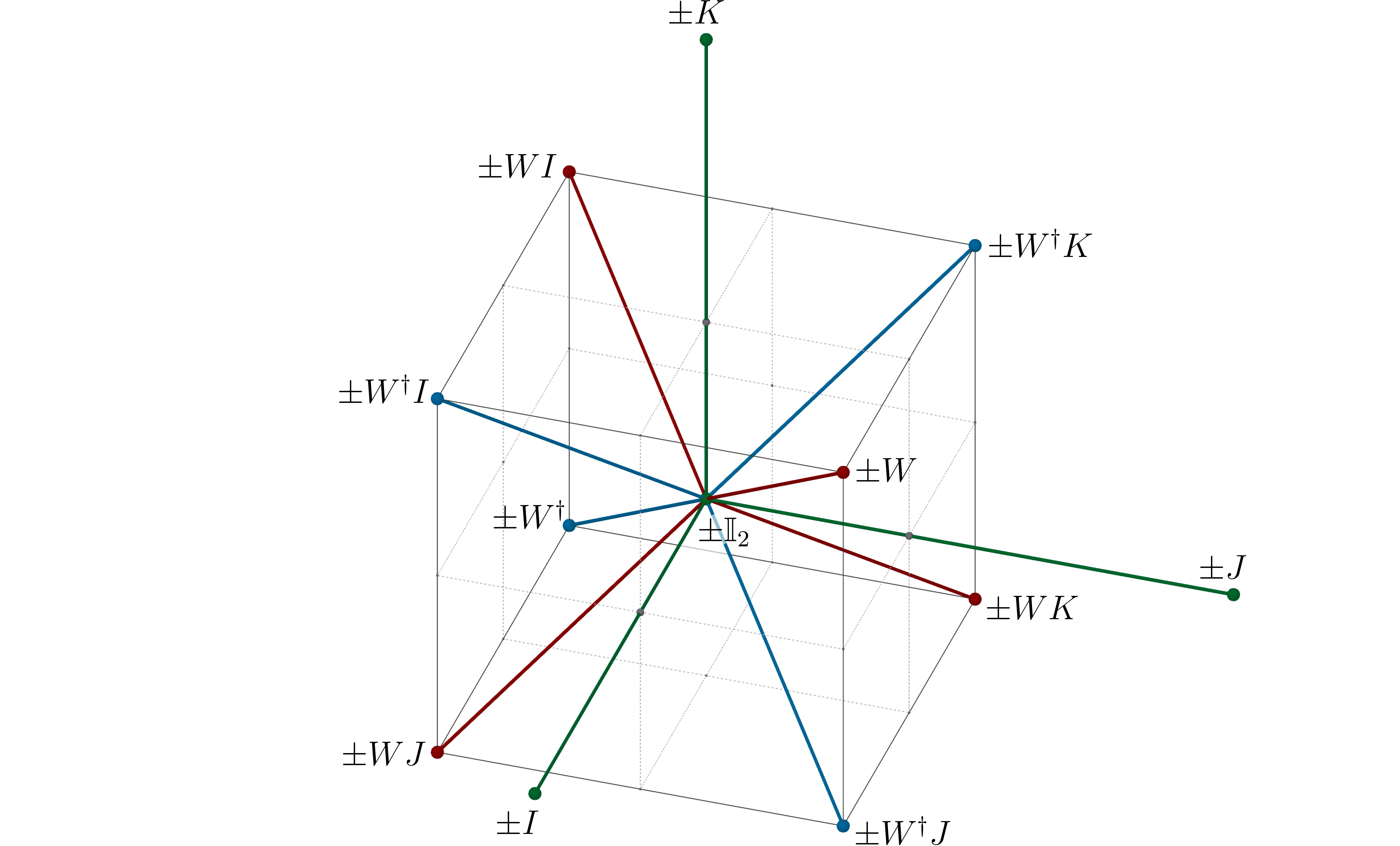}
\vspace{1pt}		
\caption[LoF entry]{--- $\SO$-\rep of $\Db=\textcolor{verte}{\Bb}\cup\textcolor{marun}{W\Bb}\cup\textcolor{bluu}{W^\dag\Bb}$.
The length of each segment is $\pi$ for the elements of $\Bb$ (\vertepalla), and $2\pi/3$ for those in the cosets $W\Bb$ (\marunpalla) and $W^\dag\Bb$ (\bluupalla).
For each main diagonal of $\mathbb{R}^3$, the two antipodal vertices are split between the cosets: one lies in $\mathscr R(W\Bb)$ and the other in $\mathscr R(W^\dag\Bb)$.
}
\label{fig:des}
\end{figure}

For the eight matrices in $\Bb=\{\pm \id_2,\pm I_i\}$ we have
\begin{align}
\mathscr R(\pm\id_2)=\id_3\,,
&&
\mathscr R\qty(\pm I_i)=R_i(\pi)\,.
\end{align}
In this way, $ \id_2$ (order 1) and $-\id_2$ (order 2) are both represented by $\id_3$ (the only $3\times 3$ matrix of order 1), whereas each $\pm I_i$ (order 4) is represented by the rotation by $\pi$ about the $i$-th coordinate direction, which is also the inversion of the coordinate plane orthogonal to $\vb e_i$ (an involution, order 2). 
When rotations are described via their axis--angle vectors, $\mathscr R(\Bb)$ is represented by $\{\vb*0,\pi \vb e_i\}$. This is the vertex set of the standard 3-simplex (i.e., the convex hull of the null vector and the standard basis vectors of $\mathbb R^3$), scaled by a factor $\pi$ (see~\figref{fig:des}).

From Sec.~\secref{sec:Norma} we know that the 16 matrices in $W\Bb \cup W^\dag \Bb$ can be put in the form $\pm V_{\vb* n}$, with $V_{\vb* n}$ defined in Eq.~\eqref{eq:Vn}. As a consequence, Eq.~\eqref{eq:RW} generalizes to
\begin{align}
\mathscr R (\pm V_{\vb* n})= R\qty(\frac{2\pi}{3}\vb*n)\,,
&&
\mathscr R 
\qty(\pm V^\dag_{\vb* n})
= R\qty(-\frac{2\pi}{3}\vb*n)\,.
\end{align}
In terms of axis--angle vectors, $\mathscr R(W\Bb \cup W^\dag \Bb)$ is the vertex set of the cube inscribed in the $2$-sphere of radius $2\pi/3$ with edges parallel to the coordinate axes (see~\figref{fig:des}).

Remarkably, each of the sets $\mathscr R(W\Bb)$ and $\mathscr R(W^\dag \Bb)$ admits a full-fledged geometric interpretation on its own. Indeed, we know from Sec.~\secref{sec:Norma} that $W\Bb$ and $W^\dag\Bb$ are distinguished by the parity of the number of negative components of the unit vector $\vb*n$: even for $W\Bb$, odd for $W^\dag\Bb$. As a result, switching to axis--angle vectors, $\mathscr R (W\Bb)$ is represented by the 4 points $(+,+,+)$, $(+,-,-)$, $(-,+,-)$, $(-,-,+)$, scaled by a factor $2\pi/(3\sqrt 3)$, and $\mathscr R(W^\dag\Bb)$ by their antipodes. In other words, $\mathscr R \qty(W\Bb)$ and $\mathscr R (W^\dag \Bb)$ are the vertex sets of the two regular tetrahedra in the cube with vertex set $\mathscr R (W\Bb \cup W^\dag \Bb)$ (see~\figref{fig:des}). The compound of the two regular tetrahedra in the cube is the classical \emph{stella octangula}, a nonconvex 3-\D configuration with the same symmetries as the cube.

\subsubsection{Interpretation on $\mathbb{S}^3$.} \label{sec:3S}
Now we turn to the 4-\D or quaternionic viewpoint, where each of the sets $\Bb$, $W\Bb$, $W^\dag\Bb$, $W\Bb\cup W^\dag \Bb$, and $\Db$ is the vertex set of some regular polytope (a highly symmetric convex configuration that can be regarded as a 4-\D analogue of a Platonic solid) inscribed in the 3-sphere. Moreover, in addition to their geometric symmetries, two of these sets are endowed with an algebraic structure, as subgroups of the unit quaternions~\cite{coxeter}.

Let us start with $\Bb$, whose elements are the eight points on the 3-sphere with integer coordinates (i.e., lying in the \emph{integer lattice} $\mathbb{Z}^4$), which are exactly the eight intersections of $\mathbb{S}^3$ with the 4 coordinate axes, that is, the four antipodal pairs
\begin{align}\label{eq:NS}
(\pm 1,0,0,0) 
\cong\pm 1\,,
&\;\;&
(0,\pm 1,0,0) 
\cong \pm \I \,,
&\;\;&
(0,0,\pm 1,0) 
\cong \pm \J\,,
&\;\;&
(0,0,0,\pm 1) 
\cong \pm \K\,.
\end{align}
From a geometric perspective, they are the vertices of the \emph{cross polytope} inscribed in the 3-sphere, that is, the inscribed \emph{16-cell} (the 4-\D analogue of the regular octahedron)  with vertices on the coordinate axes. From an algebraic standpoint, $\Bb$ consists of the eight elements of $\Db$ whose order is not a multiple of 3, and is a realization of the \emph{quaternion group} $Q_8$ (i.e., the subgroup of the unit quaternions generated by any two imaginary units).

Next, the elements of $W\Bb \cup W^\dag \Bb$ are the sixteen points of the 3-sphere 
with half-integer coordinates (i.e., lying in the \emph{half-integer lattice} $(\frac{1}{2},\frac{1}{2},\frac{1}{2},\frac{1}{2})+\mathbb{Z}^4$), or, equivalently, the intersections of $\mathbb{S}^3$ with
the eight main diagonals of $\mathbb{R}^4$ (collectively described by $\abs{s} = \abs{x} = \abs{y} = \abs{z}$),
\begin{align}\label{eq:cubo}
(s,x,y,z) \cong s +x\I +y \J+z \K\,,
&&
s,x,y,z=\pm\frac{1}{2}\,.
\end{align}
Geometrically, these points come in antipodal pairs, and are the vertices of the \emph{8-cell} or \emph{tesseract} (the 4-\D analogue of the cube) inscribed in the 3-sphere with edges parallel to the coordinate axes. 
Algebraically, $W\Bb \cup W^\dag \Bb$ consists of the sixteen elements in $\Db$ whose order is a multiple of 3; in particular, neither $W\Bb \cup W^\dag \Bb$ nor any of its nonempty subsets is closed under multiplication. 

The elements of $W\Bb$ and $W^\dag \Bb$ can be distinguished by the parity of the number of negative components: even for $W\Bb$, odd for $W^\dag\Bb$ (a pattern which is immediately clear from~\tabref{tab:tabula}). This criterion is equivalent to the one discussed in the previous section. Indeed, by Eq.~\eqref{eq:EulQuat},
\begin{equation}
V_{\vb*n}\cong\frac{1 }{2}+ \frac{\sqrt 3}{2}\,(n_x \I +n_y \J+n_z \K)
= q(\vb*n)\,,
\end{equation}
hence the negative components of the 3-\D unit vector $\vb*n$ are one-to-one with the negative components of the unit quaternion $q(\vb*n)$. Therefore we have $\pm V_{\vb*n} \cong \pm q(\vb*n)$, and the parity of the number of negative components of opposite quaternions is clearly the same.
Observing that all points of the form~\eqref{eq:cubo} are characterized by $sxyz=\pm 1/16$, this situation can be summarized in a very simple rule: the point is in $W\Bb$ when the product of all its components is positive, and in $W^\dag\Bb$ when negative. Note, in particular, that antipodal points are always in the same set.

Remarkably, each of the cosets $W\Bb$ and $W^\dag \Bb$ has a relevant geometric interpretation on its own. Explicitly, $W\Bb$ consists of  $(+,+,+,+)/2$, $(-,-,-,-)/2$, and the six permutations of $(+,+,-,-)/2$, whereas $W^\dag\Bb$ is made up of the four permutations of $(+,+,+,-)/2$, and the four of $(+,-,-,-)/2$. 
These two parity classes are the vertex sets of the two \emph{16-cells} in the tesseract (the two %``checkerboard'' classes, or 
\emph{demitesseracts}). The partition of the 8-cell's vertices into those of two 16-cells can be regarded as the 4-\D analogue of partitioning a cube's vertices into those of two tetrahedra.
As a result, the cosets $\Bb$, $W\Bb$, $W^\dag\Bb$ are pairwise congruent under some element of $\SO[4]$ (i.e., they go into each other under 4-\D rotations).
%mutually unbiased bases?

Finally, $\Db=\Bb \cup W\Bb\cup W^\dag \Bb$ consists of all points on the 3-sphere with integer or half-integer coordinates (i.e., lying in $\tfrac{1}{2}\mathbb{Z}^4$, the union of the integer and the half-integer lattice), or, equivalently, the intersections of $\mathbb{S}^3$ with the four coordinate axes and the eight main diagonals of $\mathbb R^4$. These points are singled out by Eqs.~\eqref{eq:NS} and~\eqref{eq:cubo}, or, more compactly, by
\begin{align}\label{eq:16cell}
(s,x,y,z) \cong s +x\I +y \J+z \K\,,
&\;\;& 
s^2+x^2+y^2+z^2=1\,,
&\;\;& 
s,x,y,z\in \qty{0,\pm\frac{1}{2}, \pm 1}\,,
\end{align}
and, once again, come in antipodal pairs. From a geometric standpoint, $\Db$ is the vertex set of a \emph{24-cell} (an inherently 4-\D regular polytope, with no proper analogue in any other number of spatial dimensions).
From an algebraic standpoint, $\Db$ is the (inner) semidirect product
\begin{equation}\label{eq:2T}
\Db
= \Bb \rtimes \left\langle -W \right\rangle.
\end{equation}
Indeed, we already observed that $\Bb$ is a normal subgroup of $\Db$, and $\Db$ is the union of the cosets $\Bb$, $\Bb (-W)=W\Bb$, and $\Bb (-W)^2=W^\dag \Bb$. This is a realization of the \emph{binary tetrahedral group}, denoted by $2 \sf T$, which is isomorphic to the semidirect product ${\sf Q}_8 \rtimes_\varphi {\sf C}_3$, where ${\sf C}_3=\left\langle c \right\rangle$ is a cyclic group of order 3, and its action on ${\sf Q}_8$, $ \varphi \colon {\sf C}_3 \to {\rm Aut}({\sf Q}_8)$, is determined by the prescription that $\varphi (c)$ cyclically permutes the imaginary units as $\I\mapsto\J\mapsto \K\mapsto\I$. By Eq.~\eqref{eq:conjWIi}, and the projective nature of conjugation, this is precisely the action by conjugation of $-W$ on $I$, $J$, and $K$.

\subsubsection{Relation with the Clifford group.}
Let us briefly relate the structures discussed above to the Clifford group. The one-\q Pauli group is the subgroup of $\U[2]$ generated by the Pauli matrices. 
It therefore consists of all Pauli operators multiplied by the fourth roots of unity, $\sqrt[4]{1}=\qty{\pm 1,\pm \iu}$; it is customary to refer to each of its 16 elements simply as a Pauli. From a purely group-theoretic perspective, the one-\q Clifford group is defined as the normalizer of the Pauli group in $\U[2]$, that is, the subgroup of all unitaries whose action by conjugation maps Paulis to Paulis, or, equivalently, permutes the Pauli matrices $(X,Y,Z)$ --- or their ``special unitary versions'' $(I,J,K)$ --- up to sign.
	
Since any scalar unitary $\e^{\iu \phi}\id_2$ commutes with all unitaries, the Clifford group is infinite. However, when we restrict attention to $\SU$, the relevant setting for this work, the situation becomes particularly transparent: the subgroup $\Bb$ is the intersection of the Pauli group with $\SU$, and its normalizer in $\SU$, denoted $\Clif$, is the intersection of the Clifford group with $\SU$. 
It is immediate that $Q\in \SU$ belongs to $\Clif$ if and only if conjugation by $Q$ permutes $(I,J,K)$ up to sign. From Eq.~\eqref{eq:conjWIi} we know that $W$ and $W^\dag$ satisfy this condition; therefore, $\Db$ is a subgroup of $\Clif$. 
	
In the language of quaternions, a unit quaternion $q$ is Clifford if and only if $( q\I \bar q, q\J \bar q, q\K \bar q)$ is a permutation of $(\I,\J,\K)$ up to sign. A straightforward computation shows that the solutions are precisely the unit quaternions whose components take values in $\{0,\pm 1/2,\pm 1/\sqrt 2,\pm 1\}$ --- in the specific combinations compatible with unit norm. In particular, we recover the 24 integer and half-integer solutions of Eq.~\eqref{eq:16cell}, corresponding to $\Db$, together with 24 additional solutions having exactly two nonzero components of modulus $1/\sqrt 2$. 
	
Consequently, $\Clif$ has 48 elements. The 24 elements of $\Clif\setminus \Db$  are of two distinct types: the six antipodal pairs $\pm (1\pm I_i)/\sqrt 2$, corresponding to rotations by $\pm \pi/2$ about the $i$-th axis, and the six antipodal pairs $\pm (I_i\pm I_j)/\sqrt 2$, with $i\ne j$, corresponding to rotations by $\pm \pi$ about the bisectors of the $(i,j)$ coordinate plane. Loosely speaking, elements of the first type --- which we already encountered in Sec.~\secref{sec:SpeciNorma} when discussing the 2-\des in~\cite{Bennett} --- can be regarded as ``square roots’’ of the inversions about the coordinate axes, whereas those of the second type are essentially generalizations of the Hadamard matrix $H=(X+Z)/\sqrt 2$.
	
Two final observations are in order. First, the above result should not come as a surprise to readers already familiar with the quaternionic description of rotations: the condition that $( q\I \bar q, q\J \bar q, q\K \bar q)$ be a permutation of $(\I,\J,\K)$ up to sign has a clear geometric meaning, singling out precisely all rotations that permute the standard basis vectors of $\mathbb R^3$ again up to sign. Second, in quantum information theory unitary gates are regarded up to an overall phase; in particular, antipodal pairs in $\SU$ correspond to the same physical gate, resulting in 24 Clifford gates.

\begin{figure}[!tb]
	\centering
	\begin{subfigure}[b]{.675\textwidth}
		\centering
		\includegraphics[width=\textwidth]{44FullUp.png}
		\vspace{15pt}
		\caption{} \label{fig:4const}
	\end{subfigure}
	\hfill
	\begin{subfigure}[b]{.280\textwidth}	
		\centering	
		\includegraphics[width=.9\textwidth]{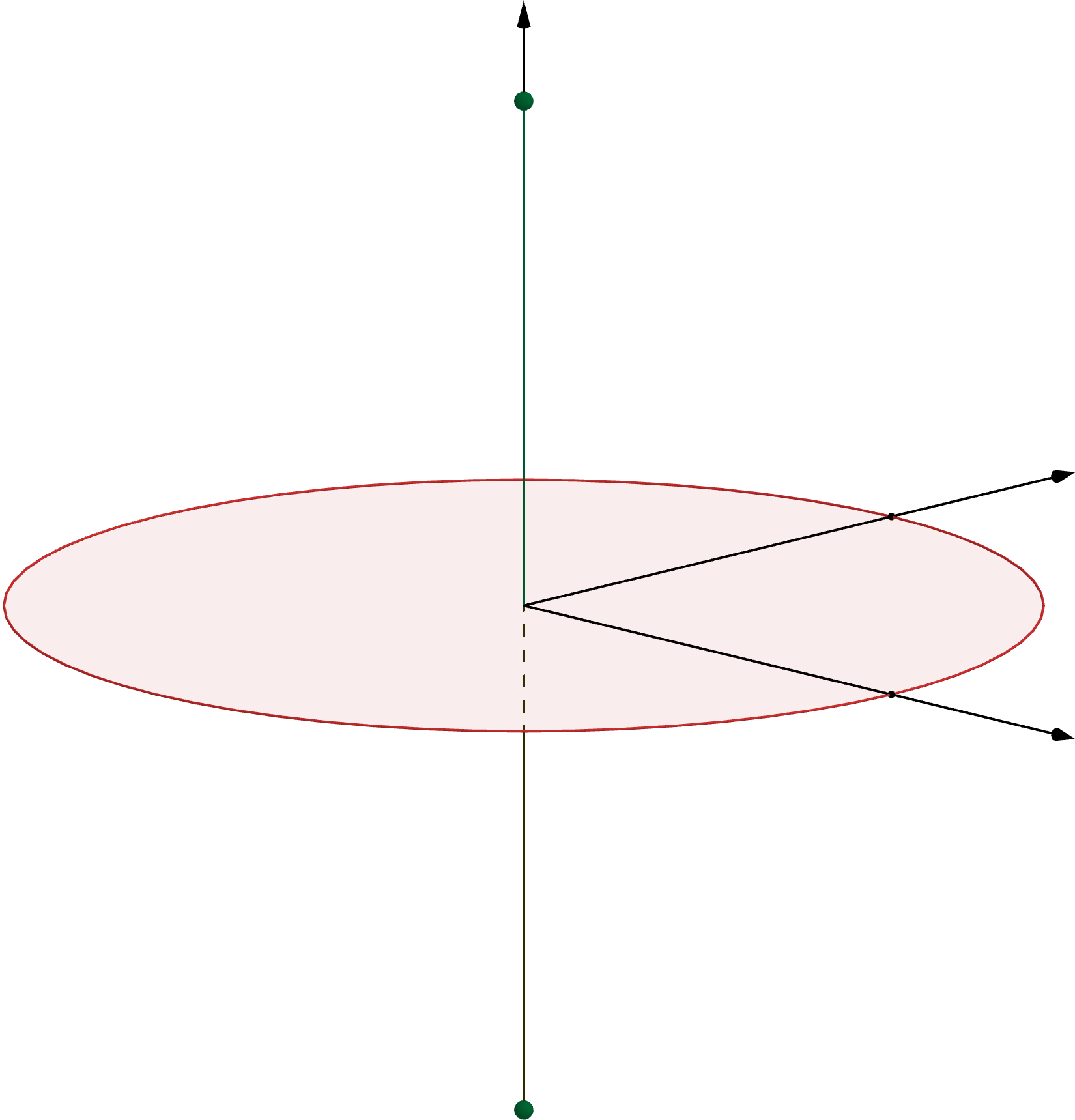}	
		\caption{} \label{fig:4asse}		
		\vspace{5pt}
		\includegraphics[width=.9\textwidth]{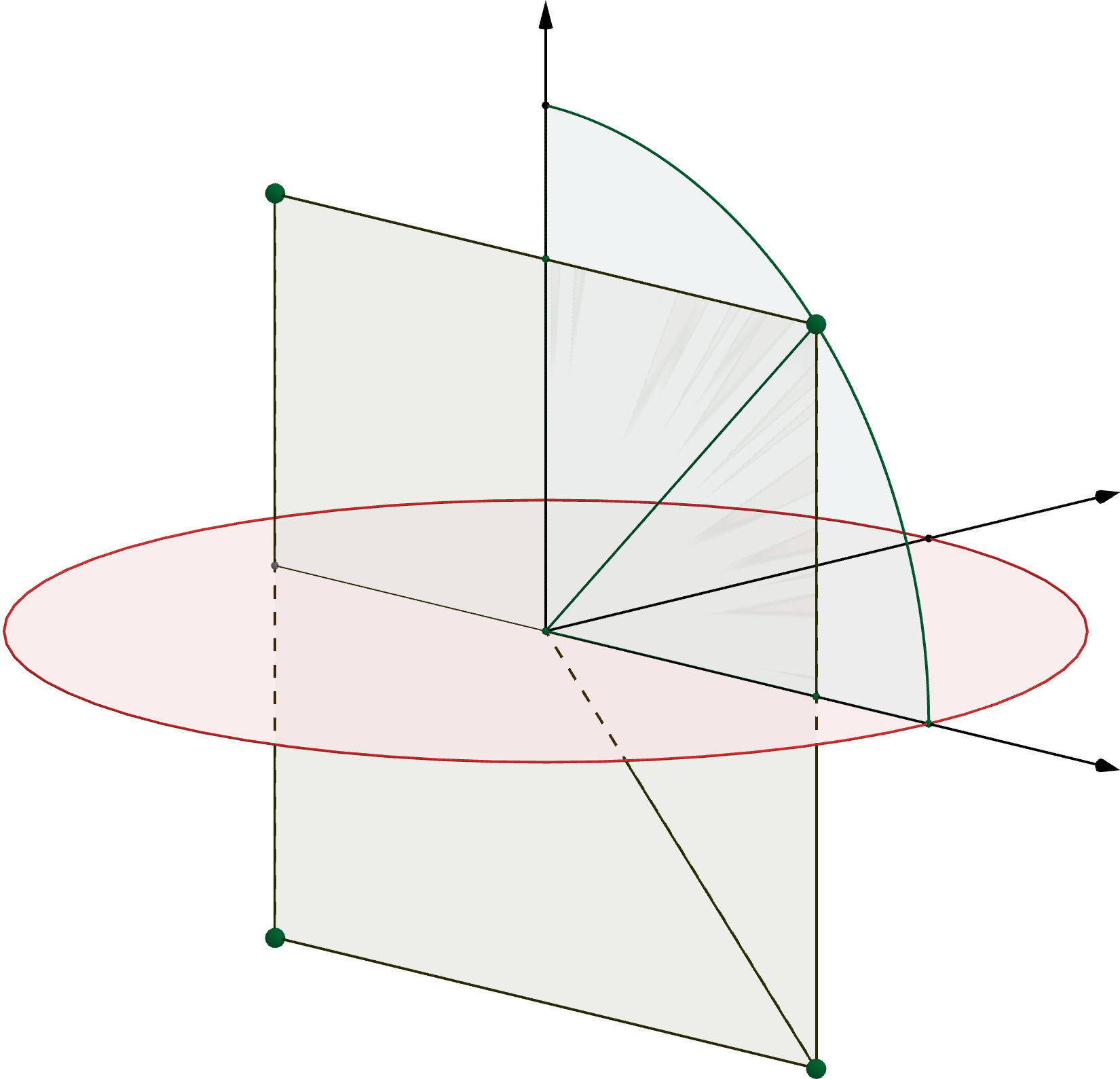}
		\caption{} \label{fig:4quadro}
		\vspace{5pt}	
		\includegraphics[width=.9\textwidth]{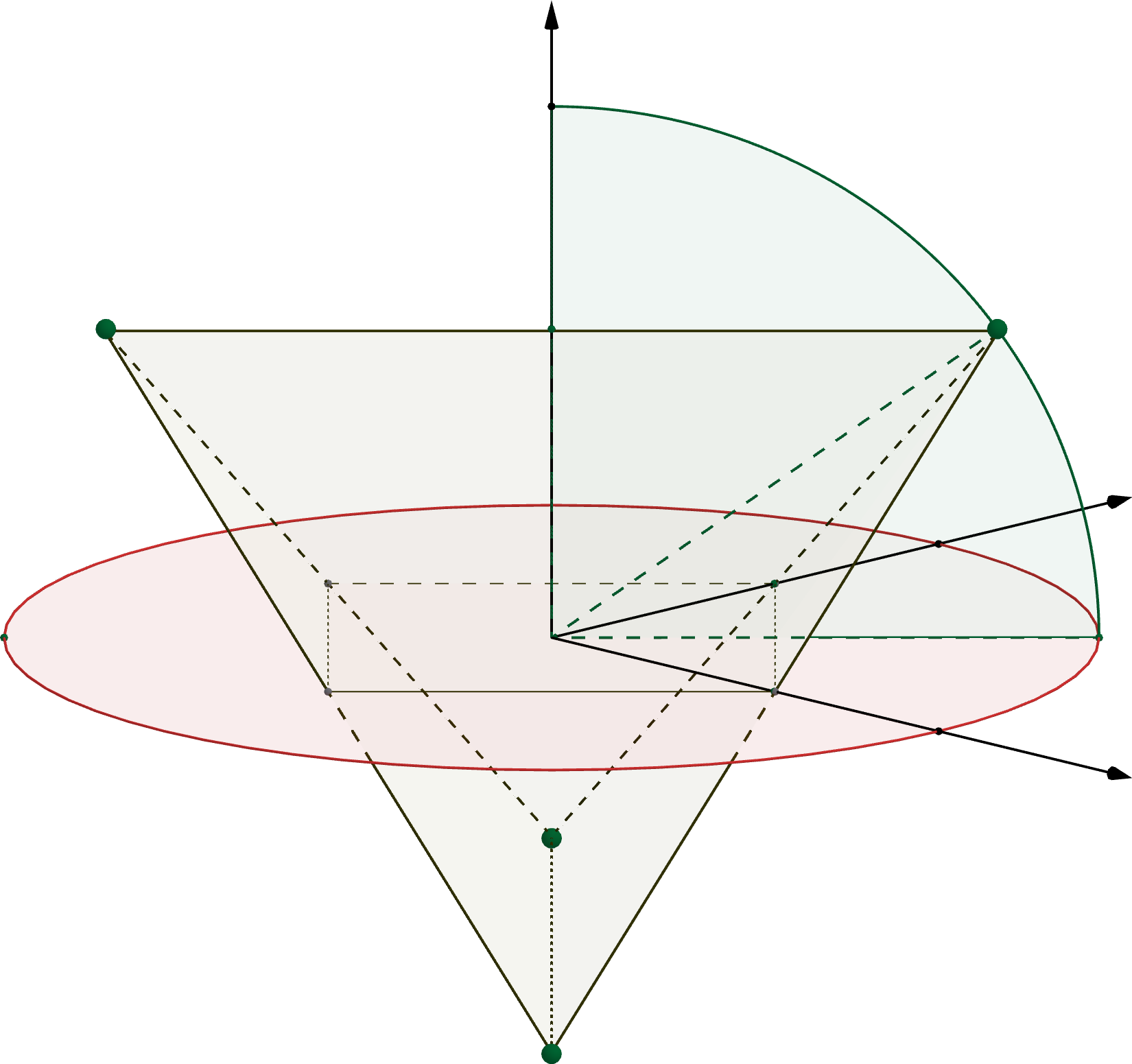}	
		\caption{} \label{fig:4tetra}
	\end{subfigure}		
\caption[LoF entry]{--- Constellation on the Bloch sphere generated by conjugation of a one-\q pure state by the four Pauli operators. Left panel: construction. Conjugation by $X,Y,Z$ reflects the Bloch vector $\vb*r$ about the corresponding coordinate axes. The four resulting vectors (\vertepalla) form a tetrahedron in a rectangular parallelepiped inscribed in the sphere (equatorial plane \equa), with edges parallel to the coordinate axes. The barycenter lies at the origin. Right panels: special symmetric configurations for selected states. For $\vb*r=(0,0,1)$ (i.e., $\rho=\dyad{0}$) the four points collapse pairwise to opposite endpoints of the diameter along the $z$-axis. For $\vb*r = (1,0,1)/\sqrt{2}$ they are the vertices of the square inscribed in the unit circle of the $xz$-plane, with sides parallel to the coordinate axes. For  $\vb*r = (1,1,1)/\sqrt{3}$ they are the vertices of a regular tetrahedron in the cube inscribed in the unit sphere with edges parallel to the coordinate axes.	
}\label{fig:4}
\end{figure}

\begin{figure}[!tb]
	\centering
	\begin{subfigure}[b]{.675\textwidth}
		\centering
		\includegraphics[width=\textwidth]{122FullUp.png}
		%\vspace{1pt}
		\caption{} \label{fig:12const}
	\end{subfigure}
	\hfill
	\begin{subfigure}[b]{.280\textwidth}	
		\centering	
		\includegraphics[width=\textwidth]{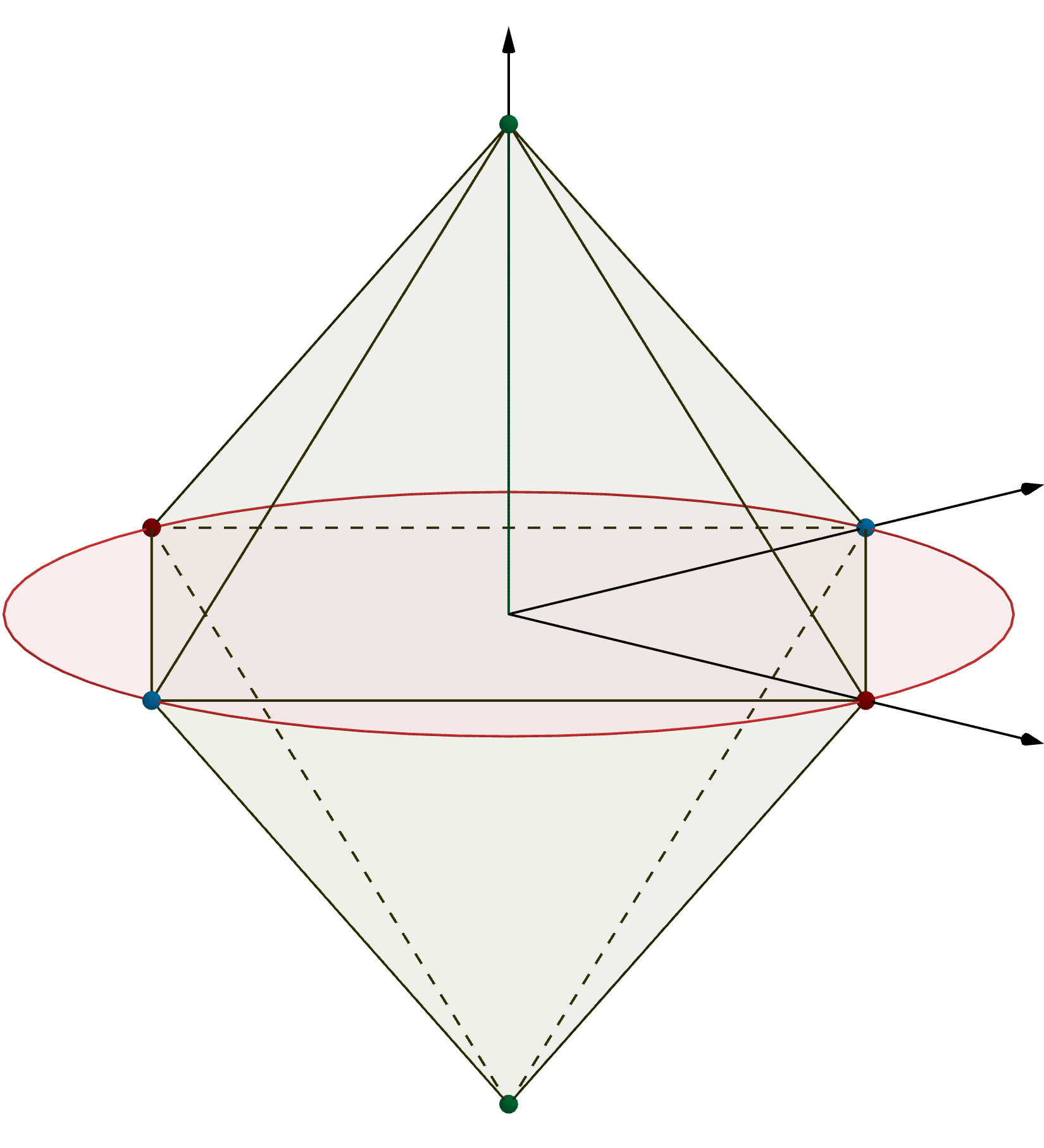}	
		\caption{} \label{fig:12octa}
		\vspace{20pt}	
		\includegraphics[width=\textwidth]{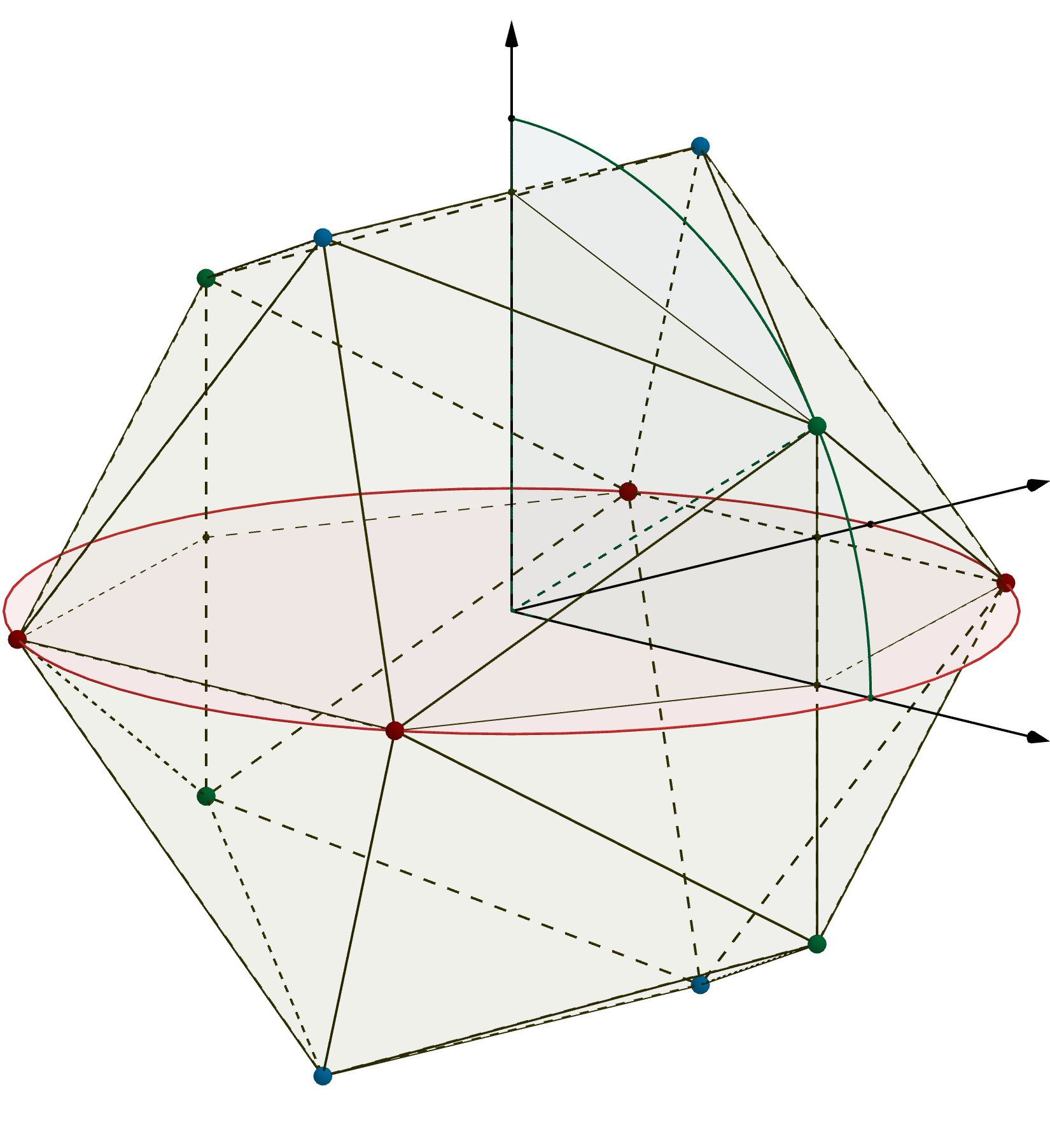}	
		\caption{} \label{fig:12ico}
	\end{subfigure}		
\caption[LoF entry]{--- Constellation on the Bloch sphere generated by conjugation of a one-\q pure state by the twelve unitaries in the 2-\des $\DD=\textcolor{verte}{\B}\cup\textcolor{marun}{W\B}\cup\textcolor{bluu}{W^\dag\B}$. Left panel: the two four-point sets $\vb*r',\vb*x',\vb*y',\vb*z'$ (\marunpalla) and $\vb*r'',\vb*x'',\vb*y'',\vb*z''$ (\bluupalla), obtained from $\textcolor{marun}{W\B}$ and $\textcolor{bluu}{W^\dag\B}$, arise by rotating the Pauli constellation $\vb*r,\vb*x,\vb*y,\vb*z$ (\vertepalla, cf.~\figref{fig:4const}) by $\pm 2\pi/3$ about the oriented axis $(1,1,1)/\sqrt3$ (\aranlinea). Each subset has barycenter at the origin. As in the Pauli case, highly symmetric configurations occur only for selected states (right panels). For $\vb*r=(0,0,1)$ (i.e., $\rho=\dyad{0}$), the points of each subset collapse pairwise to opposite endpoints of a coordinate diameter, yielding the vertices of the inscribed regular octahedron with vertices on the coordinate axes. For $\vb*r = (1,0,\varphi)/\sqrt{1+\varphi^2}$, $\varphi=(1+\sqrt5)/2$ being the golden ratio, they are the vertices of the inscribed regular icosahedron with golden rectangles in the coordinate planes. For any choice of north pole, the pyramidal structure of the icosahedron is governed by the cyclic permutation $\vertepalla\mapsto\marunpalla\mapsto\bluupalla\mapsto\vertepalla$.}
	\label{fig:12}
\end{figure}

\subsubsection{Constellations on the Bloch sphere.}	
Let us consider a one-\q state with Bloch form $\rho = (\id_2 + \vb*r \cdot \vb*X)/2$. Conjugation of $\rho$ by a unitary $U$ amounts to a rotation of the Bloch vector. More precisely, if $U'$ is any normalization of $U$ in $\SU$ (i.e., if $U'=\omega^\ast U$, with $\omega^2=\det U$), then $\vb*r$ is mapped to $\mathscr{R}(U')\vb*r$. 
As a result, the 12 unitaries in $\DD$ and the 24 special unitaries in $\Db$ generate the same constellation of at most 12 points in the unit ball, the only difference being the multiplicity of each point (the multiplicity for $\Db$ being twice that for $\DD$). Analogous considerations apply to $\B$ and $\Bb$. Without loss of generality, we can restrict our attention to pure states, that is, to constellations on the Bloch sphere.
Since $\Bb$ and $\Db$ are regular polytopes in four dimensions, we can ask whether the constellations produced by $\B$ and $\DD$ on the Bloch sphere inherit any form of geometric regularity from $\Bb$ and $\Db$. In other words, we wonder whether regularity descends from $\SU \cong \mathbb S^3$ to the Bloch sphere $\mathbb S^2$.

The answer is shown in~\figrefs[fig:4][fig:12]. In the Bloch representation, the defining property of a 1-\des is that the barycenter of the resulting constellation coincides with the center of the sphere. The constellation generated by $\B$ (\figref{fig:4}) is the vertex set of a tetrahedron inside a rectangular parallelepiped with edges parallel to the coordinate axes, ensuring that the barycenter lies at the origin. For the constellation generated by $\DD$ (\figref{fig:12}) we observe additional structure. In particular, the twelve-point constellation decomposes into three four-point subconstellations: the Pauli constellation and two rotated copies thereof. Each individual barycenter lies at the origin, in accordance with the fact that a 2-\des is also a 1-\des. Moreover, the points can be grouped into four three-point subconstellations lying on parallel planes. In both cases, highly symmetric configurations, such as a Platonic solid, arise only for a discrete set of states.
Loosely speaking, in descending from $\mathbb S^3$ to $\mathbb S^2$ we are discarding one continuous parameter, and therefore a significant portion of the information encoded in the special unitary. Essentially, while a rotation in three dimensions is determined by its action on two orthogonal unit vectors, here we observe its action on a single one. As a simple illustration, suppose we know that a unit vector is fixed by a rotation: then the rotation axis is completely determined (up to sign), but the rotation angle remains completely unspecified.

\subsubsection{Remarks.}\label{sec:AlgeBack}
The most important outcome of the previous sections is that both geometric and algebraic (more precisely, group-the\-o\-ret\-ic) features have emerged. Specifically, once $\DD$ is normalized in $\SU$, and then considered together with its antipodal set, we end up with a regular polytope and a subgroup of $\SU$ --- analogous considerations apply to $\B$. This result is quite interesting, since our construction of $\DD$, as a completion of $\B$ to a 2-\des, has been carried out without imposing any geometric or algebraic constraints.

Both aspects have been relevant to the theory of unitary designs from the very beginning. 
Already in~\cite{gross}, we find a precise formalization of the (geometric) idea %--- after which~\cite{gross} is titled --- 
that unitary designs are sets of ``evenly distributed unitaries''  which motivates the (algebraic) strategy of looking for them among the images of unitary representations of finite groups.
Then we may ask: how well do our findings overlap with these concepts? The honest answer would be: only to a limited extent. Most of our considerations are better seen as peculiar aspects of the specific cases we examined, rather than as structural features of the general problem.

In the first place, the geometric framework on which we relied has to be regarded as exceptional,
essentially because we worked with $d=2$, and with simple finite sets of unitaries, which could be easily phase-shifted to $\SU$. 
Regarding the first point, since $\SU$ is isometric to the 3-sphere, any finite subset of $\SU$ can be interpreted as the vertex set of some inscribed convex polytope. This allows for a powerful geometric interpretation in terms of Euclidean geometry in four dimensions. This feature is a specificity of the case $d=2$, as the only spheres that can be homeomorphic to a compact Lie group are $\mathbb{S}^1$ and $\mathbb{S}^3$. In other words, what is special here is not only the appearance of regular polytopes, but the appearance of polytopes at all. Regarding the second point, phase-shifting an arbitrary set $\Ss \subset \U[2]$ to $\SU$ may be problematic already in the finite case.
Letting $\pi \colon \U[2]\to\PU$ denote the quotient map, in Sec.~\secref{sec:Norma} we defined a normalization of $\Ss$ as the image of any one-to-one map $f \colon \Ss\to\SU$ such that $f(U)\in \pi^{-1}(\pi(U)) \cap \SU$ for all $U\in\Ss$. When $\Ss$ has no two proportional elements (as in the cases we examined), mapping each unitary to the preimage in $\SU$ of its projective class yields injectivity, and there are exactly $2^{\abs{\Ss}}$ such maps. Here the problem is abundance: the choice of a normalization is arbitrary, and the only ``canonical'' object is $\bar\Ss = \pi^{-1}(\pi(\Ss))\cap\SU$, the full preimage of $\Ss$ in $\SU$, with twice the size of $\Ss$. 
On the other hand, as soon as $\Ss$ has three proportional elements, no such map exists: $\Ss$ admits no normalization in $\SU$. From an algebraic standpoint, this difficulty could be bypassed by switching to multisets, but geometrically this would mean allowing vertices with multiplicities.

Even though the results of the previous sections are far from showing a clear path to a geometric characterization of designs, they do offer some hints. All the vertex sets of regular polytopes that we have encountered are designs. At the same time, even after restricting our attention to $\SU$, we encountered designs that are not vertex sets of regular polytopes. For instance, we know that any normalization of $\DD$ is a 2-\des. That is, the full vertex set of a 24-cell and any of the $2^{12}$ sets obtained by selecting one vertex from each antipodal pair are equally good geometric configurations, even though the latter sets are far less symmetric. Reasonably enough, the high-level symmetry considerations involved in the notion of a regular polytope (such as vertex-, edge-, face- and cell-tran\-si\-tiv\-i\-ty) cannot play a leading role. Moreover, observe that  $\Bb$, $W \Bb $, $W^\dag \Bb $, $W \Bb \cup W^\dagger \Bb$, and $\Db$ are all 1-\dess, but only $\Db$ is a 2-\des. This suggests that the geometric characterization we are after may have different realizations at order 1 and 2, with only $\Db$ meeting the latter and stronger condition.

%Incidentally, we may easily concede that the vertices of a regular polytope match our intuitive idea of an ``even distribution'', perhaps on the ground of vertex-tran\-si\-tiv\-i\-ty. However a regular polytope is much more than that: it is also edge-, face-, and cell-tran\-si\-tive, hence only six such polytopes exist in four dimensions. By contrast, it turns out that, unitary $t$-design of $\U$ \emph{of almost any size} exists, for all integers $t$ and $d$. 

In fact, the relevant notion of ``evenly distributed unitaries'' arises from a completely different perspective: a metric characterization of unitary designs, which sits (and is easily obtained) within the framework of quantum channels. Here the word ``geometric'' carries a rather different meaning: it no longer refers to Euclidean geometry, but to the quadratic structure of a finite-dimensional (complex) Hilbert space.
Two unitaries $U,V\in \U$ may be thought of as ``repelling'' each other, and the repulsive energy \emph{of order $t$} is given by $\abs {\inner{U}{V}}^{2t}$, the $t$-th power of the squared modulus of their Hilbert--Schmidt overlap. For a finite set $\Ss \subset \U$, its frame potential of order $t$ is the average of this repulsive energy over all pairs in $\Ss$.
This positive quantity is bounded below by the frame potential of the entire unitary group $\U$, defined by replacing the arithmetic mean over all pairs in $\Ss$ with the corresponding Haar average. Crucially, the lower bound is attained if and only if $\Ss$ is a $t$-\des of $\U$, and, more generally, the gap between the frame potentials of $\Ss$ and $\U$ gives the squared distance between their respective twirling channels of order $t$. In other words, the frame potential not only provides a metric characterization of unitary designs, but also quantifies how close a finite set of unitaries is to being one.
%This notion makes sense for any finite set of unitaries, and depends from the start on the choice of the integer parameter $t$. 

Let us now see how the frame potential ``seamlessly takes on an algebraic role''~\cite{gross}. If $\varrho\colon G \to U(d)$ is a $d$-\D unitary representation of a finite group $G$, the frame potential of its image $\varrho(G)$ (a finite subgroup of $\U$) is just the average of $G\ni g\mapsto \abs{\tr (\varrho(g))}^{2t}=\abs{\chi_\varrho(g)} ^{2t}$, the $t$-th power of the squared modulus of the character of $\varrho$ --- observe also that the computation further simplifies to a weighted average over the conjugacy classes of $G$, as characters are class functions. 
One then compares the result with the universal lower bound, and, if equality holds, $\varrho (G)$ is a unitary $t$-\des. 
Finally, if the center ${\rm Z}(\varrho(G))$ of $\varrho(G)$ is nontrivial, one can construct a $t$-\des of size $\abs{\varrho(G)} / \abs{{\rm Z}(\varrho(G))}$ by first quotienting $\varrho(G)$ by ${\rm Z}(\varrho(G))$, and then selecting one representative from each projective class.

Therefore, a ``more standard'' way to obtain a 2-\des of $\U[2]$ would be along the following steps. We first observe that the frame potential of order 2 of $\U[2]$ is 2.  Then we start examining the frame potentials, of the same order, of the images of 2-\D representations of some finite groups. Eventually, we end up considering
\begin{align}
\varrho\qty(\I,c^\ell)=I (-W)^\ell\,,
&&
\varrho\qty(\J,c^\ell)=J (-W)^\ell\,,
&&
\varrho\qty(\K,c^\ell)=K (-W)^\ell\,,
\end{align}
a unitary representation of the binary tetrahedral group ${\sf Q}_8 \rtimes_\varphi {\sf C}_3$. The image of $\varrho$ is exactly $\Db = \Bb \rtimes \left\langle -W \right\rangle$, whose frame potential of order 2 is 2, hence is a 2-\des. The center of $\Db$ is ${\rm Z}(\Db)=\qty{\pm \id_2}$, and the quotient group is
\begin{multline}
\faktor{\Db}{{\rm Z}\qty(\Db)}
\cong \left\{\pi(\id_2),\,\pi(X),\,\pi(Z),\,\pi(Y),\right.\\
	 \left.\pi(W),\, \pi(W X),\,\pi(W Y),\,\pi(W Z),\,\pi (W^\dag),\,\pi(W^\dag X),\,\pi(W^\dag Y),\,\pi(W^\dag Z) \right\},
\end{multline}
where the right-hand side is a subgroup of $\PU$. The semidirect product~\eqref{eq:2T} descends to the quotient, because its factors, $\pi(\Bb)=\pi(\B)=\{\pi(\id_2),\pi(X_i)\}$ and $\pi(\left\langle -W\right\rangle) =\{\pi(\id_2), \pi(W),\allowbreak \pi(W^\dag)\}$, have a trivial intersection. Now $\pi(\B)$ is a realization of the Klein group (the symmetry group of a rectangle), namely, ${\sf V}=\left\langle a,b\mid a^2=b^2=(ab)^2=e\right\rangle$. Therefore $\pi(\Db)=\pi(\DD)$ realizes the tetrahedral group ${\sf T}\cong {\sf V} \rtimes_\psi C_3$, where  ${\sf C}_3=\left\langle c \right\rangle$ is the cyclic group of order 3, whose action on ${\sf V}$, $\psi \colon {\sf C}_3\mapsto {\rm Aut}({\sf V})$, is determined by prescribing that $\psi(c)$ cyclically permute the nonidentity elements as $a\mapsto b\mapsto ab\mapsto a$. Finally, $\DD$ is obtained by lifting $\pi(\id_2)\mapsto \id_2$, $\pi(X_i)\mapsto X_i$, and so on.

\section{Conclusions}
In this work we have shown how the simplest instances of unitary designs, namely, 1- and 2-\dess of $\U[2]$, can be handled through the theory of angular momentum. This approach offers an obvious advantage: its simplicity. On the one hand, it makes the subject accessible to a broader audience; on the other, it helps build intuition by allowing us to visualize the relevant objects in terms of more familiar ones. Within this framework, we have obtained, in a constructive manner, the $2$-\des originally presented by Bennett, DiVincenzo, Smolin, and Wootters~\cite{Bennett}.

Essentially all the results discussed for the 1-\dess of $\U[2]$ generalize straightforwardly to arbitrary dimension. In particular, the 1-\dess of minimum size in $d$ dimensions can still be characterized as the orthogonal bases of the $d$-\D operator algebra consisting entirely of unitary matrices.

Our construction extending the Pauli basis to a 2-\des of $\U[2]$ presents several remarkable aspects. In particular, as a completion procedure, it is reminiscent of Dankert's proof that the $n$-qubit Clifford group yields a 2-\des of $\U[2^n]$~\cite{Dankert}.  Moreover, even though it is carried out without imposing any geometric or algebraic constraints, it leads to the emergence of both geometric and algebraic features.

\section*{Acknowledgments}
This work was supported by the European Union -- Next Generation EU, under the PNRR MUR project PE0000023 ``National Quantum Science and Technology Institute (NQSTI)'' --- \textit{Missione 4, Componente 2, Investimento 1.3, D.D.\ MUR n.\ 341 del 15.03.2022}.

\printbibliography[heading=bibintoc]\nocite{*}

@book{saku,
	title = {Modern Quantum Mechanics (Revised Edition)},
	author = {Sakurai, J.J.},
	year = {1994},
	publisher = {Addison-Wesley}
}

@book{cohen,
	title={Quantum Mechanics, Volume 2: Angular Momentum, Spin, and Approximation Methods},
	author={Cohen-Tannoudji, C. and Diu, B. and Lalo{\"e}, F.},
	year={2019},
	publisher={Wiley}
}

@book{biede,
	title={Quantum Theory of Angular Momentum: A Collection of Reprints and Original Papers},
	author={Biedenharn, L.C. and Van Dam, H.},
	series={Perspectives in Physics: a Series of Reprint Collections},
	year={1965},
	publisher={Academic Press}
}

@book{watrous,
	title={The Theory of Quantum Information},
	author={Watrous, J.},
	year={2018},
	publisher={Cambridge University Press}
}

@book{renes,
	title = {Quantum Information Theory},
	author = {Renes, J.M.},
	publisher = {De Gruyter Oldenbourg},
	year = {2022},
}

@book{Handbook,
	title={Handbook of Combinatorial Designs},
	author={Colbourn, C.J. and Dinitz, J.H.},
	series={Discrete Mathematics and Its Applications},
	year={2006},
	publisher={Taylor \& Francis}
}

@misc{Preskill,
	author = {Preskill, J.},
	title = {Lecture Notes on Quantum Computation},
	year = {2022},
	url={www.preskill.caltech.edu/ph229/},
	organization = {Caltech},
}

@article{gross,
	author = {Gross, D. and Audenaert, K. and Eisert, J.},
	title = {Evenly distributed unitaries: On the structure of unitary designs},
	journal = {Journal of Mathematical Physics},
	volume = {48},
	number = {5},
	pages = {052104},
	year = {2007},
	month = {05},
	day={17},
}

@misc{gross2,
	author = {Gross, D. and Audenaert, K. and Eisert, J.},
	title = {Evenly distributed unitaries: On the structure of unitary designs},
	version = {2},
	year = {2007},
	month = {05},
	day={13},
	archivePrefix={arXiv},
	eprint={quant-ph/0611002v2},
}

@article{Bannai2009,
	journal = {European Journal of Combinatorics},
	volume = {30},
	number = {6},
	pages = {1392--1425},
	year = {2009},
	note = {Association Schemes: Ideas and Perspectives},
	author = {Bannai, E.~[Eichi] and Bannai, E.~[Etsuko]},
	title = {A survey on spherical designs and algebraic combinatorics on spheres},
}

@article{Bannai2022,
	author = {Bannai, E. and Nakata, Y. and Okuda, T. and  Zhao, D.},
	title = {Explicit construction of exact unitary designs},
	journal = {Advances in Mathematics},
	volume = {405},
	pages = {108457},
	year = {2022},
}

@article{Bennett,
	title = {Mixed-state entanglement and quantum error correction},
	author = {Bennett, C.H. and DiVincenzo, D.P. and Smolin, J.A. and Wootters, W.K.},
	journal = {Phys. Rev. A},
	volume = {54},
	issue = {5},
	pages = {3824--3851},
	numpages = {0},
	year = {1996},
	month = {11},
	publisher = {American Physical Society},
}

@article{DiVincenzo,
	author={DiVincenzo, D.P. and Leung, D.W. and Terhal, B.M.},
	journal={IEEE Transactions on Information Theory}, 
	title={Quantum data hiding}, 
	year={2002},
	volume={48},
	number={3},
	pages={580--598},
}

@article{Werner,
	title = {Quantum states with Einstein-Podolsky-Rosen correlations admitting a hidden-variable model},
	author = {Werner, R.F.},
	journal = {Phys. Rev. A},
	volume = {40},
	issue = {8},
	pages = {4277--4281},
	numpages = {0},
	year = {1989},
	month = {10},
	publisher = {American Physical Society},
}

@misc{Dankert,
	title={Efficient Simulation of Random Quantum States and Operators}, 
	author={Dankert, C.},
	year={2005},
	version = {2},
	archivePrefix={arXiv},
	eprint={quant-ph/0512217v2},
}

@article{Dankert2009,
	title = {Exact and approximate unitary 2-designs and their application to fidelity estimation},
	author = {Dankert, C. and Cleve, R. and Emerson, J. and Livine, E.},
	journal = {Phys. Rev. A},
	volume = {80},
	issue = {1},
	pages = {012304},
	numpages = {6},
	year = {2009},
	month = {7},
	publisher = {American Physical Society},
}

@article{Scott,
	year = {2008},
	month = {1},
	publisher = {},
	volume = {41},
	number = {5},
	pages = {055308},
	author = {Scott, A.J.},
	title = {Optimizing quantum process tomography with unitary 2-designs},
	journal = {Journal of Physics A: Mathematical and Theoretical},
}

@article{RoyScott,
	author = {Roy, A. and Scott, A.J.},
	title = {Unitary designs and codes},
	journal = {Designs, Codes and Cryptography},
	number = {1},
	pages = {13--31},
	volume = {53},
	year = {2009},
	month={10},
	day={1},
}

@article{Emerson_2005,
	year = {2005},
	month = {09},
	publisher = {},
	volume = {7},
	number = {10},
	pages = {S347},
	author = {Emerson, J. and  Alicki, R. and Życzkowski, K.},
	title = {Scalable noise estimation with random unitary operators},
	journal = {Journal of Optics B: Quantum and Semiclassical Optics}
}

@article{Zhu,
	title = {Multiqubit Clifford groups are unitary 3-designs},
	author = {Zhu, H.},
	journal = {Phys. Rev. A},
	volume = {96},
	issue = {6},
	pages = {062336},
	numpages = {7},
	year = {2017},
	month = {12},
	publisher = {American Physical Society},
}

@misc{ZhuGracefully,
	title={The Clifford group fails gracefully to be a unitary 4-design}, 
	author={Zhu, H. and Kueng, R. and Grassl, M. and Gross, D.},
	year={2016},
	archivePrefix={arXiv},
	eprint={quant-ph/1609.08172},
}

@article{Webb,
	author = {Webb, Z.},
	title = {The Clifford group forms a unitary 3-design},
	year = {2016},
	issue_date = {November 2016},
	publisher = {Rinton Press, Incorporated},
	volume = {16},
	number = {15–16},
	journal = {Quantum Info. Comput.},
	month = {11},
	pages = {1379--1400},
	numpages = {22},
}

@misc{Gottesman1998,
	title={The Heisenberg Representation of Quantum Computers}, 
	author={Gottesman, D.},
	year={1998},
	eprint={quant-ph/9807006},
	archivePrefix={arXiv},
}

@article{seymour,
	author = {Seymour, P.D. and Zaslavsky, T.},
	title = {Averaging sets: A generalization of mean values and spherical designs},
	journal = {Advances in Mathematics},
	volume = {52},
	number = {3},
	pages = {213--240},
	year = {1984},
	url = {www.sciencedirect.com/science/article/pii/0001870884900227},
}

@article{Chau,
	author={Chau, H.F.},
	journal={IEEE Transactions on Information Theory}, 
	title={Unconditionally secure key distribution in higher dimensions by depolarization}, 
	year={2005},
	volume={51},
	number={4},
	pages={1451--1468},
}

@article{Aravind,
	title = {The ``twirl'', stella octangula and mixed state entanglement},
	journal = {Physics Letters A},
	volume = {233},
	number = {1},
	pages = {7-10},
	year = {1997},
	author = {P.K. Aravind},
}

@article{Haferkamp2022,
	title={Efficient Unitary Designs with a System-Size Independent Number of Non-Clifford Gates},
	volume={397},
	number={3},
	journal={Communications in Mathematical Physics},
	publisher={Springer Science and Business Media LLC},
	author={Haferkamp, J. and Montealegre-Mora, F. and Heinrich, M. and Eisert, J. and Gross, D. and Roth, I.},
	year={2022},
	month={11}, 
	pages={995–1041} 
}

@book{coxeter,
	author = {Coxeter, H.S.M.},
	publisher = {Dover Publications},
	title = {Regular Polytopes},
	year = {1973},
}

@article{stair,
	doi = {10.1088/2058-9565/ae1d4b},
	url = {https://doi.org/10.1088/2058-9565/ae1d4b},
	year = {2025},
	month = {12},
	publisher = {IOP Publishing},
	volume = {11},
	number = {1},
	pages = {015022},
	author = {Scala, Giovanni and Bera, Anindita and Sarbicki, Gniewomir},
	title = {Entanglement detection via third-order local invariants from randomized measurements},
	journal = {Quantum Science and Technology}
}

\end{document}